\documentclass[a4paper,11pt]{article}
\pdfoutput=1 

\usepackage{jheppub} 

\usepackage[T1]{fontenc} 
\usepackage{xcolor}
\usepackage{epigraph}
\usepackage{comment}

\usepackage{amssymb}
\usepackage{amsmath}
\usepackage{graphicx}
\usepackage{framed}
\usepackage{xltabular}
\usepackage{float}
\usepackage{caption}
\usepackage{subcaption}
\newcolumntype{C}{>{$}c<{$}}

\usepackage{changepage}

\usepackage{multirow}

\usepackage{booktabs}

\usepackage{tabstackengine}
\setstacktabbedgap{1ex}

\newwrite\todofile
\immediate\openout\todofile=\jobname.tdo

\newcounter{todocounter}

\newcommand{\printtodos}{
        \section*{To-Do List}
        \immediate\closeout\todofile
        \input{\jobname.tdo}
}

\usepackage{xparse}

\ExplSyntaxOn

\NewDocumentCommand{\gr}{m}
{
  \grayscale_process:n { #1 }
}

\cs_new_protected:Nn \grayscale_process:n
{
  \peek_meaning_ignore_spaces:NTF ] 
    {{{;#1}} } 
    {{{#1;}}~} 
}

\ExplSyntaxOff

\newcommand{\la}[1]{\label{#1}}
\newcommand{\eq}[1]{\eqref{#1}}

\def\[{\left[}
\def\]{\right]}
\def\({\left(}
\def\){\right)}
\def\d{\partial}
\newcommand{\beq}{\begin{equation}}
\newcommand{\eeq}{\end{equation}}
\newcommand\beqa{\begin{eqnarray}}
\newcommand\eeqa{\end{eqnarray}}

\newcommand{\phipar}{\Phi_{||}}
\newcommand{\phiperp}[1]{\Phi_\perp^{#1}}

\def\stID(#1,#2,#3,#4,#5){{}_{#1}[0\;#2\;#3\;#4]_{#5}}

\title{Probing Line Defect CFT with Mixed-Correlator Bootstrability}

\emailAdd{andrea.cavaglia$\bullet$unito.it} 
\emailAdd{nikolay.gromov$\bullet$kcl.ac.uk}
\emailAdd{julius.julius$\bullet$phys.ens.fr}
\emailAdd{michelangelo.preti$\bullet$stonybrook.edu}
\emailAdd{nika.sokolova$\bullet$kcl.ac.uk}

\author{Andrea Cavagli\`a$^{[00]}$}

\author{Nikolay Gromov$^{[02]}$}

\author{Julius Julius$^{[20]}$}

\author{Michelangelo Preti$^{[00],[01]^{-},[01]^{+}}$}
\author{Nika Sergeevna Sokolova$^{[02]}$}

\affiliation{
$^{[00]}$Department of Physics, University of  Turin, Via P. Giuria 1, 10125, Turin, Italy
}

\affiliation{
$^{[02]}$Mathematics Department, King's College London,
The Strand, London WC2R 2LS, UK
}

\affiliation{
$^{[20]}$Laboratoire de Physique de l'\'Ecole Normale Sup\'erieure, Universit{\'e} Paris Sciences et Lettres, Centre National de la Recherche Scientifique, Sorbonne Universit{\'e}, Universit{\'e} Paris Cit{\'e}, 24 rue Lhomond, 75005 Paris, France}

\affiliation{$^{[01]^{-}}$C. N. Yang Institute for Theoretical Physics, Stony Brook University, Stony Brook, New York 11794, USA
}

\affiliation{$^{[01]^{+}}$Simons centre for Geometry and Physics, Stony Brook University, Stony Brook, New York 11794, USA
}

\abstract{
We continue our study of the defect CFT on a Maldacena--Wilson line in 
$\mathcal{N}=4$ Super--Yang--Mills theory using Bootstrability---the conformal 
bootstrap supplemented with exact integrability data. In this paper, we extend 
this program to charged sectors of the theory, considering a mixed-correlator 
setup first introduced by Liendo, Meneghelli, and Mitev. The exact spectrum in all 
channels is given by integrability at any coupling. Additionally, we use exact 
expressions for some structure constants fixed by localisation and leverage the 
exact discrete symmetries of the theory. We analyse the remaining data with the 
numerical bootstrap, developing an algorithm optimised to scan over a large 
multidimensional space of OPE coefficients and carve the allowed region with the \textquotedblleft cutting surface\textquotedblright{} procedure. We compute upper and lower bounds 
for 12 OPE coefficients for several values of the coupling. Our results are 
sharp for the lowest states in each sector but become quite wide for the excited 
states due to their near degeneracy. This highlights the need for studying the 
system with non-protected external states and for further input from integrability 
in the form of integrated correlators of non-protected operators.
}

\begin{document} 
\maketitle
\flushbottom

\section{Introduction and summary of results}
\label{sec:intro} 

With the advance of integrability methods, more and more exact or very precise non-perturbative data becomes available for $\mathcal{N}=4$ Super-Yang-Mills (SYM) theory in the planar limit.
In particular, the spectrum is the most well understood part of the theory, and the methods based on Quantum Spectral Curve (QSC) \cite{Gromov:2013pga, Gromov:2014caa} 
can readily generate the spectrum analytically at weak coupling~\cite{Marboe:2017dmb,Marboe:2018ugv}, or numerically at finite coupling~\cite{Gromov:2015wca,Hegedus:2016eop,Gromov:2023hzc,Ekhammar:2024rfj}.
Concerning OPE coefficients, whereas they are well understood at weak coupling thanks to the hexagon approach~\cite{Basso:2015zoa}, the finite coupling regime for finite length operators remains challenging.
The methods based on Separation of 
Variables~\cite{Jiang:2015lda,Gromov:2016itr,Cavaglia:2018lxi,Giombi:2018hsx,Cavaglia:2019pow,Gromov:2019wmz,
Cavaglia:2021mft,Gromov:2022waj,Bercini:2022jxo, Ekhammar:2023iph} or hybrid approaches combining the QSC data with the hexagon method~\cite{Basso:2022nny} could lead to very promising results, but they are still in the development stage. 
There has also been some recent progress in understanding the OPE coefficients of finite length operators at strong coupling~\cite{Alday:2022uxp,Alday:2022xwz,Gromov:2023hzc,Alday:2023mvu,Fardelli:2023fyq,Julius:2023hre,Julius:2024ewf}.

On the other hand, the OPE coefficients are not completely independent from the spectrum, and conformal bootstrap methods can be used to constrain them.
In this paper we continue our program of Bootstrability~\cite{Cavaglia:2021bnz,Cavaglia:2022qpg,Cavaglia:2023mmu}, where we try to answer the question of to 
which extent the full CFT data in theories like $\mathcal{N}=4$ SYM can be obtained by knowing only the spectrum of the theory itself and its deformations.
This question currently is being under active investigation by other groups in different setups~\cite{Caron-Huot:2022sdy,Caron-Huot:2024tzr} and methods~\cite{Niarchos:2023lot, Trenta:2024nni}. Furthermore, the information about the spectrum can also be very useful for the analytic bootstrap at strong coupling, where it helps to check or further constrain the assumptions on an analytic ansatz for 4-point functions~\cite{Alday:2024xpq}.

As $\mathcal{N}=4$ SYM is a gauge theory, its observables are not limited to the local operators, but also include other gauge-invariant observables like the Wilson loops. Strictly speaking, the Wilson loops are the only non-trivial observables in the strictly planar limit in addition to the two-point functions of local operators, as the OPE coefficients of local operators are $1/N_c$ suppressed. 

In particular, the supersymmetric Maldacena-Wilson loops,  defined with a specific coupling to a scalar field of the theory and on contours conformally equivalent to straight lines, are the simplest and the most understood.
One can study them using the point of view of a defect CFT living on the Wilson line~\cite{Drukker:2006xg,
Cooke:2017qgm,Giombi:2017cqn}
with a variety of non-perturbative approaches such as supersymmetric localisation~\cite{Giombi:2018qox,Giombi:2018hsx},
integrability~\cite{Correa:2018fgz, Grabner:2020nis,Julius:2021uka}
and
conformal bootstrap~\cite{Liendo:2018ukf,Ferrero:2021bsb}.

The states in the defect CFT correspond to the insertions of local operators on the Wilson line, and their OPE coefficients
are not suppressed even in the planar limit.
The integrability of this class of Wilson lines~\cite{Correa:2012at, Correa:2012hh, Correa:2012nk, Drukker:2012de} manifests in the access to the high precision spectral data~\cite{Gromov:2015dfa, Grabner:2020nis,Julius:2021uka,Cavaglia:2021bnz,QSClineToAppear}. A rich source of analytic data beyond the spectrum is also available from the four-point correlation functions, which were computed analytically perturbatively up to certain orders at both weak ~\cite{Kiryu:2018phb,   Barrat:2021tpn, Barrat:2022eim, Artico:2024wut} and strong ~\cite{Liendo:2018ukf, Giombi:2009ds, Ferrero:2021bsb, Ferrero:2023znz, Ferrero:2023gnu} coupling, which one can use to test other methods.

In this paper, we continue the Bootstrability program for this defect CFT,  using the numerical conformal bootstrap  \cite{Rattazzi:2008pe, Poland:2018epd, Simmons-Duffin:2015qma} supplemented with exact integrability data for the spectrum.   
 We focus on the mixed-correlator setup introduced in \cite{Liendo:2018ukf}. The main challenge of the numerical conformal bootstrap in this setup is that we need to scan a multidimensional space of CFT data to find a consistent solution.
As the dimensionality of the space of OPE coefficients can grow to as high as the number of states exchanged in the system, 
the problem becomes computationally challenging and requires a special algorithm to scan the space efficiently.

Let us summarize the main results of this paper. First, we discuss the consequences of discrete symmetries \cite{Cavaglia:2023mmu} of the theory for the mixed-correlator setup. We show a significant simplification in the  bootstrap problem 
 compared to that formulated  originally in \cite{Liendo:2018ukf}, which reduces the search space of OPE coefficients by roughly half. In addition, we obtained a new 
analytical result for one of the OPE coefficients involving three BPS operators,
which are non-trivial, exact functions of the 't Hooft coupling, derived
using localisation arguments.  Furthermore, by exploiting new spectral data for charged sectors, we perform a numerical bootstrap study of the mixed-correlator system using the spectrum of low lying states as input and leveraging all the discrete symmetries. To achieve this, we use a bespoke algorithm, largely inspired by~\cite{MiniCourse2023}, for scanning efficiently over a large set of OPE coefficients. As an outcome, we obtain new rigorous bounds for several new OPE coefficients involving charged non-protected operators.

The paper is organized as follows. In section \ref{sec:CFT} we review the CFT living on the Maldacena-Wilson line. In particular, we specify which operators are present in the theory and what are their correlation functions. In sections \ref{sec:Integrability} and \ref{sec:QSC} we focus on non-protected operators in long multiplets and outline how their dimensions can be found using integrability. We provide the spectrum of first low-lying operators used in the bootstrap.
In section \ref{sec:Bootstrability} we describe the mixed-correlator setup and discuss how discrete symmetries simplify it. 
In section \ref{sec:Algorithm} we describe our algorithm for scanning over OPE coefficients. 
In section \ref{sec:Results} we present the results of the numerical bootstrap study. 
Finally, in section \ref{sec:Conclusions} we conclude and discuss future directions. Supplementary material is provided in appendices.

\section{Maldacena-Wilson defect CFT}
\label{sec:CFT}
As reviewed in the introduction, in this paper we study the CFT data from integrability and conformal bootstrap perspective for the 
one-dimensional defect superconformal field theory (dCFT) living on the Maldacena-Wilson line (MWL) \cite{Maldacena:1998im, Erickson:2000af} in $\mathcal{N}=4$ Super-Yang-Mills theory. 
Below we review the setup and specify which observables we focus on.

\paragraph{The Maldacena-Wilson line.} The MWL is defined as ${\rm Tr}\;W_{-\infty}^{+\infty}$, where
\begin{equation}
\label{eqn:MWLdef}
 W_{t_1}^{t_2} \equiv \text{P}\  \text{exp} \int_{t_1}^{t_2} dt \ \bigg[i\  A_{\mu}(t) \dot{x}^{\mu} + \Phi_{||}(t)|\dot{x}|\bigg]\;.
\end{equation}
The $W_{t_1}^{t_2}$ is the path-ordering $\text{P}$ (in the order: latest first) 
along the $(t_1, t_2)$ segment of the line. 
Here $x(t)$ is the parametrisation of the straight line, 
and $A_{\mu}$ and $\Phi_{||}$ are the fields of 
4D $\mathcal{N}=4$ SYM theory.
There we split the six scalar fields 
of $\mathcal{N}=4$ SYM
$\Phi^{i}$ $i = 1\dots6$
into two groups: one
$\Phi_{||}\equiv \Phi^6$ is the same scalar as appearing in the definition of the
 line, 
and then another five fields we denote as $\Phi^{i}_{\perp}$ 
with $i = 1, \dots, 5$. 

Symmetries of the full 4D $\mathcal{N}=4$ SYM 
theory are broken by the line defect, 
however, partial symmetries are still preserved. 
Let us summarise these partial symmetries:

\begin{itemize}
    \item \texttt{Spacetime symmetry}. The 4D conformal symmetry is broken in the presence of the line. There is the 1D conformal group $\mathrm{SO}(1,2)$ preserved on the line as well as the $\mathrm{SO}(3)$ group of spacetime rotations orthogonal to the line. Overall spacetime symmetry group is then $\mathrm{SO}(1,2)\times \mathrm{SO}(3)$.
    \item R-\texttt{symmetry}. Five scalar fields $\Phi^{i}_{\perp}$ with $i = 1, \dots, 5$ perpendicular to the line preserve the $\mathrm{SO}(5)$ R-\text{symmetry}.
    \item {\texttt{Supersymmetry}.} The MWL is a $\frac{1}{2}$-\text{BPS} observable which preserves a half of the supercharges. 
\end{itemize}
The full symmetry group of the 1D dCFT living on the MWL is $\mathrm{OSp}(4^*|4)$ which is a subgroup of the symmetry group $\mathrm{PSU}(2,2|4)$ of $\mathcal{N}=4$ SYM. 

\paragraph{Correlation functions of operators inserted to the line.} Let us specify on which observables of 1D dCFT we focus. We consider correlation functions of operators inserted along the line defined as 
\begin{equation}
\label{eq:corrs}
    \langle \langle O_{1}(x_1) O_{2}(x_2)\dots O_{n}(x_n)  \rangle \rangle = \frac{\langle \text{Tr}\;{\rm P}[  O_{1}(x_1) O_{2}(x_2) \dots O_{n}(x_n) W_{-\infty}^{+\infty} ] \rangle }{\langle \text{Tr}\;W_{-\infty}^{+\infty} \rangle},
\end{equation}
where the operators are path-ordered as above \eqref{eqn:MWLdef}, and $O(t)$ are composite operators constructed from the fields of the theory. Correlation functions \eqref{eq:corrs} are non-trivial functions on 't Hooft coupling $g$. 

Due to the conformal symmetry of the theory, the most fundamental quantities are the two- and three-point functions as they are the building blocks of all other correlation functions. These functions can be found to have the following form
\begin{equation}\begin{split}
\label{eq:OPEdef}
    \langle \langle O_i (x_1) \bar{O}_j (x_2) \rangle \rangle &= \frac{\delta^{ij}}{|x_{12}|^{2 \Delta_i}},\\
    \langle \langle O_i (x_1) O_j (x_2) O_j (x_3) \rangle \rangle \! &=\! \frac{C_{ijk}}{|x_{12}|^{\Delta_i + \Delta_j - \Delta_k}|x_{13}|^{\Delta_i + \Delta_k - \Delta_j}|x_{23}|^{\Delta_j + \Delta_k - \Delta_i}} \  \ x_1 \!<\! x_2 \!<\! x_3,
\end{split}\end{equation}
where we choose the standard normalisation\footnote{The $\bar O$ is defined via Hermitian  conjugation and 
reflection $\d_t\to -\d_t$, needed for positivity.}.
In \eqref{eq:OPEdef} $\Delta_i$ are scaling dimensions of the operators $O_i$ and $C_{ijk}$ are the operator product expansion (OPE) coefficients. The scaling dimension  in general depends on the coupling and can be separated as $\Delta = \Delta_0 + \gamma(g)$, where $\Delta_0$ is the bare dimension of operators and $\gamma(g)$ is the anomalous dimension. We point out 
that for the operators with non-trivial R-charges the above normalization may not be the most convenient and we describe those in case by case basis below in section~\ref{crossingOpe}.

Due to superconformal symmetry, we have further simplifications. Operators can be assembled into supermultiplets so that their anomalous dimension is the same within the multiplet. Furthermore, the $3$-point correlators of different representatives of the multiplet are related by supersymmetry.
Let us describe the details of the supermultiplets are present this CFT, relevant for this paper. 

\paragraph{Supermultiplets.} They are specified by the quantum numbers of the superprimary operators: the scaling dimension $\Delta$ of the $\mathrm{SO}(1,2)$ 1D conformal group, the transverse spin spin $s$ of $\mathrm{SO}(3)$ rotations around the line and the Dynkin labels $[a,b]$ of the $\mathrm{SP}(4) \cong \mathrm{SO}(5)$ R-symmetry.\footnote{Note that the irrep of $\mathrm{SP}(4)$ corresponding to the Dynkin labels $[a,b]$ corresponds to the irrep $[b,a]$ of $\mathrm{SO}(5)$.} For the current setup the full classification was done in \cite{Liendo:2016ymz, Agmon:2020pde}.

There are three sets of supermultiplets: \textit{short}, \textit{semi-short} and \textit{long}. 
The semi-short multiples are only relevant at zero coupling since normally become long at finite $g$, so we do not consider those separately in this paper.
The scaling dimension of operators in short multiplets is protected by the supersymmetry as they preserve a half of the supercharges. They are denoted as  $\mathcal{B}_k$
and have zero anomalous dimension and are constrained to have the following quantum numbers
\begin{equation}
    \{ \Delta, s, [a,b] \} = \{ k, 0, [0, k] \},\ \ \ \  k \in  \{1,2,3,\dots \}.
\end{equation}
In our setup, we mainly focus on the first two short multiplets  with $k=1$ and $2$, even though $3$ and $4$ will also play some minor role. The first multiplet $\mathcal{B}_1$ contains the superconformal primary operators $\Phi^{i}_{\perp}$, it has the protected scaling dimension $\Delta = 1$ and is also known as the \textit{displacement} multiplet. The second multiplet $\mathcal{B}_{2}$ contains the superconformal primary operators $\Phi_{\perp}^{\{ i} \Phi_{\perp}^{j \}} - \frac{1}{5} \delta^{i j} (\Phi_{\perp} \cdot \Phi_{\perp})$.

Long multiplets can, in principle, have any quantum numbers $\{ \Delta, s, [a,b] \}$ as long as the unitarity bound is not violated\footnote{The unitarity bound for long multiplets in this theory is $\Delta > \frac{1}{2}s + a + b + 1$}. 
In this paper we only consider the long multiplets with zero spin $s=0$, and thus we denote them as $\mathcal{L}^{\Delta}_{ [a,b]}$, with only three quantum numbers. Specifically, long multiplets which are relevant for our setup are $\mathcal{L}_{[0,0]}^{\Delta}$, $\mathcal{L}_{[0,1]}^{\Delta}$, $\mathcal{L}_{[0,2]}^{\Delta}$ and $\mathcal{L}_{[2,0]}^{\Delta}$. 
Long multiplets may become semi-short if their scaling dimension saturates the unitarity bound which usually happens at strictly $g=0$.

\subsection{Integrability Description of the Spectrum}
\label{sec:Integrability}

The 1D dCFT on the MWL is integrable~\cite{Drukker:2006xg,
Grabner:2020nis} in the planar limit, 
which means that the non-trivial scaling dimensions of the operator insertions can in principle be computed analytically.
There have been a variety of results applying integrability methods in this context, analytically in the weak coupling regime~\cite{Correa:2018fgz}, numerically in the finite coupling regime~~\cite{Grabner:2020nis,Julius:2021uka,Cavaglia:2021bnz,QSClineToAppear}, as well as in a special integrability preserving limit called the ladders limit~\cite{Cavaglia:2018lxi,Gromov:2021ahm}. 
The integrability is inherited from the parent 4D $\mathcal{N}=4$ SYM theory, 
where the non-perturbative spectrum of long multiplets can be described 
using the Quantum Spectral Curve (QSC)~\cite{Gromov:2013pga}. 
In this section, we provide detailed description of
the low-lying spectrum of long multiplets 
and discuss their description via the QSC.

First of all, the dimensions $\Delta$ of long multiplets $\mathcal{L}_{[a,b]}^{\Delta}$ become integer and degenerate both at weak ($g = 0$) and strong ($g \to \infty$) coupling.
The degeneracy of states can be computed by free fields counting~\cite{Ferrero:2023znz}. We collect part of this result in Table~\ref{tab:countingweakstrong}.

\begin{table}[t]
    \centering
    \begin{tabular}{|c||c|c|c|c|c|c|}
        \hline
         \multicolumn{7}{|C|}{\text{Degeneracy at } g=0}\\
        \hline
        $\Delta_0$ & 1 & 2 & 3 & 4 & 5 & 6 \\
        \hline
        $\mathcal{L}_{[0,0]}^{\Delta}$  &1 & 2 & 6 & 25 & 128 & 758 \\
        \hline
        $\mathcal{L}_{[0,1]}^{\Delta}$  & & 2 & 6 & 28 & 167 &  1134 \\
        \hline
        $\mathcal{L}_{[0,2]}^{\Delta}$  & &  & 3 & 12 & 76 & 588 \\
        \hline
        $\mathcal{L}_{[2,0]}^{\Delta}$  & &  & 2 & 16 & 128 & 1038 \\
        \hline
    \end{tabular}
    \qquad\qquad
    \begin{tabular}{|c||c|c|c|c|c|c|}
        \hline
         \multicolumn{7}{|C|}{\text{Degeneracy at } g=\infty}\\
        \hline
        $\Delta_{\infty}$ & 2 & 3 & 4 & 5 & 6 & 7 \\
        \hline
        $\mathcal{L}_{[0,0]}^{\Delta}$  &1 & 0 & 2 & 0 & 4 & 1 \\
        \hline
        $\mathcal{L}_{[0,1]}^{\Delta}$  & & 1 & 1 & 2 & 2 &  6 \\
        \hline
        $\mathcal{L}_{[0,2]}^{\Delta}$  & &  & 1 & 1 & 3 & 2 \\
        \hline
        $\mathcal{L}_{[2,0]}^{\Delta}$  & &  &  & 1 & 0 & 3 \\
        \hline
    \end{tabular}
    \caption{Degeneracy of states of the long multiplets exchanged in  our mixed-correlator system, at $g=0$ and $g=\infty$. Data taken from~\cite{Ferrero:2023znz}. }
    \label{tab:countingweakstrong}
\end{table}

Let us give some flavour of what are these long multiplets at weak coupling. 
For the scalar operators, the authors of \cite{Correa:2018fgz} 
successfully derived explicit form of the 1-loop dilatation operator, in the same fashion of what it was done  for the single-trace operators problem~\cite{Minahan:2002ve}. 
The problem of finding the spectrum  of long multiplets at the one-loop order in this sector 
is then equivalent to finding eigenstates and eigenvalues of an explicit Hamiltonian.
Therefore, we are able to build explicitly some of the operators. The Hamiltonian is only known for operators composed exclusively of scalars without spacetime derivatives (we denote states composed without derivatives along the MWL line at $g\to 0$ as ``highest twist'' states).
\begin{figure}[t]
\centering
    \begin{subfigure}[b]{0.45\textwidth}
    \centering
    \includegraphics[width=7cm]{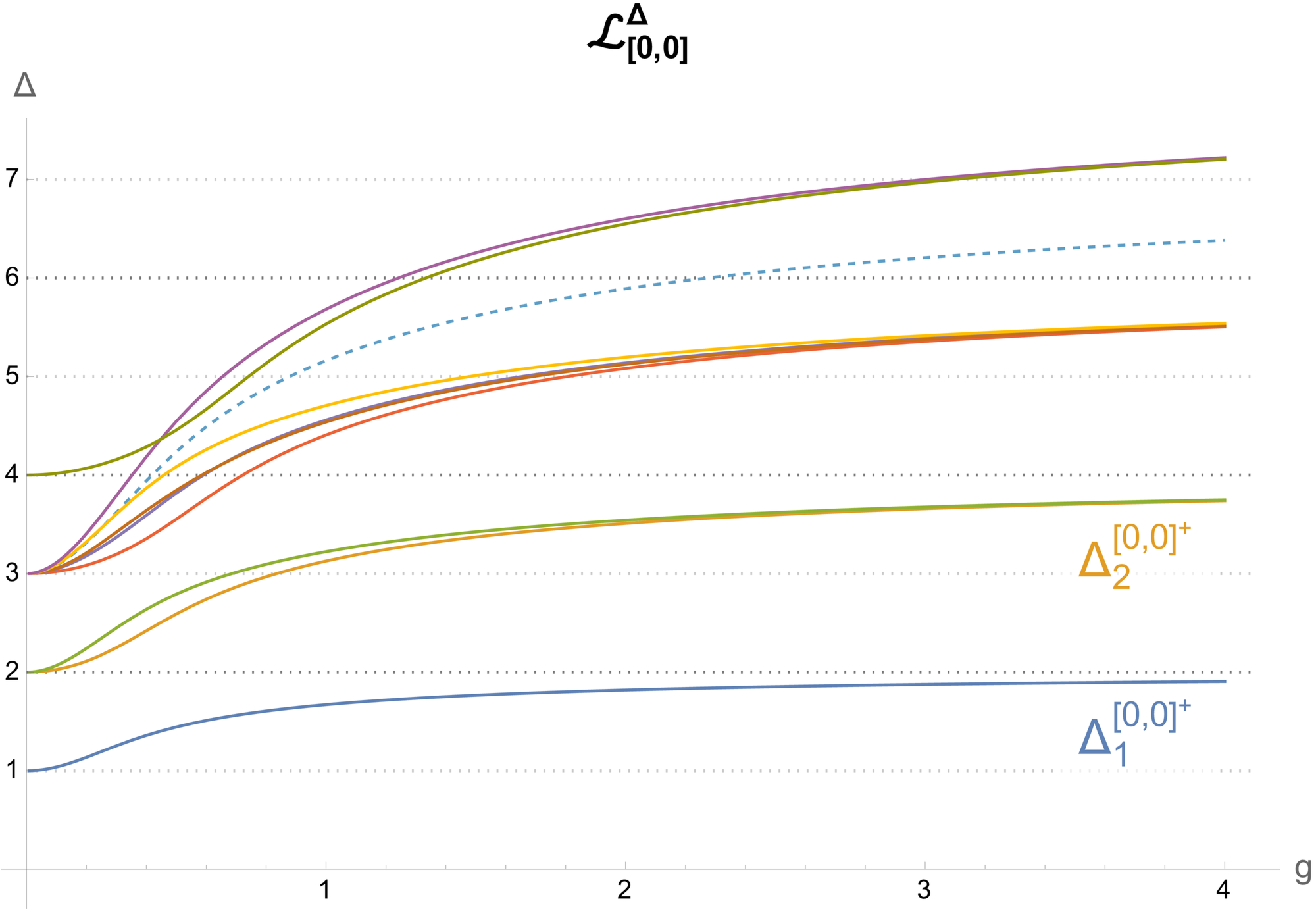}
    \end{subfigure}
\quad
    \begin{subfigure}[b]{0.45\textwidth}
    \centering
    \includegraphics[width=7cm]{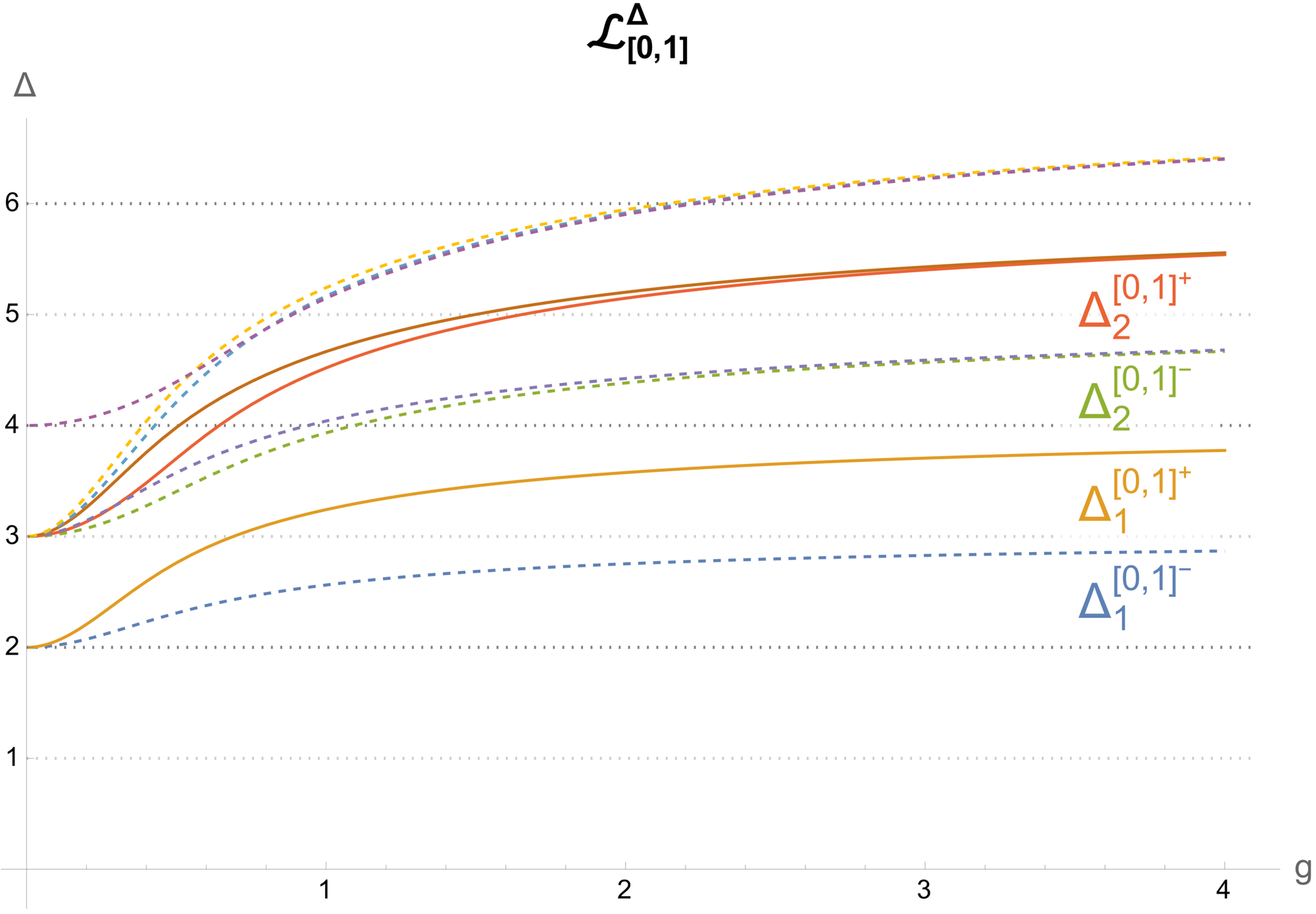}
    \end{subfigure}
\\
    \begin{subfigure}[b]{0.45\textwidth}
    \centering
    \includegraphics[width=7cm]{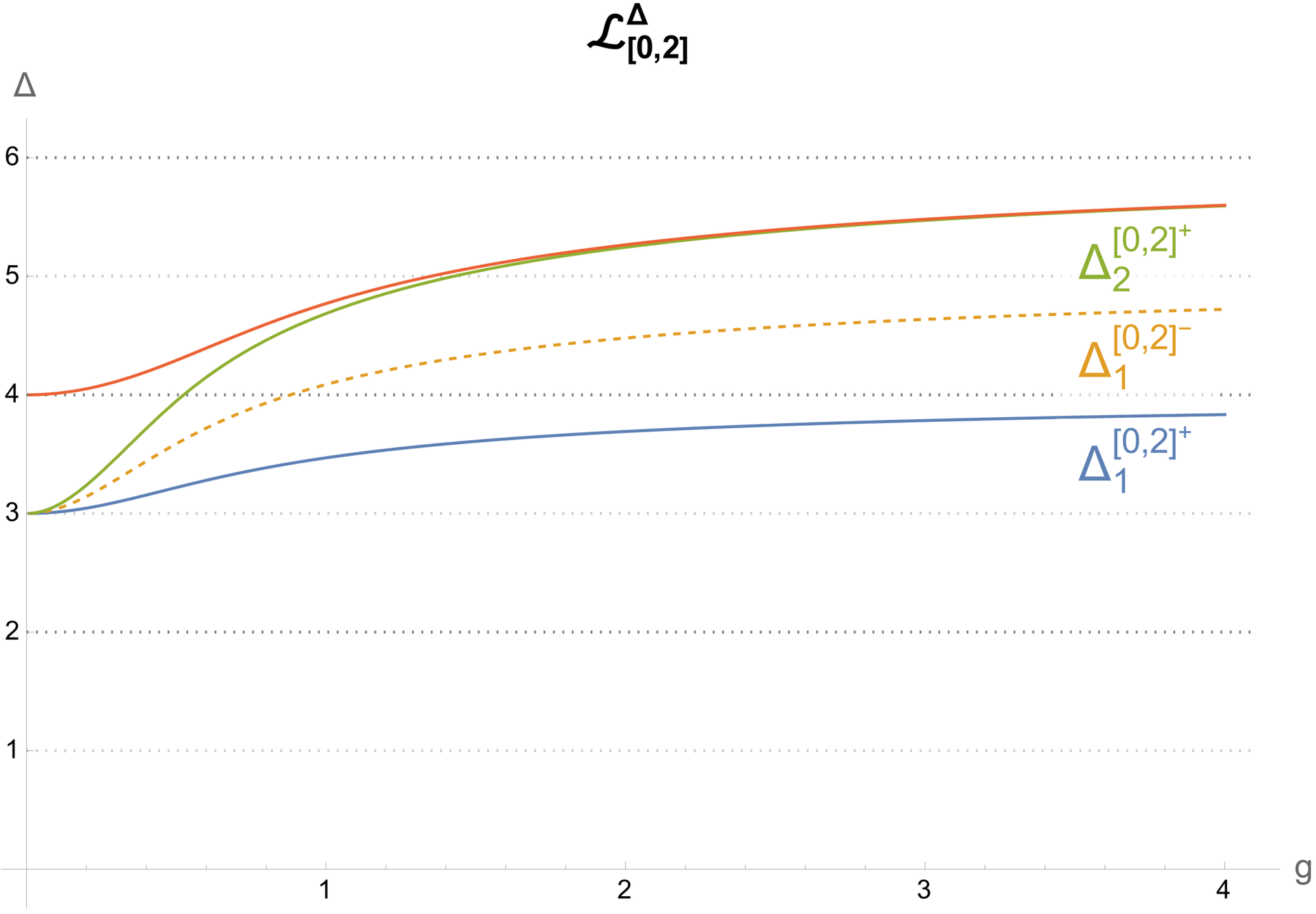}
    \end{subfigure}
\quad
    \begin{subfigure}[b]{0.45\textwidth}
    \centering
    \includegraphics[width=7cm]{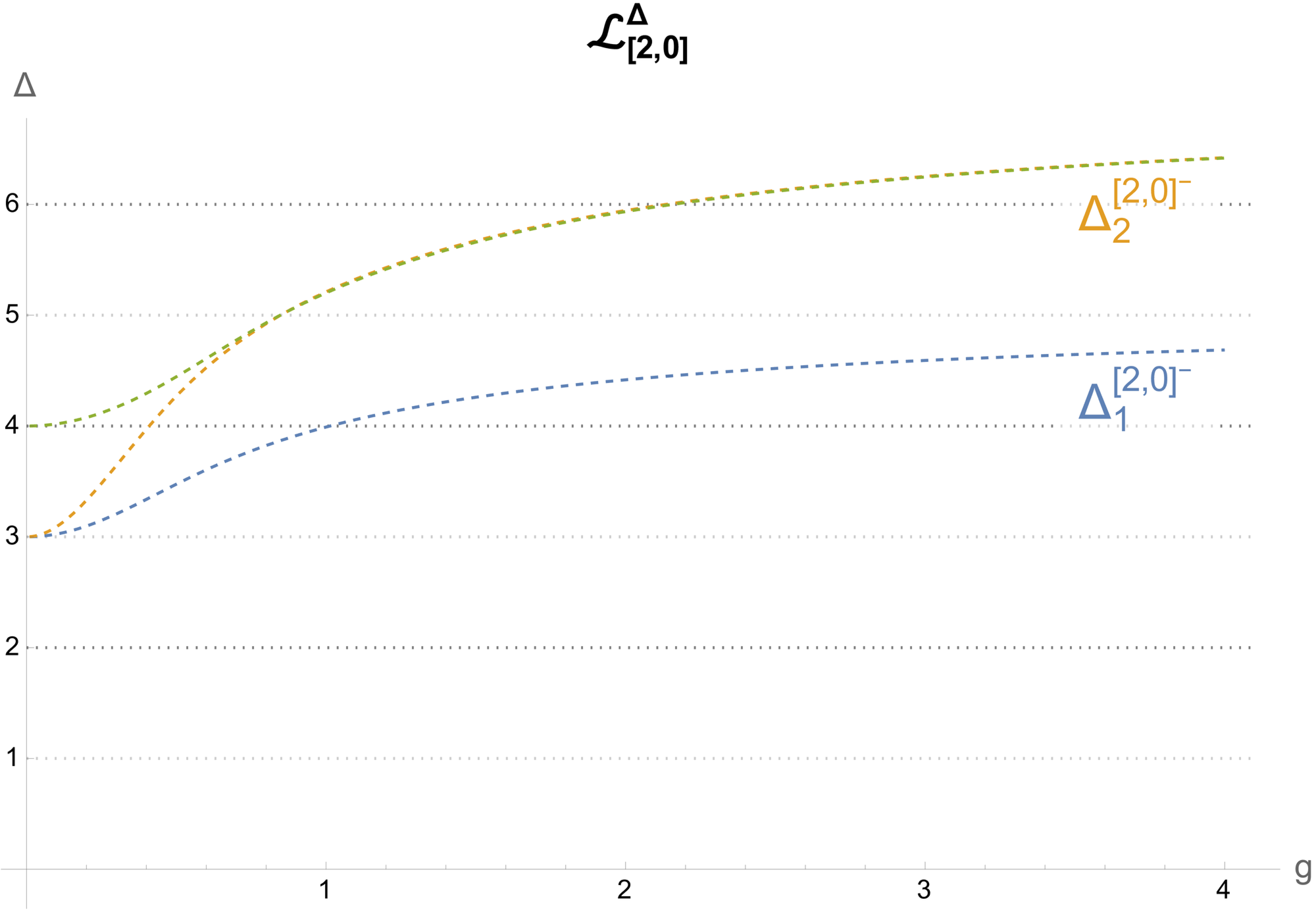}
    \end{subfigure}
\caption{Scaling dimensions $\Delta$ of long multiplets displayed in the range of coupling $g = [0,4]$. Here the plain lines correspond to parity-even operators ($\mathbb{P} = +1$), and the dashed lines to parity-odd ($\mathbb{P} = -1$), see section \ref{sec:discrete} for a discussion of this symmetry. 
Some of the lowest-lying states that will be relevant in section \ref{sec:Results} are labelled with their \texttt{Tag} defined in \eqref{tag}.}
\label{fig:spectrumplot}
\end{figure}

Even though the Hamiltonian is not known beyond this sector, it was noticed in~\cite{Cavaglia:2021bnz} that the one-loop anomalous dimensions of 
insertions can map to local operators scaling dimensions of the $\mathcal{N}=4$ SYM theory. 
The map obtained in that paper was for states that are neutral under all the global symmetries of the defect CFT, \textit{i.e.} for the superprimaries of long multiplets of type $\mathcal{L}_{[0,0]}^\Delta$. Here we generalise the map to include states with arbitrary $R$-charge, namely,
a state with quantum numbers of the 1D dCFT $\{ \Delta_{1D}, 0, a, b \}$ is mapped to the local operators of 4D  $\mathcal{N}=4$ SYM with bare dimension $2 \Delta_{1D} + 1$, Lorentz spins $[0\ 0]$ and  Dynkin labels of the $SU(4)$ R-symmetry of the bulk theory $[a\ 2b\  a]$ additionally restricted by the parity symmetry
$u\to-u$ (for the spectral parameter $u$) with the simultaneous exchange of the Left and Right wings. Based on this observation, we are able to identify all multiplets in accordance with the counting in Table~\ref{tab:countingweakstrong} and their one-loop dimensions and weak coupling fields representation.

In Table~\ref{tabledim} we provide the lowest lying spectrum of long multiplets. We give their bare dimensions $\Delta_0$, one loop anomalous dimension, and for the operators built out of scalars their wavefunction. We identify our states by defining their $\texttt{State ID}$ which looks as follows 
$$
\texttt{State ID} \equiv
\text{}_{\Delta_0}[\gr s\  a\  b\  \gr{\texttt{twist}}]_{\texttt{sol}},
$$ 
where $\Delta_0$ is the bare dimension, $s$ is the transverse spin, $[a, b]$ are the R-symmetry Dynkin labels (of the defect theory), $$\texttt{twist} \equiv \Delta_0 - \texttt{number of derivatives along the MWL at zero coupling}\;, $$
and $\texttt{sol}$ labels states with all other quantum numbers being the same. 
Furthermore, every state has an associated parity $\mathbb{P}$ as defined in \cite{Cavaglia:2023mmu}, and discussed further in Section~\ref{sec:discrete}. 
In order to simplify notation throughout the paper, we also introduce a shortened version of the state label that we call \texttt{Tag}, which is defined as
\begin{equation}\label{tag}
\texttt{Tag}\equiv \Delta_n^{[a,b]^\mathbb{P}}
\end{equation}
where $n$ identify the ordering of the states at weak coupling. As an example see Table \ref{tabledim} and for a complete list see appendix \ref{app:spectraldata}. 

The numerical non-perturbative results, obtained with Quantum Spectral Curve method for the first low-lying multiplets are presented on Figure~\ref{fig:spectrumplot}. Next we describe the technicalities of the Quantum Spectrum Curve method.

\begin{table}[t]
\begin{adjustwidth}{}{}
\begin{tabular}{|C|C|C|C|C|C|}
\hline
    \texttt{State ID} & \texttt{Tag} & \Delta_0 & \Delta_0 + \#\,g^2  &  \text{One-loop wavefunction for h.w.s} & \mathbb{P}\\ 
    \hline\hline
    \rule{0pt}{3.5ex}\text{}_ 1 \text{[\gr0 0 0 \gr1]}_ 1 & \Delta_1^{[0,0]^+} & 1 & 4 & \Phi_{||} & + 1 \\[2ex]
    \rule{0pt}{3.5ex}\text{}_ 2 \text{[\gr 0 0 0 \gr 2]}_ 1 & \Delta_2^{[0,0]^+} & 2 & 2.763932022500210 & \Phi_{||}^2 - \frac{1}{\sqrt{5}}\Phi^i_\perp\,\Phi^i_\perp & + 1 \\[2ex]
    \rule{0pt}{3.5ex}\text{}_ 2 \text{[\gr 0 0 0 \gr 2]}_ 2 & \Delta_3^{[0,0]^+} & 2 & 7.236067977499789 & \Phi_{||}^2 + \frac{1}{\sqrt{5}}\Phi^i_\perp\,\Phi^i_\perp & +1 \\[2ex]
    \hline\hline
    \rule{0pt}{3.5ex}\text{}_ 2 \text{[\gr 0 0 1 \gr 2]}_ 1 & \Delta_1^{[0,1]^-} & 2 & 2 & \Phi_{||} \Phi_{\perp}^i +  \Phi_{\perp}^i \Phi_{||}  & -1 \\[2ex]
    \rule{0pt}{3.5ex}\text{}_ 2 \text{[\gr 0 0 1 \gr 2]}_ 2 & \Delta_1^{[0,1]^+} & 2 & 6 & \Phi_{||} \Phi_{\perp}^i -  \Phi_{\perp}^i \Phi_{||}  & +1 \\[2ex]
    \hline\hline
    \rule{0pt}{3.5ex}\text{}_ 3 \text{[\gr 0 0 2 \gr 3]}_ 1 & \Delta_1^{[0,2]^+} & 3 & 4 & \multirow{3}{*}{\shortstack{\small $ \Phi_{||} (\Phi_\perp^i \Phi_\perp^j +\Phi_\perp^j \Phi_\perp^i) $\\ \small $- (\Phi_\perp^j \Phi_\perp^i +\Phi_\perp^i \Phi_\perp^j) \Phi_{||}$ \\ \small $
- \delta^{ij}\frac{2}{5} (\phipar\phiperp{k}\phiperp{k} - \phiperp{k}\phiperp{k}\phipar)$ }} & -1 \\[2ex]
    \rule{0pt}{3.5ex} & & & & & \\[2ex] 
    \hline\hline
    \rule{0pt}{3.5ex}\text{}_ 3 \text{[\gr 0 2 0 \gr 3]}_ 1 & \Delta_1^{[2,0]^-} & 3 & 2.535898384862245 & \multirow{3}{*}{\shortstack{\small $(\sqrt{3}+1)(\Phi^i_{\perp}\Phi_{||}\Phi^j_{\perp} - \Phi^j_{\perp}\Phi_{||}\Phi^i_{\perp})$\\ \small $+(\Phi^i_{\perp}\Phi^j_{\perp} \Phi_{||} - \Phi^j_{\perp}\Phi^i_{\perp} \Phi_{||})$\\ \small $ + (\Phi_{||} \Phi^i_{\perp}\Phi^j_{\perp}  -  \Phi_{||}\Phi^j_{\perp}\Phi^i_{\perp})$}}  & -1 \\[2ex]
    \rule{0pt}{3.5ex} & & & & & \\[2ex]
    \hline
\end{tabular}
\captionof{table}{
\la{tabledim}
Examples of first states in long multiplets with the following information: their $\texttt{StateID}$, their \texttt{Tag}, their bare dimension $\Delta_0$, their one-loop anomalous dimension, corresponding wavefunctions, parity.} 
\end{adjustwidth}
\end{table}

\subsection{Quantum Spectral Curve for Operator Insertions.} 

\label{sec:QSC}

We proceed now in describing the peculiarities of the QSC for the operator insertions.
Our numerical method is well-suited to track a number of states from weak to strong coupling (we collected data for $g \in [0, 4]$) such that we have plenty of spectral data available for our purposes. The procedure of finding the spectrum of operators via QSC numerically is described in many sources (see e.g.\,\cite{Gromov:2017blm}). So we focus on specifics of the current setup in what follows.

To begin with, we assign to any state a set of eight functions of the complex spectral parameter $\mathbf{P}_k(u), \mathbf{Q}_l(u)$ with $k,l = 1,\dots,4$, such as a state is described by a set of ``curves''. The $\mathbf{P}$-functions have one cut on $u$-plane for $u = [2g, -2g]$, and the $\mathbf{Q}$-functions have a infinite ladder of cuts in the lower half plane. The $\mathbf{P}$-functions can be parametrised as follows:
\begin{equation}
    \mathbf{P}_{k}(u) = x^{\texttt{powP}_k} { A}_k\left(1+\sum_{n = 1}^{\infty}\frac{d_{k,n}}{x^n}\right)\;\;,\;\;x(u)=\frac{u+\sqrt{u-2 g} \sqrt{2 g+u}}{2 g}, 
\end{equation}
where $\mathtt{powP}_k = {\left\{-a-b-\frac{5}{2},-b-\frac{3}{2},b+\frac{1}{2},a+b+\frac{3}{2}\right\}}_{k}$ is the power in the asymptotics of the $\mathbf{P}$-functions in the large-$u$ limit. Here the asymptotics provided is for the states of the 1D dCFT and depends on their quantum numbers as defined previously. The coefficients ${d}_{k,n}$ introduced above depend on $g$ and are different for each state. They are initially computed perturbatively at weak coupling providing initial ``seeds'' for the numerical procedure. The perturbative values of $\mathbf{P}$- and $\mathbf{Q}$-functions at weak coupling can be found using the Baxter equation, following \cite{Grabner:2020nis}.

The prefactor coefficients $A_{k}$ and similar coefficients $B_{l}$ for large $u$ asymptotic $\mathbf{Q}_l \sim B_l\,u^{\mathtt{powQ}_l}$ are also given in terms of the quantum numbers as follows
\begin{align}
\begin{split}
\label{eq:ABassymp}
    \mathtt{powQ}_l &= \{ \Delta + \frac{3}{2}, \Delta + \frac{1}{2}, -\Delta - \frac{3}{2}, -\Delta - \frac{5}{2} \}_l,\\
    A_{k} &= \left\{ 1, 1, -  \frac{(\Delta - b +1)(\Delta-b)(\Delta+b+2)(\Delta+b+3)}{2(a + 1)(b + 1)(a + 3b + 3)} i, \right.\\ 
    & \left. -\frac{(\Delta-a-b)(\Delta-a-b-1)(\Delta+a+b+3)(\Delta+a+b+4)}{2(a+1)(a+b+2)(a + 3b + 3)} i \right\}_k,\\
    B_{l} &= \left\{\frac{(\Delta - b + 1)(\Delta-a-b)(\Delta+b+3)(\Delta+a+b+4)}{2(\Delta+2)(2\Delta+3)}i,\right.\\
    &\left.\frac{(\Delta-b)(\Delta-a-b-1)(\Delta+b+2)(\Delta+a+b+3)}{2(\Delta+1)(2\Delta+3)}i,1,1\right\}_l.  
\end{split}
\end{align}
Given the large $u$ asymptotics, the relations between $\mathbf{P}$- and $\mathbf{Q}$- functions (QQ-relations) are used to shift them from the large-$u$ domain to close to the cuts on the real line at $u \in [-2g, 2g]$.  Once the numerical values of the functions are obtained close to the cut, we impose the ``gluing'' conditions. 
For the 1D dCFT, they differ from those of the local operators and are expressed as follows~\cite{Gromov:2015dfa}
\begin{align}
\label{eq:Gluing}
    \left(\begin{array}{c}
\tilde{\mathbf{q}}_1(u) \\
\tilde{\mathbf{q}}_2(u) \\
\tilde{\mathbf{q}}_3(u) \\
\tilde{\mathbf{q}}_4(u)
\end{array}\right)=\left(\begin{array}{cccc}
1 & 0 & 0 & 0 \\
0 & 1 & 0 & 0 \\
\alpha \sinh (2 \pi u) & 0 & 1 & 0 \\
0 & -\alpha \sinh (2 \pi u) & 0 & 1
\end{array}\right)\left(\begin{array}{l}
\mathbf{q}_1(-u) \\
\mathbf{q}_2(-u) \\
\mathbf{q}_3(-u) \\
\mathbf{q}_4(-u)
\end{array}\right)
    \;,
\end{align}
where $\mathbf{q}_i \equiv \mathbf{Q}_i/\sqrt{u}$ and  $\tilde{\mathbf{q}}_i(u)$ denotes their value at another side of the cut $[-2g,2g]$. The gluing conditions can be imposed by computing the values of ${\bf Q}_i$ at the branch cut $[-2g,2g]$, which depend, in a complicated way, on the coefficients $d_{c,n}$ and of $\Delta$ thus \eqref{eq:Gluing} provide an involved set of equations on these parameters, which can be solved numerically. The procedure to solve the QSC numerically for the given setup is described in more details in \cite{Gromov:2015wca, Gromov:2017blm, Grabner:2020nis, Gromov:2023hzc}. Due to the difference in the gluing conditions imposed, the spectrum of 1D dCFT converges to different values compared to full 4D $\mathcal{N}=4$ SYM beyond the one-loop order.

Using the above procedure, we produce the spectral plots on Figure~\ref{fig:spectrumplot}, tracking the spectrum from weak all the way to strong coupling. The spectrum condenses to integer values at $g=\infty$ and agrees with the counting of \cite{Ferrero:2023znz}, \textit{cf.} Table~\ref{tab:countingweakstrong}.
Therefore, we can be confident we have the complete spectrum up to some cut-off (which is the top level on each of the plots). In the next section, we describe how this spectrum is used in combination with the conformal bootstrap.

\section{Simple mixed-correlator system}
\label{sec:Bootstrability}
In this paper we focus on the constraints 
following from the known parts of the spectrum presented in the previous section and from the crossing condition of a system of 4-point correlators, which contain these non-protected states in their OPE expansion.

Thus we consider a system of correlators of operators belonging to the short multiplets $\mathcal{B}_1$ and $\mathcal{B}_2$, which is the smallest mixed-correlator setup as was pointed out in~\cite{Liendo:2018ukf}.
Schematically our system contains four independent correlators
\begin{equation}\begin{split}\label{eq:system}
&\langle\langle O_{\mathcal{B}_1}(x_1)O_{\mathcal{B}_1}(x_2)O_{\mathcal{B}_1}(x_3)O_{\mathcal{B}_1}(x_4) \rangle\rangle , \;\;\; \langle\langle O_{\mathcal{B}_2} (x_1)O_{\mathcal{B}_2}(x_2)O_{\mathcal{B}_2}(x_3)O_{\mathcal{B}_2}(x_4) \rangle\rangle, \\ &\langle\langle O_{\mathcal{B}_1}(x_1) O_{\mathcal{B}_2}(x_2) O_{\mathcal{B}_1}(x_3) O_{\mathcal{B}_2}(x_4) \rangle\rangle , \;\;\; \langle\langle O_{\mathcal{B}_1}(x_1) O_{\mathcal{B}_1}(x_2)   O_{\mathcal{B}_2}(x_3)O_{\mathcal{B}_2} (x_4)\rangle\rangle.
\end{split}\end{equation}
In the following, we review the  bootstrap problem following the analysis of \cite{Liendo:2018ukf}. The new ingredient added in this paper  is the implementation of the discrete parity symmetry of the theory, which, as anticipated in \cite{Cavaglia:2023mmu}, allows us to further restrict the possible values of OPE coefficients. 

In the product of two short multiplets $\mathcal{B}_1$ and $\mathcal{B}_2$, at a generic coupling $g$, only short or long multiplets are exchanged. Moreover, as outlined in \cite{Liendo:2018ukf}, their possible quantum numbers are strictly constrained leading to the following selection rules
\beq
\label{eq: D with D OPE}
\begin{split}
\mathcal{B}_1\,\times\,\mathcal{B}_1\,=\,&\mathcal{I}+\mathcal{B}_2
+\sum_{\Delta\geq 1}\mathcal{L}^{\Delta}_{[0,0]}\,,\\
\mathcal{B}_1\, \times \,\mathcal{B}_2\,=\,& \mathcal{B}_1+\mathcal{B}_3
+\sum_{\Delta\geq 2} \mathcal{L}^{\Delta}_{[0,1]}\,,\\
\mathcal{B}_2\, \times \,\mathcal{B}_2\,=\,&
\mathcal{I}+\mathcal{B}_2+\mathcal{B}_4
+\sum_{\Delta\geq 1}\mathcal{L}^{\Delta}_{[0,0]}+\sum_{\Delta\geq 3}\left(\mathcal{L}^{\Delta}_{[0,2]}+\mathcal{L}^{\Delta}_{[2,0]}\right)\,,
\end{split}
\eeq
where $\mathcal{I}$ is the identity. In the following, we explore the OPE coefficients and the corresponding conformal blocks of all multiplets exchanged above.

\subsection{Crossing equations and selection rules on OPE coefficients}\la{crossingOpe}

The operators at the top of the $\mathcal{B}_{m_i}$ supermultiplet are in the R-symmetry representation $[0, m_i]$ corresponding to traceless symmetric tensor with $m_i$ indices, which can be contracted with a polarisation 5D null vector $Y_i$.
We thus use the standard notation $O_{m_i}(x_i, Y_i )$, which encodes both spacial location of the operator on the line and the R-symmetry polarization
of the operators in $\mathcal{B}_{m_i}$. 

Following~\cite{Liendo:2018ukf}, the $4$-point correlators of these superprimary operators can be written in the following form
\begin{equation}\label{eq:superconfKA}
    \langle \langle O_{m_1}(x_1, Y_1) \dots O_{m_4}(x_4, Y_4) \rangle \rangle = K_{m_1 m_2 m_3 m_4}\, \mathcal{A}_{m_1 m_2 m_3 m_4}(x, \zeta_1, \zeta_2),
\end{equation}
where $m_1 + m_2 + m_3 + m_4$ should be even due to the R-symmetry constraint and $K_{m_1 m_2 m_3 m_4}$ is a kinematic prefactor given by 
\beq
\label{eq:kinprefactor}
K_{m_1 m_2 m_3 m_4}
= (12)^{\frac{m_1+m_2}{2}} (34)^{\frac{m_3+m_4}{2}}\!\!\left(\frac{(14)}{(24)}\right)^{\frac{m_1-m_2}{2}} \!\!\!\left(\frac{(13)}{({14})}\right)^{\frac{m_3-m_4}{2}}, \;\;\; \;
(ij) \!\equiv\! \frac{Y_i \cdot Y_j}{x_{ij}^2}.
\eeq
The amplitude $\mathcal{A}_{m_1 m_2 m_3 m_4}$ depends on the one-dimensional spacetime cross-ratio $x$, as well as on 2 independent cross-ratios $\zeta_i$ built with the polarisation vectors defined as
\begin{gather}
    x = \frac{x_{12} x_{34}}{x_{13} x_{24}},\ \ \  x_{ij} \equiv  x_i - x_j,\\
    \nonumber
\zeta_1 \zeta_2  = \frac{(Y_1\cdot Y_2)(Y_3\cdot Y_4)}{(Y_1\cdot Y_3)(Y_2\cdot Y_4)},\qquad (1 - \zeta_1)(1 - \zeta_2) = \frac{(Y_1\cdot Y_4) (Y_2\cdot Y_3)}{(Y_1\cdot Y_3)(Y_2\cdot Y_4)}.
\end{gather}
Notice that the l.h.s.  of  \eqref{eq:kinprefactor} is polynomial in $Y_i$, which is somewhat hidden in the r.h.s..
The system of correlators \eqref{eq:system} is invariant under crossing symmetry, which  can be understood as  a cyclic invariance mapping the positions $1234 \rightarrow 4123$. Under this relabelling, the cross ratios transform as follows
\beq
x \to 1-x\;\;,\;\;
\zeta_1 \to 1 - \zeta_1\;\;,\;\;
\zeta_2 \to 1 - \zeta_2\;. 
\eeq
The crossing equations implies the following constraints on the non-trivial part of the correlator\footnote{ Notice that the third crossing equation above was written differently in \cite{Liendo:2018ukf}, namely in the form
\beq
\tilde{\mathfrak{X}}^{\frac{3}{2}}\mathcal{A}_{1212}(x,\zeta_i)=\mathfrak{X}^{\frac{3}{2}}\mathcal{A}_{1212}(1-x,1-\zeta_i).
\eeq
The two crossing equations are simultaneously true, due to the parity symmetry of the 1D dCFT (related to changing orientation on the line), which we discuss next and which was analysed systematically in \cite{Cavaglia:2023mmu}. This symmetry proves explicitly that $\mathcal{A}_{1212}(x, \zeta_i) = \mathcal{A}_{2121}(x, \zeta_i) $. We will see in the next sections how it simplifies the bootstrap problem. }
\beq\label{eq:crossingeq0}
\begin{split}
\tilde{\mathfrak{X}}\, \mathcal{A}_{1111}(x,\zeta_i)&=\mathfrak{X}\,\mathcal{A}_{1111}(1-x,1-\zeta_i)\,,\\
\tilde{\mathfrak{X}}^2\mathcal{A}_{2222}(x,\zeta_i)&=\mathfrak{X}^2\mathcal{A}_{2222}(1-x,1-\zeta_i)\,,\\
\tilde{\mathfrak{X}}^{\frac{3}{2}}\mathcal{A}_{1212}(x,\zeta_i)&=\mathfrak{X}^{\frac{3}{2}}\mathcal{A}_{2121}(1-x,1-\zeta_i)\,,\\
\tilde{\mathfrak{X}}^2 \mathcal{A}_{1221}(x,\zeta_i)&=\mathfrak{X}^{\frac{3}{2}}\mathcal{A}_{1122}(1-x,1-\zeta_i),\,
\end{split}
\eeq
where 
\beq
\label{eq: definition of sX}
\mathfrak{X}\equiv \frac{x^2}{\zeta_1\zeta_2}\,,\qquad \tilde{\mathfrak{X}}\equiv\frac{(1-x)^2}{(1-\zeta_1)(1-\zeta_2)}\,.
\eeq
These additional powers come from the transformation of the kinematic prefactors in (\ref{eq:superconfKA}). 

\subsubsection{OPE coefficients of the operators in non-trivial R-symmetry irreps}
Let us introduce the definition of OPE coefficients,
following the conventions of \cite{Liendo:2018ukf}. 
In particular, this section explains the convention for stripping  off appropriate combinations of polarisation vectors $Y_i$, giving a precise definition of the OPE coefficients which we bound by the numerical bootstrap below, 
extending the definitions \eq{eq:OPEdef} for R-symmetry scalar operators.

We are interested in the OPE channels (\ref{eq: D with D OPE}). For simplicity, we will only explicitly introduce OPE coefficients relating three superconformal primaries. These contain all the information since the superconformal blocks deduced in \cite{Liendo:2018ukf} allow one to reconstruct the OPE coefficients for all other members of the supermultiplets involved.

\paragraph{Representation $[0,0]$.}
Let us start from the OPE of the channel $\mathcal{B}_m \times \mathcal{B}_m \rightarrow \mathcal{L}^{\Delta}_{[0,0]}$ (where in this paper $m=1$ or $m=2$). For the case of these correlators the SUSY 
relates the CFT data for all members of the intermediate long multiplet. We can thus focus on  the super-component with lowest scaling dimension. Such an operator is neutral under the R-symmetry, and we denote it as $O_{\Delta}^{[0,0]}$. Assuming the canonical normalization \eq{eq:OPEdef}
for the structure constant, we get 
\beq\la{Cdef00}
C_{O_m(Y_1),
O_m(Y_2),
 O_{\Delta }^{[0,0]}}  \equiv (Y_1 \cdot Y_2 )^m\; C_{m m\Delta^{[0,0]}}, 
\eeq
where we stripped the polarization dependent part, as it is the only structure consistent with R-symmetry.  This definition implies that there is the following term in the OPE expansion associated with this multiplet:
\beq
O_m(0, Y_1) \,O_m(1, Y_2) \sim  (Y_1 \cdot Y_2 )^m \; C_{mm\Delta^{[0,0]}}\; \bar O_{\Delta}^{[0,0]}(0) + \texttt{(super)-descendants} .
\eeq

\paragraph{Representations $[2,0]$ and $[0,2]$.}
In the fusion channel $\mathcal{B}_2 \times \mathcal{B}_2$, we can also produce long superprimary operators in the representations $\mathbf{r} =[0,2]$ or $\mathbf{r} = [2,0]$ of the R-symmetry. The lowest-dimension component $O_{\Delta}^{[2,0]}$ 
and $O_{\Delta}^{[0,2]}$
have a pair of R-symmetry indices, with $[0,2]$ being symmetric and traceless tensor and $[2,0]$ is anti-symmetric tensor. In order to preserve the covariance in R-symmetry, the standard normalization \eq{eq:OPEdef} is replaced by
\beqa\la{Cdef2002}
\langle\langle \bar O_{\Delta,a_1a_2}^{[0,2]}(0) \; {O}_{\Delta,b_1b_2}^{[0,2]}(1) \rangle\rangle&=&
\(\frac{1}{2}\delta_{a_1b_1}
\delta_{a_2b_2}+
\frac{1}{2}\delta_{a_1b_2}
\delta_{a_2b_1}-
\frac{1}{5}\delta_{a_1a_2}\delta_{b_1b_2}\)\;,\\
\langle\langle \bar O_{\Delta,a_1a_2}^{[2,0]}(0) \; {O}_{\Delta,b_1b_2}^{[2,0]}(1) \rangle\rangle&=&
2\(\frac{1}{2}\delta_{a_1b_1}
\delta_{a_2b_2}-
\frac{1}{2}\delta_{a_1b_2}
\delta_{a_2b_1}\)\;,
\eeqa
where the expressions in the r.h.s. are the projectors on the corresponding representation.
Notice the extra factor of $2$ for antisymmetric representation, which follows the conventions of \cite{Liendo:2018ukf}.

In addition we assume that the conjugate operator is
different only by a phase factor from the original operator (as otherwise the spectrum would be degenerate for generic $g$). We fix the remaining phase freedom in the normalization so that
\beq\la{conj02}
\bar O_{\Delta,a_1a_2}^{[0,2]} =
+O_{\Delta,a_1a_2}^{[0,2]}\,,\qquad\quad
\bar O_{\Delta,a_1a_2}^{[2,0]} =
-O_{\Delta,a_1a_2}^{[2,0]}\;.
\eeq
With these normalisations, the OPE coefficient is normalized so that the corresponding terms in the OPE expansion is
\begin{equation}\small\begin{split}\la{OPE02}
O_2(0, Y_1) O_2(1, Y_2) &\simeq 
(Y_1\cdot Y_2)\[ C_{22\Delta^{[0,2]}}  \;   O_{\Delta,a_1a_2}^{[0,2]}(0)+
C_{22\Delta'^{[2,0]}} \;   O_{\Delta',a_1a_2}^{[2,0]}(0)\]Y_1^{a_1}Y_2^{a_2}
+ \dots
\end{split}\normalsize\end{equation}
with $\dots$ representing super-descendants and operators with other dimensions.

\paragraph{Representation $[0,1]$.}
Finally, let us consider the operators  in the representation $[0,1]$ of R-symmetry in the OPE channels $\mathcal{B}_1 \times \mathcal{B}_2$ and $\mathcal{B}_2 \times \mathcal{B}_1$. Being in the fundamental representation of the $SO(5)$ R-symmetry,  they carry a single index, 
 and we denote them as $O^{[0,1]}_{\Delta,a}(x)$.  We normalize these operators in the standard way \eq{eq:OPEdef}. We 
 assume that the complex conjugation acts trivially on these operators and that we can set
\beq\la{conj01}
\bar O_{\Delta,a}^{[0,1]}= O_{\Delta,a}^{[0,1]}\;,
\eeq
to fix the phase factor. Then the OPE coefficient is defined by the following contribution 
\beqa
&&O_1(0, Y_1) \,O_2(1, Y_2) \sim  (Y_1 \cdot Y_2 )\; C_{12\Delta^{[0,1]}}\; O_{\Delta,a}^{[0,1]}(0) \, Y_2^a + \dots \;,\\
&&O_2(0, Y_1) \,O_1(1, Y_2) \sim  (Y_1 \cdot Y_2 )\; C_{21\Delta^{[0,1]}}\; O_{\Delta,a}^{[0,1]}(0) \, Y_1^a + \dots \;.
\eeqa
Some comments on our conventions are necessary. We have defined the operators such that the 2-point functions are normalized as in the previous section, eliminating the freedom to define the operators with an additional phase. However, the operators are still defined up to a sign, and consequently, our OPE coefficients are also defined up to a sign. This sign, however, is irrelevant in the OPE expansion of four-point functions, as it cancels in the combinations of the OPE presented there
\begin{equation}\begin{split}
\label{eq:BlockExpMain}
\mathcal{A}_{1111} &\simeq  \sum_{\Delta}  C_{11\Delta^{[0,0]}}^2
\times ( \texttt{\footnotesize superblock})
+\!\!\!\sum_{n=0,2} (C^{\rm BPS}_{11n})^2\times (\texttt{\footnotesize superblock})
,\\
\mathcal{A}_{2222} &\simeq   \!\!\!\sum_{\mathbf{r} \in \left\{[0,0], [2,0], [0,2]\right\}} \!\!\sum_{\Delta }\;  C_{22\Delta^{\mathbf{r}} }^2 \times ( \texttt{\footnotesize superblock})
+\!\!\!\!\!\sum_{n=0,2,4} (C_{22n}^{\rm BPS})^2\!\times\! (\texttt{\footnotesize superblock}),
\\
\mathcal{A}_{2121}&=\mathcal{A}_{1212} \simeq  \sum_{\Delta } C_{12\Delta^{[0,1]}}^2 \times ( \texttt{\footnotesize superblock}) +\!\!\!\sum_{n=1,3} (C_{12n}^{\rm BPS})^2\times (\texttt{\footnotesize superblock}),\\
\mathcal{A}_{1221} &\simeq  \sum_{\Delta }  C_{12\Delta^{[0,1]} } \; C_{21\Delta^{[0,1]}} \times ( \texttt{\footnotesize superblock}) 
+\!\!\!\sum_{n=1,3} C_{12n}^{\rm BPS} \, C_{21n}^{\rm BPS} \!\times\! (\texttt{\footnotesize superblock}),\\
\mathcal{A}_{1122} &\simeq  \sum_{\Delta }  C_{11\Delta^{[0,0]} }\; C_{22\Delta^{[0,0]}}  \times ( \texttt{\footnotesize superblock}) 
+\!\!\!\sum_{n=0,2} C^{\rm BPS}_{11n} \, C^{\rm BPS}_{22n} \!\times \!(\texttt{\footnotesize superblock})
,
\end{split}\end{equation}
where the superblocks are those found in \cite{Liendo:2018ukf} and reported explicitly in appendix~\ref{app:decomposition}. In the above relations, we introduced the OPE coefficients for the exchange of BPS multiplets $\mathcal{B}_n$ (where $n=0$ stands for the identity multiplet) according to the selection rules (\ref{eq: D with D OPE}). By definition $C^{\rm BPS}_{110} = C^{\rm BPS}_{220} = 1$. The other OPE coefficients involving the BPS multiplets entering the above equations are $C^{\rm BPS}_{112}$, $C^{\rm BPS}_{222}$, $C^{\rm BPS}_{224}$, $C^{\rm BPS}_{123}$, $C^{\rm BPS}_{121} = C^{\rm BPS}_{112}$. These are non-trivial functions of the coupling that can be computed exactly as discussed in section \ref{sec:localisation}.

In the next section we  present the selection rules which the discrete symmetries of the theory impose on the combinations of non-BPS OPE coefficients entering our system.

\subsubsection{Discrete symmetries and their impact on OPE coefficients }\label{sec:discrete}
\paragraph{Parity. }
As discussed in \cite{Cavaglia:2023mmu}, the defect CFT possesses a discrete parity symmetry, which allows us to define a parity charge $\mathbb{P}_O \in \left\{+1, -1\right\}$ associated to any primary operator $O$. By construction, the parity charge commutes with R-symmetry generators (as well as with the dilatation operator on the line). This implies that we can define a basis of operators with fixed parity, and, for operators $O_{\Delta}^{\mathbf{r}}$ with R-symmetry representations $\mathbf{r}$ such as the ones introduced above, there is a charge $\mathbb{P}_{O}$ that is independent of the R-symmetry polarization. 

The charge can be read from the perturbative construction of the operators, which, at every loop order, are built as products of fundamental fields of $\mathcal{N}$=4 SYM, all in the adjoint representation of the gauge group (these products of fields are then inserted inside the path-ordered exponential defining the Wilson line). 
On such products of fundamental fields, parity acts as follows (we assume the operator is at the point $t=0$)
\beq\label{eq:parityfields}
\texttt{ParityTransform} \left[ f_1 \cdot \dots \cdot f_n \right] = (-1)^n \left. f_n \cdots f_1 \right|_{t\rightarrow -t, \;\; \Phi_{||}\rightarrow -\Phi_{||}},
\eeq
i.e., for operators built as products of $n$ fundamental fields, we write them in reverse order, and in addition act with the discrete symmetries $\Phi_{||}\rightarrow -\Phi_{||} $ and $t \rightarrow -t$, where $t$ is the coordinate along the defect (this changes e.g. the sign of derivatives $\d_t$). The operators can be chosen
to have a definite parity charge
$\mathbb{P}=\pm 1$ which should hold non-perturbatively. The charge can also be computed in terms of integrability data at weak coupling, see \cite{Cavaglia:2023mmu}. For the bootstrap problems considered in this work, we computed the dimensions of several operators, and for all of them we also determined the parity charge, see e.g. Table \ref{tabledim} and appendix \ref{app:spectraldata}. We also point out that the BPS  multiplets $\mathcal{B}_k$ have the parity $\mathbb{P} = (-1)^k$. 

The parity symmetry has an immediate consequence for correlation functions. As noted in \cite{Cavaglia:2023mmu}, correlation functions of scalar operators satisfy the following transformation rule
\begin{equation}
\label{eq:corrparity}
    \langle \langle O_1 (t_1) \dots O_n (t_n) \rangle \rangle = \mathbb{P}_1 \dots \mathbb{P}_n \langle \langle O_n (-t_n) \dots  O_1 (-t_1) \rangle \rangle .
\end{equation}
Applied to the 4-point function  $\langle \langle O_{\mathcal{B}_1} O_{\mathcal{B}_2} O_{\mathcal{B}_1} O_{\mathcal{B}_2} \rangle \rangle$, this symmetry immediately implies the relation mentioned above among two amplitudes, namely $\mathcal{A}_{1212}(x, \zeta_i) = \mathcal{A}_{2121}(x, \zeta_i)$.

Similarly, in \cite{Cavaglia:2023mmu} 
the parity transformation for 3-point functions was analysed and the following identity for OPE coefficients 
was obtained, for the operators neutral under the global symmetries:
\begin{equation}\label{eq:Crule} C_{O_1 O_2 O_3} = \mathbb{P}_1 \mathbb{P}_2 \mathbb{P}_3 C_{O_3 O_2 O_1}\;.
\end{equation}
This relation implies that in addition to the cyclic property of the OPE coefficients, the order of the operators can be reversed taking into account an additional parity prefactor. An immediate consequence of \eqref{eq:Crule} is that for $O_2=O_1$ 
we have $C_{O_1O_1O_3}=0$ when ${\mathbb P}_3=-1$ since ${\mathbb P}^2=1$.
As in our case the external operators are not literally equal, as they may come with different polarizations, we have to generalize this selection rule.

In order to generalize the selection rule for the case with non-trivial R-symmetry structures, we have to get back to the definitions \eq{Cdef00} and \eq{OPE02}.
We notice that for the case $[2,0]$ the additional R-symmetry factor is antisymmetric in interchange of the protected operators
\beq\la{C20def3point}
\langle\langle
{\cal O}_2(t_1,Y_1)
{\cal O}_2(t_2,Y_2)
O_{\Delta,a_1a_2}^{[2,0]}(t_3)
\rangle\rangle = -2(Y_1\cdot Y_2) Y_1^{[b_1} Y_2^{b_2]}\frac{C_{22\Delta^{[2,0]}}}{t_{12}^{4-\Delta}t_{13}^{\Delta}t_{23}^\Delta}\;,
\eeq
leading to the following selection rules
\begin{equation}\begin{split}\label{eq:parityselection}
C_{m m \Delta^{[0,0]}} &= 0 \quad\rm{ if }\quad \mathbb{P}_{O_{\Delta}} = -1,\\
C_{m m \Delta^{[0,2]}} &= 0 \quad\rm{if}\quad\mathbb{P}_{O_{\Delta}} = -1,\\
C_{m m \Delta^{[2,0]}} &= 0\quad\rm{if}\quad\mathbb{P}_{O_{\Delta}} = +1\;.
\end{split}\end{equation}

\paragraph{Reality. }
The other symmetry analysed in \cite{Cavaglia:2023mmu} concerns complex conjugation. It was shown that correlation functions under complex conjugation are related to correlators of conjugated operators $\bar{O}$ as follows,
\beq\label{eq:conjcorr}
\overline{\langle\langle O_1(t_1) \dots O_n(t_n) \rangle\rangle }  = \mathbb{P}_1 \cdots \mathbb{P}_n \langle\langle \bar{O}_1(t_1) \dots \bar{O}_n(t_n) \rangle\rangle ,
\eeq
where $\bar{O}$ is defined by taking the Hermitian conjugate of $O$ seen as an $N \times N$ matrix, and in addition changing the sign of the coordinate parallel to the defect, i.e. $t-t_i \rightarrow t_i-t$ (affecting possible sign of $\d_t$ in each of the operators). 
This definition is such that $\langle O \bar{O} \rangle \geq 0$, due to the reflection-positivity principle. 
For the structure constants, defined as $3$-point function of a normalized operators \eq{eq:OPEdef}, this has the following implication
\beq\la{eq:conjcor}
\bar C_{O_1O_2O_3}= 
{\mathbb P}_1
{\mathbb P}_2
{\mathbb P}_3
C_{\bar O_1\bar O_2\bar O_3}\;.
\eeq
Under the conjugation, the operators in our system transform in a simple way as we have already imposed a convention on the normalization of the non-protected operators, so that their conjugation acts as in \eq{conj02} and \eq{conj01}. Furthermore, the protected operators has an explicit, protected operatorial form which implies that
\begin{equation}\label{eq:conjO}
\overline{O_m(x, Y) } = O_m(x, \bar{Y})\;.
\end{equation}
Then applying \eq{eq:conjcor} for the case \eq{C20def3point} we get
\beq\la{C20def3pointCnj}
{\mathbb P}_\Delta
\langle\langle
{\cal O}_2(t_1,\bar Y_1)
{\cal O}_2(t_2,\bar Y_2)
\[-O_{\Delta,a_1a_2}^{[2,0]}(t_3)\]
\rangle\rangle = -2(\bar Y_1\cdot \bar Y_2) \bar Y_1^{[b_1} \bar Y_2^{b_2]}\frac{\bar C_{22\Delta^{[2,0]}}}{t_{12}^{4-\Delta}t_{13}^{\Delta}t_{23}^\Delta}\;,
\eeq
and now, taking into account the selection rule 
\eq{eq:parityselection},  we conclude that 
$C_{22\Delta^{[2,0]}}$ is always real for our conventions. Similarly, we can repeat almost identically the argument and prove reality of the other OPE coefficients exchanged in $\mathcal{B}_2 \times \mathcal{B}_2$ and $\mathcal{B}_1 \times \mathcal{B}_1$. In total,
\beq\label{eq:Creal}
C_{m m\Delta^{\mathbf{r}}}\in \mathbb{R}, \;\;\;\; \mathbf{r} \in \left\{[0,0], [0,2], [2,0]\right\}\;, \;\; m=1,2.
\eeq
For the OPE coefficients in the 
$\mathcal{B}_1 \times \mathcal{B}_2$ channel, we have more interesting properties. Applying 
\eq{eq:conjcor}, repeating the same steps as before, we get
\beq\label{eq:barC12}
\overline{ C_{12\Delta^{[0,1]}}}= \mathbb{P}_2 \mathbb{P}_1 \mathbb{P}_{O_{\Delta}} \; C_{12\Delta^{[0,1]}} ,
\eeq
and since $\mathbb{P}_2 = 1$, $\mathbb{P}_1 = -1$, this implies that $C_{12\Delta^{[0,1]}}$ is either purely real or purely imaginary, depending on the parity of the operator,
\beq\label{eq:C12reality}
C_{12\Delta^{[0,1]}}\;\;\;\; 
\in
\;\;\;\;\;\;\left\{ \begin{array}{cc}
 \mathbb{R} \text{ for } \mathbb{P}_{O_{\Delta}} = -1 ,\\
 i \mathbb{R} \text{ for } \mathbb{P}_{O_{\Delta}} = +1 
\end{array} \right.\;.
\eeq
The reality properties (\ref{eq:Creal}) and (\ref{eq:C12reality}) yield positivity constraints for the combinations of OPE coefficients appearing in the decompositions \eqref{eq:BlockExpMain}. In particular, (\ref{eq:Creal}) implies that
\beq\label{eq:positivesimple}
C_{11\Delta^{[0,0]}}^2, \, C_{22\Delta^{[0,0]}}^2, \, C_{22\Delta^{[2,0]}}^2, \, C_{22\Delta^{[0,2]}}^2 \;\;\;\;\geq 0 ,
\eeq
while (\ref{eq:C12reality}) yields
\beq\label{eq:C12positivity}
C_{12\Delta^{[0,1]}}^2 
\;\;\;\;\;\;\left\{ \begin{array}{cc}
\geq 0 \text{ for } \mathbb{P}_{O_{\Delta}} = -1 ,\\
\leq 0 \text{ for } \mathbb{P}_{O_{\Delta}} = +1 
\end{array} \right. .
\eeq
From (\ref{eq:C12reality}) we also deduce that one particular combination, the one appearing in the decomposition of $\mathcal{A}_{1221}$, is always positive,
\beq\label{eq:C12squared}
C_{12\Delta^{[0,1]}}C_{21\Delta^{[0,1]}} = |C_{12\Delta^{[0,1]}}|^2 \geq 0 .
\eeq
It was observed in \cite{Liendo:2018ukf} that in the defect CFT the reality of OPE coefficients such as $C_{12\Delta^{[0,1]}}$ is not guaranteed, and they were treated as generic complex numbers in the bootstrap analysis of that paper. 
Here we showed  that, taking into account the parity symmetry of the operators, the result is sharper, as these OPE coefficients are either purely real or purely imaginary, thus reducing by half the space of OPE coefficients to bootstrap. The selection rules and the final form of the OPE decompositions are summarized in appendix~\ref{app:decomposition}.

\subsubsection{Crossing equations 
 }
Combining the crossing conditions (\ref{eq:crossingeq0}) in conjunction with the OPE decomposition, one obtains relations depending on various R-symmetry polarisation structures. It was shown in \cite{Liendo:2018ukf} that these can be reduced to a few independent constraints which involve only the cross-ratio $x$. Keeping into account the selection rules and reality properties we have described above,
there are in total seven independent constraints.\footnote{This is one less than those written down in \cite{Liendo:2018ukf}, since in this earlier work the reality property (\ref{eq:C12reality}) was not imposed. }
To write the equations taking into account the selection rules, we distinguish the following types of spectra that are playing a distinguished role in the system
\beq\label{eq:spectra}
\texttt{S}^{[0,0]^+}, \;\;\;\texttt{S}^{[2,0]^-},\;\;\;\;\texttt{S}^{[0,2]^+}, \;\;\;\; \texttt{S}^{[0,1]^{-}} ,
\;\;\;\; \texttt{S}^{[0,1]^{+}} ,
\eeq
where the $\pm$ superscripts denote the values of the parity charges $\mathbb{P}$. We can forget about $\texttt{S}^{[0,0]^-}$, $\texttt{S}^{[0,2]^-}$ and $\texttt{S}^{[2,0]^+}$ since we know from the selection rule that these states have vanishing  OPE coefficients in the channels we consider. 

The system of independent relations can be written as\footnote{In detail, this is a system of 7 equations since $V$ and $\tilde{V}$ are vectors with 7 components each. The first equation rewrites the 1st crossing constraint (\ref{eq:crossingeq0}), the second to fourth relations are equivalent to the 2nd crossing constraint in (\ref{eq:crossingeq0}), the fifth relation is equivalent to the 3rd crossing constraint (\ref{eq:crossingeq0}), while the last crossing constraint is expressed by the last two relations of (\ref{eq: final crossing equations 1a}). }
\begin{equation}
\label{eq: final crossing equations 1a}
0 =\!\!\!\!\!\!\sum_{\Delta \in  \texttt{S}^{[0,0]^+}\cup \texttt{S}^{[2,0]^-}\cup \texttt{S}^{[0,2]^+}}\!\!\!\! \left(\!\begin{array}{cc} C_{11\Delta^{\mathbf{r}} }& C_{22\Delta^{\mathbf{r}}}\end{array}\!\right) V_{\Delta^{\mathbf{r}}}\left(\!\begin{array}{c}C_{11\Delta^{\mathbf{r}} }\\C_{22\Delta^{\mathbf{r}}} \end{array}\!\right) \!+ \!\!\!\!\sum_{\Delta \in \texttt{S}^{[0,1]^{\pm}}}  \!\!\!\!\!|C_{12\Delta^{[0,1]}} |^2 \tilde{V}_{\Delta^{[0,1]^{\pm}}}\!+\mathcal{B}_{\text{BPS}},
\end{equation}
where with a slight abuse of notation we denote with $\mathbf{r}\in \left\{ [0,0], [2,0], [0,2]\right\}$ the R-symmetry representation of the state exchanged in the first sum, and the 7-component vectors $V$ and $\tilde{V}$ are defined as
\beq
\label{eq: define V and tilde V}
V_{\Delta^{\mathbf{r}}}=\left(\begin{array}{c}
\theta(\mathbf{r})\left(\begin{array}{cc}[x f_{1,\Delta^\mathbf{r}}(x)]_s& 0\\ 0 & 0\end{array}\right)\\
\left(\begin{array}{cc} 0 & 0\\ 0 & [ f_{1,\Delta^\mathbf{r}}(x)]_a\end{array}\right)\\
\left(\begin{array}{cc} 0 & 0\\ 0 & [f_{2,\Delta^\mathbf{r}}(x)]_s\end{array}\right)\\
\left(\begin{array}{cc} 0 & 0\\ 0 & \left[f_{3,\Delta^\mathbf{r}}(x)\right]_a\end{array}\right)\\
\left(\begin{array}{cc}0& 0\\ 0 & 0\end{array}\right)\\
\frac{\theta(\mathbf{r})}{2}\left(\begin{array}{cc}0& [ x f_{1,\Delta^\mathbf{r}}(x)]_s\\ {}[ x f_{1,\Delta^\mathbf{r}}(x)]_s & 0\end{array}\right)\\
\frac{\theta(\mathbf{r})}{2}\left(\begin{array}{cc}0& [ x f_{1,\Delta^\mathbf{r}}(x)]_a\\ {}[ x f_{1,\Delta^\mathbf{r}}(x)]_a & 0\end{array}\right)
\end{array}\right)\,, 
\quad
 \tilde{V}_{\Delta^{[0,1]^{\pm}}}=\left(\begin{array}{c}
  0 \\
 0 \\
0 \\
0 \\
\mp[f^{1,1}_{\Delta}]_s \\
\;\;[ f^{1,-1}_{\Delta} ]_s \\
-[ f^{1,-1}_{\Delta} ]_a
\end{array}\right)\, ,
\eeq
in terms of the reduced superblocks from \cite{Liendo:2018ukf} whose definition is reviewed in Appendix \ref{app:Superblocks}.
In \eqref{eq: define V and tilde V} we use the following notations
\beq
\left[ f(x) \right]_s \equiv f(x) + f(1-x), \;\;\; \left[ f(x) \right]_a \equiv f(x) - f(1-x) ,
\eeq
and 
\beq\label{[00]requirement}
\theta(\mathbf{r}) = \left\{\begin{array}{cc} 1 & \text{if }\mathbf{r} = [0,0] , \\
0 & \text{otherwise }
\end{array} . \right.
\eeq
Finally, in (\ref{eq: final crossing equations 1a}) the sums run over superconformal primaries of the exchanged non-protected multiplets, while the contribution of all BPS states and the identity in the OPE is collected in
\begin{equation}\begin{split}
\label{eq:BPScompactcrossing0}
\mathcal{B}_{\text{BPS}} &\equiv \left(\begin{array}{cc}1 & 1\end{array}\right)V_{\mathcal{I}}\left(\begin{array}{c}1\\1\end{array}\right)+(\tilde{V}_{\mathcal{B}_3})_{11} (C^{\rm BPS}_{123})^2 +(V_{\mathcal{B}_4})_{22}(C^{\rm BPS}_{224})^2 \\
&\qquad\qquad+\left(\begin{array}{cc}C^{\rm BPS}_{112} & C^{\rm BPS}_{222}\end{array}\right)\left[V_{
\mathcal{B}_2}+\left(\begin{array}{cc}(\tilde{V}_{\mathcal{B}_1})_{11} & 0\\0 & 0\end{array}\right)\right]\left(\begin{array}{c}C^{\rm BPS}_{112}\\C^{\rm BPS}_{222}\end{array}\right),
\end{split}\end{equation}
where we used that $C^{\rm BPS}_{121} = C^{\rm BPS}_{112}$, and where the vectors $V_{\mathcal{B}_k}$, $\tilde{V}_{\mathcal{B}_k}$, containing the superblocks for the exchange of short multiplets, are reviewed in appendix \ref{app:details crossing}. 
 Using a further relation from \cite{Liendo:2018ukf}, that we report below in \eqref{eq: mixed equations minibootstrap}, we can conveniently eliminate this coefficient and rewrite the above quantity to obtain
\begin{equation}\begin{split}
\label{eq:BPScompactcrossing}
\mathcal{B}_{\text{BPS}} &\equiv \left(\begin{array}{cc}1 & 1\end{array}\right)V_{\mathcal{I}}\left(\begin{array}{c}1\\1\end{array}\right)+(\tilde{V}_{\mathcal{B}_3})_{11} +(V_{\mathcal{B}_4})_{22}(C^{\rm BPS}_{224})^2 \\
&+\left(\begin{array}{cc}C^{\rm BPS}_{112} & C^{\rm BPS}_{222}\end{array}\right)\left[V_{
\mathcal{B}_2}+\left(\begin{array}{cc}(\tilde{V}_{\mathcal{B}_1})_{11} & 0\\0 & 0\end{array}\right)+\left(\begin{array}{cc}-(\tilde{V}_{\mathcal{B}_3})_{11} & \frac{(\tilde{V}_{\mathcal{B}_3})_{11}}{2}\\
\frac{(\tilde{V}_{\mathcal{B}_3})_{11}}{2} & 0\end{array}\right)\right]\left(\begin{array}{c}C^{\rm BPS}_{112}\\C^{\rm BPS}_{222}\end{array}\right).
\end{split}\end{equation}
As we present in the next section, all $C^{\rm BPS}$ are computable as exact functions of the coupling. Thus, $\mathcal{B}_{\text{BPS}}$ collects all explicitly known pieces of the system of bootstrap equations and sources as an inhomogeneous term in the crossing equations on the unknown structure constant of unprotected operators \eqref{eq: final crossing equations 1a}.

\subsection{Constraints from localisation: topological structure constants}
\label{sec:localisation}

Structure constants for exchange of the BPS multiplets are determined by the vev of the circular Wilson loop and its derivatives. The logic behind this statement strongly relies on the fact that, for an infinite family of 1/8 BPS Wilson loops defined on curves of arbitrary shape on a two-sphere \cite{Drukker:2007yx,Drukker:2007qr}, there exists an exact localisation to 2d Yang-Mills (YM) theory \cite{Pestun:2009nn}. Given the invariance under area-preserving diffeomorphisms of 2d YM theory, the vev of the 1/8 BPS Wilson loop depends only on the area $A$ of the region enclosed by the loop and it reads
\beq
\label{WLgeneral}
w(\lambda')=\frac{2}{\sqrt{\lambda'}}I_1(\sqrt{\lambda'}), \qquad \lambda' \equiv \frac{A(4\pi-A)}{4\pi^2}\lambda \,,
\eeq
where for $A=2\pi$, the loop reduces to the maximal circle on the sphere (1/2 BPS) with vev given by $\lambda'=\lambda$ \cite{Erickson:2000af}.

The localisation to the 2d theory holds also in presence of 1/2 BPS operators \cite{Giombi:2009ds,Giombi:2012ep,Bonini:2014vta,Bonini:2015fng}. In these cases, superconformal primary operators of $\mathcal{N}=4$ SYM are associated to the Hodge-dual of the field strength of 2d YM. The key observation of \cite{Correa:2012at,Giombi:2017cqn} is that small deformations of the path of the Wilson loop (and consequently of its area) correspond to the insertion of a field strength (displacement operator) along the contour. Consequently, exploiting the relation to 2d YM, derivatives of \eqref{WLgeneral} respect to the area $A$ are equivalent to insertions of 1/2 BPS operators along the loop. In our case, the $n$-point function of appropriately normalised topological operators $O_{\mathcal{B}_1}$ of the simplest short supermultiplet can be written as the $n$-th derivative of \eqref{WLgeneral} as follows 
\begin{equation}
\langle\langle O_{\mathcal{B}_1}^n \rangle\rangle = \frac{\partial_A^n w\left(\frac{A(4\pi-A)}{4\pi^2}\lambda\right)}{w(\lambda)}\Biggl|_{A\rightarrow 2\pi}
\end{equation}
where, after setting the area to be half the surface of the two-sphere, only derivatives of the 1/2 BPS Wilson loop $w(\lambda)$ will appear.

This leads to a set of relations that exactly fix three-point functions as explicit functions of the coupling. Indeed, 
the localisation formula for the OPE coefficient corresponding to the $\mathcal{B}_1\times \mathcal{B}_1\rightarrow \mathcal{B}_2$ fusion rule was derived in \cite{Giombi:2018qox,Liendo:2018ukf} and reads
\beq
\label{eq: localisation formula for C112}
(C_{112}^{\rm BPS})^2=-1+3\frac{w\,w''}{w'^2}\,.
\eeq
Formula \eqref{eq: localisation formula for C112} can also be derived considering solely observables accessible with integrability \cite{Cavaglia:2022qpg} exploiting the same techniques presented in \cite{Cavaglia:2022yvv} to prove the integrated correlator constraints.

In the current setup, we need to consider additional topological OPE coefficients. They are related to the selection rules $\mathcal{B}_1\times \mathcal{B}_2\rightarrow \mathcal{B}_3$, $\mathcal{B}_2\times \mathcal{B}_2\rightarrow \mathcal{B}_2$, 
and $\mathcal{B}_2\times \mathcal{B}_2\rightarrow \mathcal{B}_4$ and we denoted them as $C_{123}^{\rm BPS}$, $C_{222}^{\rm BPS}$ and $C_{224}^{\rm BPS}$ respectively. In \cite{Liendo:2018ukf}, the authors obtain  $C_{222}^{\rm BPS}$ in terms of $C_{112}^{\rm BPS}$ as follows 
\beq
\label{eq: localisation formula for C222}
C_{222}^{\rm BPS}\!=\!\frac{w' C_{112}^{\rm BPS}}{(w'^2-3 w\, w'')^2}\Big[15 w^2\, w^{(3)}+2 w'^3
-9 w\, w'\, w''\Big],
\eeq
and a relation between $C_{123}^{\rm BPS}$ and the previous two that reads
\beq
\label{eq: mixed equations minibootstrap}
1+C_{112}^{\rm BPS}C_{222}^{\rm BPS}=(C_{112}^{\rm BPS})^2+(C_{123}^{\rm BPS})^2\,.
\eeq
In order to obtain the missing topological structure constants $C_{224}^{\rm BPS}$, we follow the same logic. Using the OPE decomposition \eqref{eq: D with D OPE} and the orthonormality of the operators we have
\begin{equation}\label{OB2}
\frac{\langle\langle O_{\mathcal{B}_2}^4 \rangle\rangle}{\langle\langle O_{\mathcal{B}_2}^2 \rangle\rangle^2}
=1+(C_{222}^{\rm BPS})^2+(C_{224}^{\rm BPS})^2\;,
\end{equation}
where $O_{\mathcal{B}_2}$ is the element of the supermultiplet $\mathcal{B}_2$. Finally, using $O_{\mathcal{B}_2}=\frac{1}{C_{112}^{\rm BPS}}(\tilde{O}_{\mathcal{B}_1}^2-1)$ where $\langle\langle\tilde{O}_{\mathcal{B}_1}^n\rangle\rangle=\langle\langle{O}_{\mathcal{B}_1}^n\rangle\rangle/\langle\langle{O}_{\mathcal{B}_1}^2\rangle\rangle^{n/2}$ and plugging everything into \eqref{OB2} we obtain
\begin{equation}
\begin{split}
(C_{224}^{\rm BPS})^2=\frac{3 \left(35 w^3 \,w^{(4)}
-w'^4
-20 w^2 \,w^{(3)} w'+6 w \,w'^2 \,w''\right)}{\left(w'^2-3 w \,w''\right)^2}-1-(C_{222}^{\rm BPS})^2\,.
\end{split}
\end{equation}
that completes the set of BPS structure constants appearing in \eqref{eq:BPScompactcrossing}. In the following we present the last ingredient of our setup, i.e. the exact integral constraints on the simplest four-point function.

\subsection{Integrable deformations and integrated correlators} 
The 1D dCFT is embedded into a larger 4D theory, as such it carries some memory about the bulk theory.
In particular, some specific correlation functions in the 1D dCFT  can be used to parametrise local deformations of the line within the 4D space. For instance,
to the first order, the infinitesimal deformations of the line are encoded by the insertions of the operators of the $\mathcal{B}_1$ multiplet, which is called the displacement multiplet. The top component of this multiplet -- known as the ``tilt'' operators~\cite{Billo:2016cpy} i.e. $\Phi_{\perp}^i$ --, generate an infinitesimal local deformation in the R-space i.e. changes scalar field running in the MWL. Another member of the same supermultiplet, i.e. the conformal primary with the largest dimension, generates an infinitesimal transverse bump in the contour. Thus, up to some complication of the additional contact terms, which have to be analysed carefully, any infinitesimal deformation of the contour in the R-space or in the 4D bulk space can be recast as an integrated correlator of the members of this multiplet in the original 1D dCFT. 

In addition, there is a class of deformations which break the 1D conformal symmetry but preserve integrability. In particular, a very important class of deformations consists of acting with a global rotation (in actual space as well as internal couplings space) on the half of the line, thus forming a \emph{cusp}. The cusp is associated to a regularisation-independent  anomalous dimension, which can be computed with integrability~\cite{Drukker:2012de,Correa:2012hh} with the help of a specially adapted version of the QSC~\cite{Gromov:2015dfa}. 

We are particularly interested in the case where the cusp is formed by a rotation of an angle $\theta$ in R-space in the polarization of the scalars on half of the line. The expansion of the cusp anomalous dimension in the limit of small $\theta$ reads as follows
\beq\label{gammacusp}
\Gamma^{\text{cusp}}(\theta) = \mathbb{B}(g)\sin^2\theta  + \frac{1}{4} (\mathbb{B}(g) + \mathbb{C}(g) ) \sin^4\theta +O(\sin^6\theta ),
\eeq
where we introduced two functions of the coupling $\mathbb{B}(g)$ and $\mathbb{C}(g)$. The Bremsstrahlung function $\mathbb{B}(g)$ has a meaning of the normalisation of the metric of the defect conformal manifold. It was first computed with localisation arguments in \cite{Correa:2012at} and it is given by 
\beq
\mathbb{B}(g) = 8 g^2 \partial_{g}\,\text{log}\,( w(g)),
\eeq
where $w(g)$ is the vev of the circular Wilson loop given by \eqref{WLgeneral} for $A=2\pi$ and $\lambda=16\pi^2g^2$. The second quantity appearing on the r.h.s. of \eqref{gammacusp} is the curvature function $\mathbb{C}(g)$ computed analytically from the QSC in terms of a double integral representation in~\cite{Gromov:2015dfa}. These integrals were evaluated perturbatively and numerically in \cite{Cavaglia:2022qpg}. In \cite{Cavaglia:2022yvv} we provided an attached Mathematica notebook to evaluate them numerically at high-precision for a given value of the coupling $g$.
These two functions can be related with integrated correlators in the original dCFT, providing non-trivial additional constraints on its CFT data.

This is achieved through a relation involving two integrated correlator identities, first found in~\cite{Cavaglia:2022qpg} and then rigorously derived in \cite{Drukker:2022pxk,Cavaglia:2022yvv}.
It gives the two above functions in terms of 
the 4-point function $\langle\langle  O_{\mathcal{B}_1}(x_1) O_{\mathcal{B}_1}(x_2) O_{\mathcal{B}_1}(x_3) O_{\mathcal{B}_1}(x_4) \rangle\rangle$ integrated with a certain measure.
The set of constraints can be compactly written in terms of the reduced correlator $f(x)$\footnote{The reduced correlator is introduced in Appendix \ref{app:decomposition}, and it just a part of the simplified expression for $\langle\langle  O_{\mathcal{B}_1} O_{\mathcal{B}_1} O_{\mathcal{B}_1} O_{\mathcal{B}_1} \rangle\rangle$.} as follows
\begin{equation}
\int_{0}^{1/2}
     \left(f(x) - x +\frac{ (C^{\rm BPS}_{112})^2 }{2} x^2\right)\mu_a(x) dx +\mathcal{K}_a(g) = 0\;\;,\;\;a=1,2\; ,\label{eq:constr12}
\end{equation}
where the integration measures $\mu_a(x)$ are 
\beqa
\label{eq:measures}
\mu_1(x) &=& \frac{1}{x-1}+\frac{(x-1)^2}{x^3}, \\
\mu_2(x) &=& \frac{2 x - 1}{x^2} ,
\eeqa
and the constants $\mathcal{K}_a$, $a = 1,2$ are
\beqa
\mathcal{K}_1(g) &=&  \frac{\mathbb{B} - 3 \mathbb{C} }{8 \mathbb{B}^2} + (C^{\rm BPS}_{112})^2 \left(\log(2)-\frac{7}{8} \right) + \log(2),\\
\mathcal{K}_2(g) &=& -\frac{\mathbb{C} }{4 \mathbb{B}^2 } - \frac{7}{8}(C^{\rm BPS}_{112})^2 + 1 + \log(2) ,
\eeqa
where $C^{\rm BPS}_{112}$ is given explicitly in 
\eq{eq: localisation formula for C112}.

Note that the two constraints (\ref{eq:constr12}) can be recast as the following sum rules:
\beq
\sum_{\Delta_n^{[0,0]} \in \texttt{S}^{[0,0]^+}} C_{11\Delta^{[0,0]}_n}^2 \; b_a(\Delta^{[0,0]}_n ) + \texttt{RHS}_a = 0, \text{  with  }a = 1,2,\label{eq:intsumrule}
\eeq
where $b_a(\Delta) \equiv \int_0^{\frac{1}{2}} \mu_a(x) F_{\Delta}(x) dx$ are defined in terms of the superblock $F_{\Delta}(x)$ and the measures $\mu_a(x)$, and $\texttt{RHS}_a(g)$ are explicit functions of the coupling. Expressions for these building blocks are presented in appendix \ref{app:blocksintconstraints}.

In closing this section, we comment that these are certainly not the only integrated correlator constraints one can derive by similar arguments. In fact, one can compute from integrability not just the standard cusp anomalous dimension, but also an infinite spectrum of excitations obtained by inserting local operators at the cusp location. They are simply additional solutions to the QSC equations, and their dimensions are expected to be related to additional integrated correlator constraints, for different correlation functions which may involve also non-protected operators. 
It would be interesting to deduce such extra conditions on the theory and exploit them in the bootstrap program.

\section{Numerical bootstrap setup}
\label{sec:Algorithm}
In this section we describe how we impose the crossing constraints (\ref{eq: final crossing equations 1a}) with the numerical bootstrap in the current setup. Since the spectrum is accessible with the QSC, we will treat the crossing equations as constraints on the possible values of OPE coefficients for non-protected operators and obtain bounds on this missing data.

\subsection{\texttt{SDPB} implementation}\label{sec:SDPB}

Numerical bootstrap algorithms allow one to bound the values of conformal data compatible with a given system of crossing equations \cite{Rattazzi:2008pe, Kos:2014bka, Simmons-Duffin:2015qma, Kos:2015mba, Kos:2016ysd, Chester:2019ifh, Reehorst:2021ykw, Liu:2023elz}, also see reviews of the main developments of these methods \cite{Poland:2018epd, Rychkov:2023wsd}.

In the approach taken below, which follows the main logic of  the \textit{cutting-surface} algorithm of  \cite{Chester:2019ifh},
we try to find the boundaries of the allowed region by asking a sequence of questions, where each time the bootstrap operates in \textit{oracle-mode}: we test some given CFT data with an option to exclude it if \texttt{SDPB} can prove it to be inconsistent. For simplicity we refer to those outputs as \texttt{True}/\texttt{False} where \texttt{False} means that the data is rigorously excluded. Since every negative answer actually excludes a whole region and not just a point, by scanning over the data appropriately we can get a better and better determination of the allowed region. 
 While the main ideas follow \cite{Chester:2019ifh}, our setup is different since we have exact knowledge of the spectrum. We will also adopt some different technical solutions as described below. More modern methods such as the navigator \cite{Reehorst:2021ykw} or the skydiving \cite{Liu:2023elz} methods adopt a different logic and might be advantageous to study our theory as well, but they were not necessary for the present setup.
Let us start by describing the elementary oracle-mode step. 

\paragraph{Setup of the oracle. } 

Generally, the data of which we want to inspect the consistency are the scaling dimensions  for a number of low-lying operators, and the corresponding OPE coefficients. Schematically, for $a,b$ being either $1$ or $2$ for our system,
\beq\label{eq:data}
\texttt{candidate data}  =\left. \left\{ \left\{ \Delta_{O}, \;\;\; C_{a b \Delta_O}\right\} , \;\;  {O} \in \mathcal{B}_a \times \mathcal{B}_b \right\}\right|_{\Delta < \texttt{cutoff}}, \nonumber
\eeq
where we keep only the operators below a certain threshold in $\Delta$. 
More precisely, the spectra we exchange are naturally divided into the five groups (\ref{eq:spectra}), where we drop the $\pm$ superscript for $\texttt{S}^{[0,0]^+}$, $\texttt{S}^{[2,0]^-}$ and $\texttt{S}^{[0,2]^+}$ since we know from parity considerations that these states are the only ones with non-vanishing OPE coefficients.
We consider independent spectral cutoffs for the 5 different types of spectra entering our system denoting them as
\beq\label{eq:cutoffs}
\Delta_{\text{gap}}^{[0,0]}, \;\; \Delta_{\text{gap}}^{[0,2]}, \;\; \Delta_{\text{gap}}^{[2,0]}, \;\; \Delta_{\text{gap}}^{[0,1]^-}, \;\; \Delta_{\text{gap}}^{[0,1]^+}.
\eeq
 Concretely, in our Bootstrability setup each $\Delta_{\text{gap}}$ is taken to coincide with the last spectral level computed from the QSC in each  channel. In this way we maximise the information on the spectrum fed into the algorithm.  
 
In our setup, the scaling dimensions of low-lying states coming from the QSC can be considered exact.\footnote{The errors in these data coming from the numerical solution of the QSC can be controlled easily. In our studies we used data with $\sim 20$ digits for the spectrum, and we estimate that this uncertainty has negligible effects on the scale of the bounds we report for the OPE coefficients.}  
Therefore, the only actual unknowns in the \texttt{candidate data} are the values of the OPE coefficients below each spectral cutoff. In addition, we do not make any assumptions, apart from the reality and parity selection rules described above, about what happens above the cutoffs. Using well known semi-definite programming methods available in \texttt{SDPB}~\cite{Simmons-Duffin:2015qma,Landry:2019qug}, we can build an oracle which, given a CFT data, is sometimes able to exclude it by proving it to be inconsistent. 
\begin{framed}
\beq\label{eq:oracle}
\underbrace{\texttt{candidate data} }_{\texttt{QSC spectrum}+\texttt{candidate OPE coeff's} }\!\!\!\!\!\!\!\!\!\!\Rightarrow \;\text{allowed?}\;\Rightarrow \;\texttt{SDPB}\;\Rightarrow\; \text{maybe/no}.  
\eeq
\end{framed}
Here , `maybe' means that the \texttt{SDPB} algorithm (operating with some search parameters), is not able to rigorously exclude the data by manifesting their inconsistency. Thus, we can say that when we get `no' for an answer, the data are surely inconsistent (provided we control effectively any systematic errors), while the data passing the test of a specific \texttt{SDPB} run  might still potentially be excluded, as we show below, by asking a more refined question, where we input more spectral data, increase the resolution of the algorithm (controlled by the cutoff $\Lambda$, see below) and/or scan over a larger OPE space.

Let us review how this is done. First, we split the crossing equations (\ref{eq: final crossing equations 1a}) as
\beq\label{eq:crossingsplit}
\underbrace{ \mathcal{B}_{\text{BPS}}  + \sum_{\Delta < \Delta_{\text{gap}}} \left( \dots \right)  }_{\equiv \mathcal{F}( \texttt{candidate data}) } +  \sum_{\Delta \geq \Delta_{\text{gap}}} \left( \dots \right) = 0 ,
\eeq
where the ellipses denote that we just rearrange the terms of (\ref{eq: final crossing equations 1a}) separating the spectral sums in the low-lying part below our chosen cutoff(s). We call $\mathcal{F}$ the combination of these truncated sums together with $\mathcal{B}_{\text{BPS}}$. It is a 7-components vector of functions of the cross ratio, depending explicitly on the \texttt{candidate data}. We give its form explicitly in \eqref{eq: define V and tilde V} and appendix \ref{app:details crossing}. 

The strategy is now to build an exclusion plot for the data. To do that  we search for an appropriate linear functional that, acting on the system of relations (\ref{eq: final crossing equations 1a}), manifests an inconsistency. As standard in the numerical bootstrap, this linear functional will be represented as a linear combination of derivatives of the equations (\ref{eq: final crossing equations 1a}) with respect to the cross ratio, at the crossing-symmetric point $x = \frac{1}{2}$, i.e. such a functional acts on vectors of functions of the cross ratio $x$ as
\beq
\alpha: \;\;\;\; \left( \begin{array}{c} W_1(x) \\
W_2(x) \\
\vdots \\
W_7(x) \end{array} \right) \rightarrow \sum_{k=1}^7 \sum_{n=1}^{\Lambda} \alpha_{k,n} \left. \frac{d^n}{dx^n} W_k(x) \right|_{x = \frac{1}{2} } \in \mathbb{R} , \label{eq:alphaaction}
\eeq
where $\Lambda$ is a truncation parameter which can be increased in order to get more and more precise bounds.

Our \texttt{candidate data}  will be excluded if a functional with the following two properties exists:
\begin{itemize}
\item \texttt{Normalization}: 
\beq\label{eq:normalize}
\alpha\left[ \mathcal{F}\left( \texttt{candidate data} \right) \right] = 1 .
\eeq
\item \texttt{Semidefinite positivity}: When acting on the blocks with $\Delta $ above the cutoff(s), the functional returns a semi-definite positive result (which is a $2 \times 2$ matrix in the case of the $V$-type blocks and a single element for $\tilde V$-type blocks \eqref{eq: final crossing equations 1a}). Precisely, the conditions are
\begin{equation}\begin{split}\label{eq:positive2}
\alpha\left[ V_{\Delta^{[0,0]}} \right] &\succcurlyeq 0 , \;\;\;\;\quad \Delta \geq \Delta_{\text{gap}}^{[0,0]},\\
\alpha\left[ V_{\Delta^{[2,0]}} \right] &\succcurlyeq 0 , \;\;\;\;\quad \Delta \geq \Delta_{\text{gap}}^{[2,0]} , \\
\alpha\left[ V_{\Delta^{[0,2]}} \right] &\succcurlyeq 0 , \;\;\;\;\quad \Delta \geq \Delta_{\text{gap}}^{[0,2]}, \\
\alpha\left[ \tilde{V}_{\Delta^{[0,1]^{\pm}}}\right] &\geq 0 , \;\;\;\;\quad \Delta \geq \Delta_{\text{gap}}^{[0,1]^\pm} .
\end{split}\end{equation}
 \end{itemize}
 In semidefinite programming approaches, the way this kind of semidefinite positivity conditions on a semi-infinite interval can be imposed is by using a rational-type approximation for the blocks as functions of $\Delta$ \cite{Poland:2011ey}. This approximation is easy to find and we review the details in appendix \ref{app:Deltaapprox}. As we discuss there, the precision of this approximation of the blocks depends on one truncation parameter which we denote as $N_{\text{poles}}$. 
This is a potential source of systematic errors in the bootstrap bounds which should be controlled. Upon varying this parameter, we estimate that its impact is not visible on the scale of our bounds.

 Given the reality properties for the OPE coefficients entering the system, it is clear that, if a functional satisfying these conditions is found, the candidate data is excluded, since otherwise we would have the following impossible equation
\beq\label{eq:crossingsplit2}
0 = \alpha\left[ \text{ (\ref{eq:crossingsplit}) }\right] = \alpha[\mathcal{F}(\texttt{candidate data})] + \sum_{\Delta\geq \Delta_{\text{gap}}} \alpha[ (\dots )] = 1 + {\rm positive} \geq 1 .
\eeq
We have therefore an elementary `oracle' working as depicted in (\ref{eq:oracle}): given some candidate data, the \texttt{SDPB} algorithm searches for a linear functional proving their inconsistency.

\paragraph{Exclusion region after an \texttt{SDPB} run. } 

\begin{figure}
    \centering
    \includegraphics[width=\columnwidth]{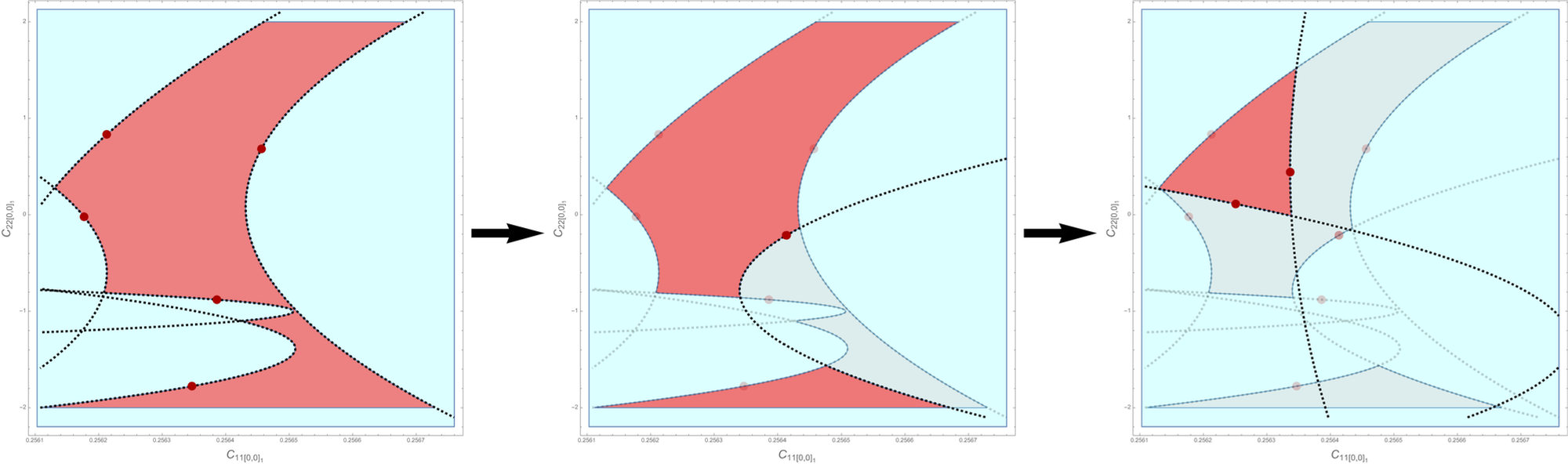}
    \caption{The allowed 2-dimensional region (in red) for the structure constants $C_{11\Delta^{[0,0]}_1}$ (x-axis) and $C_{22\Delta^{[0,0]}_1}$ (y-axis) for $g=1/5$, $\Lambda=60$ and $N_{\text{poles}}=30$. Red dots are the points evaluated to be inconsistent by the `oracle' while dashed black lines are the related quadratic equations cutting the surface as in \eqref{eq:quadratic}. In this sequence, the non-convex region on the left is divided in two unconnected ones by a single cut and at the end reduced to a single region by two subsequent iterations.}
    \label{fig:NonConnected}
\end{figure}

As mentioned above, when a run of \texttt{SDPB} excludes a given point in OPE space, then it automatically excludes a region defined by an inequality, quadratic in the OPE coefficients~\cite{Chester:2019ifh}.

In fact, suppose that, testing some candidate data encoded in the vector $\mathcal{F}_{\ast}$, we find a functional $\alpha_{\ast}$ excluding it. Then we have by construction
\beq
\alpha_{\ast}\left[ \mathcal{F}_{\ast} \right] = 1 ,
\eeq 
but the \emph{same} functional can also be used to exclude all the candidate data such that $\alpha_{\ast}\left[ \mathcal{F}\left( \texttt{candidate data} \right) \right] > 0 $. 
Turning this around, CFT data should satisfy
\beq\label{eq:quadratic}
\alpha_{\ast}\left[ \mathcal{F}\left( \texttt{candidate data} \right) \right] \leq  0 ,
\eeq
which is valid \emph{in general}, even if the functional $\alpha_{\ast}$  is found by $\texttt{SDPB}$ in order to exclude some specific candidate OPE coefficients. 

Since $\mathcal{F}$, defined explicitly  in (\ref{eq:Fdef}), is a quadratic combination of the OPE coefficients below the cutoffs, this means that for each functional $\alpha_{\ast}$ found by \texttt{SDPB}  we gain a new quadratic inequality (\ref{eq:quadratic}) delimiting the allowed values of OPE coefficients as in Figure \ref{fig:NonConnected}.

The path is now clear: we want to perform a sequence of runs of \texttt{SDPB} that each time collect new constraints and progressively shrink the allowed space of OPE coefficients. 
This is non-trivial, since we are exploring a higher-dimensional space. The most challenging step of the procedure is finding the next point to use as candidate data after each \texttt{SDPB} run. At certain point \texttt{SDPB} will fail to find a new excluding functional, and at this point we start exploring the bounds of the allowed region as we describe in the next sections.

\paragraph{Pointwise positive conditions: skipping states. } 

So far we described the situation where we
 treat all states below the gaps on the same footing, with their OPE coefficients as data we want to test. 
We search for a  functional satisfying the positivity conditions which, if  found, excludes a particular choice of OPE data for all these states. This method exploits the knowledge of the exact spectrum below the gaps, but it has the disadvantage that, to build the boundaries of the allowed region, we need to scan over a large number of parameters, i.e. all the OPE coefficients for the states below the gaps.

Alternatively we can focus on a smaller subset of the OPE coefficients, but still take advantage of the sharp knowledge of the spectrum of multiple low-lying states. To do this we want to impose positivity of the functional only where it is actually required, relaxing this condition for the values of $\Delta$ where we are certain there are no physical states.

This method provides a certain compromise, as it avoids the need to scan over a large space of OPE coefficients, but still allows us to use a large number of spectral data as input.

It works as follows. We divide the spectral levels in two groups:  $\mathbb{D}_{\text{pick}}$ and $\mathbb{D}_{\text{skip}}$, i.e.
\beq
\!\!\texttt{levels below gaps}\! = \!\!\left. \left\{\! \Delta  \!\in \texttt{S}^{[0,0]}\!\cup \texttt{S}^{[2,0]}\!\cup\texttt{S}^{[0,2]}\!\cup \texttt{S}^{[0,1]^{-}}\!\!\!\cup \texttt{S}^{[0,1]^{+}}\! \right\}\right|_{\Delta<\texttt{gaps}}  \!\!\!\!\!\!\!\!= \mathbb{D}_{\text{pick}} \!\cup \mathbb{D}_{\text{skip}} ,
\eeq
where the states in $\mathbb{D}_{\text{pick}}$ are the ones of which we want to test the OPE data,
\beq
\texttt{candidate data}' = \left. 
\left\{ \{\Delta_O,\; C_{a b \Delta_O}\} ,  \;\;\; O \in \mathcal{B}_a \times \mathcal{B}_b
\right\} \right|_{\Delta \in \mathbb{D}_{\text{pick}}} 
\; .\label{eq:reduceddata}
\eeq
To exclude these data the positivity of the functional is imposed as before for  states above the gaps \eqref{eq:positive2}, as well as pointwise for the states in $\mathbb{D}_{\text{skip}}$:\footnote{By default \texttt{SDPB} imposes positive definitiveness on a semi-infinite interval. But one can use various tricks, known in the literature, to also impose positivity on a discrete set or on a finite intervals as e.g. described in~\cite{Cavaglia:2023mmu}.}
\begin{equation}\begin{split}\label{eq:positive21}
\alpha\left[ V_{\Delta^{\mathbf{r}}} \right] &\succcurlyeq 0 , \;\;\;\;\quad \Delta^{\mathbf{r}}  \in \mathbb{D}_{\text{skip}}\quad\text{and}\quad\mathbf{r}\in \left\{[0,0],[2,0],[0,2]\right\},\\
\alpha\left[ \tilde{V}_{\Delta^{[0,1]^{\pm}}} \right] &\geq 0 , \;\;\;\;\quad  \Delta^{[0,1]^{\pm}}  \in \mathbb{D}_{\text{skip}}.
\end{split}\end{equation}

If we want to bound only a few OPE coefficients, this method is more economical. However, scanning over a larger OPE space can be expected to lead to tighter bounds. This is what was found in \cite{Kos:2016ysd} (see also \cite{Chester:2019ifh,Chester:2020iyt,Rychkov:2023wsd}), where it was shown that, for mixed-correlator bootstrap problems, the scan over OPE coefficients can lead to tighter bounds on the rest of the CFT data (in that case, the spectrum). In that setup, this gain is due to the fact that  scanning over the OPE coefficients that enter the equations with a non-trivial matrix structure allows one to relax conditions on the functionals searched by \texttt{SDPB}, and thus allows one to prove the inconsistency of more data. 

 As we discuss in section \ref{sec:Results},  also in our system, scanning over a larger OPE space leads to tighter bounds. A combination of the two techniques is also very useful in practice to narrow the allowed region for the OPE coefficients gradually before increasing the  dimensionality of the parameters to scan over, a strategy which gives a significant speed gain. 

\paragraph{Including integrated correlators. } We have described the core \texttt{SDPB} step of the algorithm using only the crossing equations as constraints. As done in \cite{Cavaglia:2022qpg}, one can easily also include the two integrated correlators conditions  (\ref{eq:constr12}). This is simple to do since these two equations are already written in the form of sum rules (\ref{eq:intsumrule}) involving the exchange of the $\texttt{S}^{[0,0]}$ spectrum only. Thus, they can simply be added to the system of  constraints obtained as derivatives of the crossing equations. Concretely, the linear functionals searched by \texttt{SDPB} will then act on a space with two extra directions, one for each integrated correlator constraint (see \cite{Cavaglia:2022qpg} for details).
 The inclusion of the integrated correlators makes the size of the allowed region shrink dramatically. 

\subsection{Scanning algorithm for OPE coefficients}\label{sec:scanningalg}

When scanning for the allowed region of OPE coefficients, 
we need to find the next point to test after each \texttt{SDPB} run. 
The new point should be chosen in such a way that it is roughly in the 
centre of the remaining allowed region, in order to shrink it quickly.

To achieve this, we have developed an algorithm that we describe here.
There are a number of more advanced algorithms~\cite{Reehorst:2021ykw} available, which we use for inspiration (for a comprehensive review see \cite{MiniCourse2023}).
However, our goal is to develop an algorithm specifically for the large number of dimensions of the OPE coefficients that we are considering, and to make it as simple as possible to use in our setup.

\paragraph{Basic idea and challenges.} 
We start by choosing an initial \textit{scanning area} for the values of the OPE coefficients we want to inspect. This is done by making an educated guess on  an area bounding the actual allowed island, based on previously obtained data.  In particular, we make the following choice
\begin{equation}\label{scanarea}
-2\leq C_{22\Delta^{[0,0]}}\leq 2\qquad\qquad 0\leq\;\begin{matrix}
C_{11\Delta^{[0,0]}} \\
C_{22\Delta^{[0,2]}} \\
C_{22\Delta^{[2,0]}} \\
C^2_{12\Delta^{[0,1]^{-}}}
\end{matrix}\;\leq 2\qquad\qquad -2\leq C^2_{12\Delta^{[0,1]^{+}}} \leq 0 ,
\end{equation}
and this also takes into account that OPE coefficients can have a negative sign as in \eqref{eq:C12positivity}. Note that we can always pick the arbitrary sign in the normalisation of the non-protected operators to ensure the validity of the first $3$ inequalities in the middle column of (\ref{scanarea}); but we do not have freedom left to fix the signs of $C_{22\Delta^{[0,0]}}$, which is why we keep them arbitrary.

As explained in \ref{sec:SDPB}, after each run with some candidate data (which are represented as a \emph{point} in the region), the answer `no' excluding the data means that \texttt{SDPB} finds a linear functional satisfying certain positivity constraints,
 which defines a new quadratic inequality on the OPE data. This is added to the list of previously obtained inequalities, and thus the allowed region shrinks further. The inequalities defining the allowed space are of the general form $C_i C_j M^{ij}>M_0$,
where $M^{ij}$ is some symmetric constant matrix and $M_0$ is a constant and $C_i$'s are the OPE coefficients we are scanning over. 

Given that at a generic iteration the allowed domain is defined by a large set of inequalities, there are two main tasks:
\begin{enumerate}
    \item Finding a point approximately in the center of the domain, to be used as candidate data in the next iteration.
    \item Finding the maximal and minimal allowed values  of any given OPE coefficient $C_{ij\Delta_k}$ (we call these {\it shadows}, and they are simply the projections of the allowed domain on the axis corresponding to a particular OPE coefficient). 
\end{enumerate} 
The main challenge is that, especially in multidimensional space, there are not many efficient and reliable algorithms  for these tasks.
Partly, this is due to the fact that the region of allowed data is in general non convex (see Figure \ref{fig:NonConnected}). Non-convex optimisation problems are hard to solve in general and NP-hard in the worst case. 
On top of this,  the domain is defined by a large number of inequalities and the number of dimensions of the scan may also in general be large.
Thus, the range of tools we can use is rather limited. 
One thing which works well is reducing the set of inequalities along a  line in the multidimensional space.
We found that the built-in \verb"Reduce" function in Mathematica is a good tool for this task. 

\paragraph{Finding a point in the centre of the domain.}
For the first task, we adopted an algorithm known before (see e.g. \cite{MiniCourse2023}). 
First of all, given the set of inequalities, we need to find a point inside the domain they define. For that, we can, for example, generate a large number of random points in our original scanning area and check if they satisfy all the inequalities.
Even though this is quite an efficient and fast method, depending on the domain, sometimes it still takes a long time to find a point inside. 
To optimise the process, we sample the random points within the hyper-rectangle defined by the shadows of the region (we will explain shortly how the shadows are computed). Just to give a simple example: imagine the domain is a sphere; then the probability of finding a point inside in the smallest box containing the sphere in the 10D case is already quite small $\sim 0.002$, becoming $\sim 2\times 10^{-8}$ in the 20D.

\begin{figure}
    \centering
\includegraphics[width=\columnwidth]{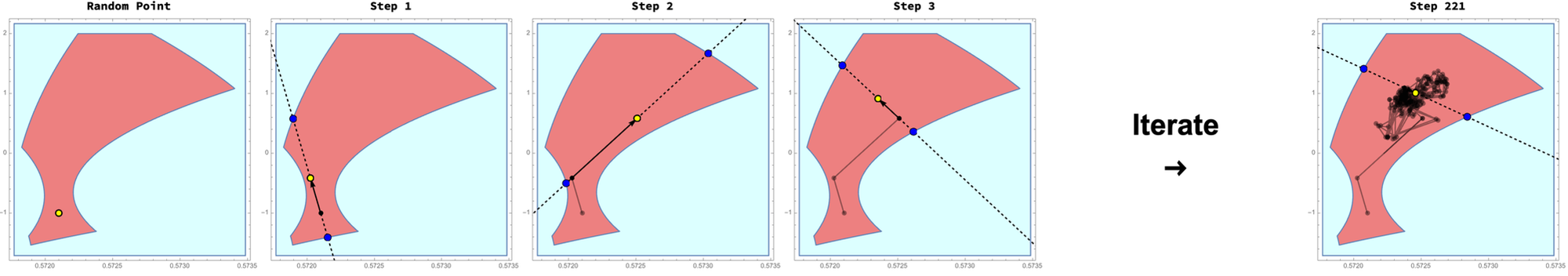}
    \caption{Example of search of a point roughly at the centre of the domain of OPE coefficients $C_{11\Delta^{[0,0]}_1}$ (x-axis) and $C_{22\Delta^{[0,0]}_1}$ (y-axis) at $g=3/2$, $\Lambda=60$ and $N_{\text{poles}}=30$. The Yellow dot is the point found at any iteration and used as starting point for the next. Blue dots represent the intersections between the random line (dashed) and the boundary of the allowed region in red. At each step, the algorithm moves (black arrow) to the mid-point between the blue dots. In the current case, the search stops after 221 iterations obtaining \texttt{centricity}>0.95.}
    \label{fig:centroid}
\end{figure}

Having found a point inside the domain, we pick a random line passing through this point.
The direction is chosen randomly and the space is rescaled by rescaling the hyper-rectangle defined by the shadows to the unit hypercube, to 
ensure that the randomisation of the direction of the line probes all directions in the space equally.
Then, we reduce the set of inequalities so that they become linear inequalities for points of this line.  
Considering this system of one-dimensional inequalities, we choose the point in the center of the intersection of the line with the domain. In the most general case, the inequalities reduce to a set of disconnected intervals; in this case, we pick the biggest interval and select its center.
Then we repeat this process several times, changing the direction of the random line each time. The convergence is estimated by how central the previous point is along the random line, giving a \texttt{centricity}
rating between 0 and 1 ($1$ is given when the starting points happen to be on the center of the random line). Then we average the \texttt{centricity} over several iterations to 
determine the exit condition.
After this process, we have a point approximately in the center of the domain allowed by all the inequalities hitherto collected.
The process is illustrated in Figure \ref{fig:centroid}.

Now, we use this point as the candidate data for the test by \texttt{SDPB}. 
If this new point is excluded, we gain a new
inequality excluding some portion of the space. We then have an updated allowed domain. We now describe how to recompute the shadows of the new domain. 
\begin{figure}
    \centering
    \includegraphics[width=\columnwidth]{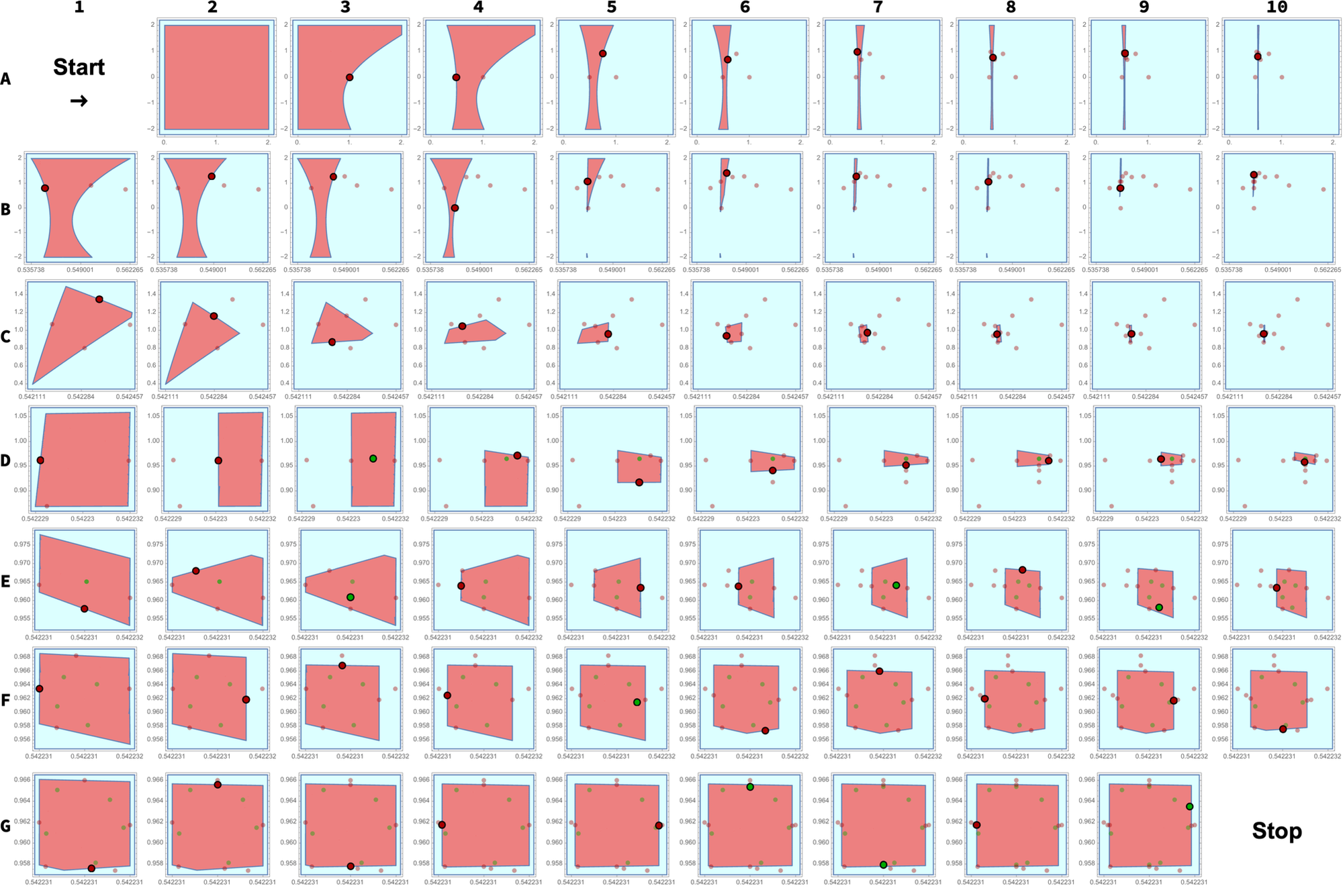}
    \caption{Example of reduction of the allowed domain of OPE coefficients $C_{11\Delta^{[0,0]}_1}$ (x-axis) and $C_{22\Delta^{[0,0]}_1}$ (y-axis) at $g=1$, $\Lambda=60$ and $N_{\text{poles}}=30$. The sequence of pictures starts in A2 and continues from left to right up to G9. Since they become too small in the original scale, each region in column 10 is rescaled to the unit hypercube at the beginning of each row (column 1). The starting region A2 is bounded by our initial choice for the OPE coefficients \eqref{scanarea}. In this example we are not using parallelisation, then at any step the algorithm chooses one point (highlighted with black border) to be analysed by the oracle. 
    If the point is excluded (red), the domain is reduced by means of the quadratic inequality as in \eqref{eq:quadratic}. After few steps (28 in this case in D3), the first \texttt{True} point (green) is found. From then on, the goal is to find the size of the smallest hypercube containing the allowed region. Its size bound the possible values of the individual OPE coefficients. The algorithm stops in G9 when the rescaled distances between allowed points and the boundary are all less than a target value we call \texttt{stopper}, in this case it is 1/100 assuring that the error on the bounds data is less than 1$\%$ of the rescaled distance.}
    \label{fig:reduce}
\end{figure}

\paragraph{Finding the shadows.}

We call {\it shadows} the extreme values of the structure constants in the potentially allowed region defined by the quadratic inequalities.
To the first approximation, one can use the built-in \verb"Minimize" and \verb"Maximize" functions in Mathematica.
However, this method fails quite quickly as the number of dimensions and inequalities grows and also is quite slow.
Thus we adopted a different method.
We start again from some point inside the domain.
We then linearise the inequalities around this point (by re-centering around these values of the OPE coefficients and dropping quadratic terms). 
After that, we can extremise the simplified inequalities with a built-in highly efficient linear optimisation. In most situations, this gives a rather good approximation right away, as many of the original quadratic inequalities typically have huge coefficients and are effectively
linear. However, sometimes the point extremising one of the OPE coefficients within the linearised set of equations may appear outside the domain defined by the initial exact inequalities; in this case, we connect this point by a straight line to the previous point and find the extremum along the line, ensuring the point is at the boundary of the domain.
After that, we repeat the linearisation procedure again. 
Various variations of this strategy include limiting the range by which the new point is moved from the previous one, to avoid the situation when the point is moved too far away from the linearisation point and the linear approximation is not valid anymore.
Or we can re-run the process several times with different random starting points to ensure the robustness of the result.
Of course, there is no method which is guaranteed to work in all cases, but the method we adopted works well in most of the cases we tested on real data.

One challenge is that the domain may become disconnected like on Figure \ref{fig:NonConnected}.
In this case, the determination of the shadow may find the extremum in one of the domains only.
However, we observed that in most cases the disconnected domain gets removed after several iterations, as the domain is reduced to a single connected region.

\paragraph{Estimating the bounds for the OPE coefficients.}

As long as the oracle keeps excluding  the points we test, i.e. as long as it keeps finding new functionals $\alpha$ satisfying all semidefinite positivity constraints, the allowed region for the OPE data keeps shrinking. 
This allowed region is delimited by the set of inequalities we are collecting along the runs.

At some point, we  pick a point to test inside this region for which the functional cannot be found. Within the parameters of fixed OPE scan space and fixed truncation $\Lambda$, this point belongs to the island of allowed data. For simplicity, we refer to such a point as a \texttt{True} point.
Now our task becomes finding the actual size of the island, so we want to find \texttt{True} points lying as close as possible to the boundary of the  region delimited by the inequalities 
(see Figure~\ref{fig:reduce}).

Let us explain how this is done. Suppose we have found a number of \texttt{True} points. 
We connect the one  with the smallest/biggest value of a given OPE coefficient
with the point  minimising/maximising the value of this OPE coefficient according to the hitherto-known inequalities. 
Then we used randomly several strategies - a) hit at the centre of this line with the oracle test
b) cut the domain with a plane going through the \texttt{True} point and hit the centre of the domain given by the intersection of the plane with the allowed region
c) a combination of a) and b) where we find the centre of the line connecting the \texttt{True} point with the extremum and then 
hit the centre of the portion of the plane intersecting the allowed region orthogonal to this point.
In this way we either find a new \texttt{True} point, closer to the boundary, or find a new constraint, thus reducing the domain further and reducing the distance between the boundary and the \texttt{True} point.
The process is illustrated in the last $4$ rows of Figure~\ref{fig:reduce} (from the D3 tile and upwards).

Iterations are terminated when all the rescaled distances between \texttt{True} and $\texttt{False}$ points reaches a pre-defined threshold we name \texttt{stopper}. This is equivalent to assure that the relative error on the \textit{shadows} is less than \texttt{stopper}.

\paragraph{Parallelisation.}
Whereas \texttt{SDPB} can use multiple cores, its productivity does not grow linearly, at least in our setup, with the number of cores.
So we introduced a higher-order parallelisation. For example, instead of finding the centre of the domain, we first divide the domain roughly into $n$ chunks and apply the method in parallel for each of the sub-domains,
generating $n$ inequalities simultaneously. 
Or at the stage after finding the allowed point, we can use the method described above to proceed in parallel reduction of the domain in various projections, reducing the error of the determination of the allowed range for different 
OPE coefficients at the same time.

\section{Numerical results}
\label{sec:Results}

In this section we present the analysis of our numerical results for several OPE coefficients in our mixed-correlator setup which are reported fully in Appendix \ref{app:bounds}.

\subsection{Harvesting the data}\label{sec:harvesting}

The mixed-correlator system introduced in Section \ref{sec:Bootstrability} allows one for the simultaneous computation of multiple OPE coefficients. Exploiting the algorithm described in section \ref{sec:Algorithm} together with the standard semi-definite programming tool \texttt{SDPB} \cite{Simmons-Duffin:2015qma,Landry:2019qug}, we are able to scan a multidimensional allowed region in the space of the structure constants. From this compact region, one can extract upper and lower bounds for each individual OPE coefficient.

Our data is obtained by incorporating in the algorithm the two integrated correlators \eqref{eq:constr12}, the parity selection rule \eqref{eq:parityselection} and the reality condition \eqref{eq:C12positivity}. For the \texttt{SDPB} input we use a truncation to $\Lambda=60$ derivatives, and
a polynomial approximation of the blocks with polynomials of degree $N_{\text{poles}} = 30$.
We also integrate the
knowledge of the dimensions of all the lowest-lying states available for the spectra participating to the mixed system as described in section \ref{sec:Bootstrability}. In particular, considering the parity/reality discrete symmetries, we have used 9 parity even states for the spectrum $\texttt{S}^{[0,0]}$, 3 parity even states for the spectrum $\texttt{S}^{[0,2]}$, 3 parity odd states for the spectrum $\texttt{S}^{[2,0]}$ and 3 parity even and 6 parity odd states for the spectrum $\texttt{S}^{[0,1]^\pm}$, which are depicted in Figure \ref{fig:spectrumplot}.
The local operators' dimensions computed with QSC have at least 20 digits of precision. The systematic error coming from the truncation involved in the approximation of the blocks and the numerical uncertainty of the spectral data is negligible with respect to the significant digits of the numerical bootstrap bounds.

For each type of spectrum exchanged in our crossing equations, in order to maximise the information injected from the integrability data, we always set $\Delta_{\text{gap}}$ equal to the largest scaling dimension available from the QSC. Based on the states included in our construction from integrability and considering the choice for $\Delta_{\text{gap}}$, the structure constants that can be analysed by our algorithm are the following 
\begin{equation}\small\begin{split}\label{Clist}
&\qquad\qquad\quad\qquad\qquad\qquad\quad\texttt{S}^{[0,0]^+} \begin{array}{c} 
\nearrow C_{11\Delta^{[0,0]}_n} \\
\searrow C_{22\Delta^{[0,0]}_n}
\end{array} \quad n=1,...,8\,,\\
&\texttt{S}^{[0,2]^+} \rightarrow C_{22\Delta^{[0,2]}_n} \quad\;\;\; n=1,2\,,\qquad\qquad\quad\quad
\texttt{S}^{[2,0]^-} \rightarrow C_{22\Delta^{[2,0]}_n} \quad\;\;\; n=1,2\,,\\
&\texttt{S}^{[0,1]^-} \rightarrow C^2_{12\Delta^{[0,1]^{-}}_n}\quad\;
n=1,..,5\,,\qquad\qquad\quad
\texttt{S}^{[0,1]^+} \rightarrow C^2_{12\Delta^{[0,1]^{+}}_n}\quad\; n=1,2\,.
\end{split}\normalsize\end{equation}
We select a subset of the OPE coefficients appearing in \eqref{Clist} to be scanned by our algorithm forming a $D$-dimensional space of parameters where $D$ is the number of structure constants included. This means that we consider these states on the spectra to be in the set $\mathbb{D}_{\text{pick}}$,  with the notation introduced in the previous section \ref{sec:SDPB}. All the remaining states up to $\Delta_{\text{gap}}$ are included in $\mathbb{D}_{\text{skip}}$ and implemented in our system as presented in \ref{sec:SDPB}.
It is important to note that in the crossing equations \eqref{eq: final crossing equations 1a}, once we include a state in the spectrum $\texttt{S}^{[0,0]}$,  the OPE coefficients $C_{11\Delta^{[0,0]}_n}$ and $C_{22\Delta^{[0,0]}_n}$ are always computed simultaneously due to \eqref{[00]requirement}. As a result, the dimension of the scanned space, $D$, increases by 2. In contrast, for states in other sectors, only a single structure constant needs to be considered.

 \begin{figure}[t]
\centering
    \includegraphics[width=0.8\columnwidth]{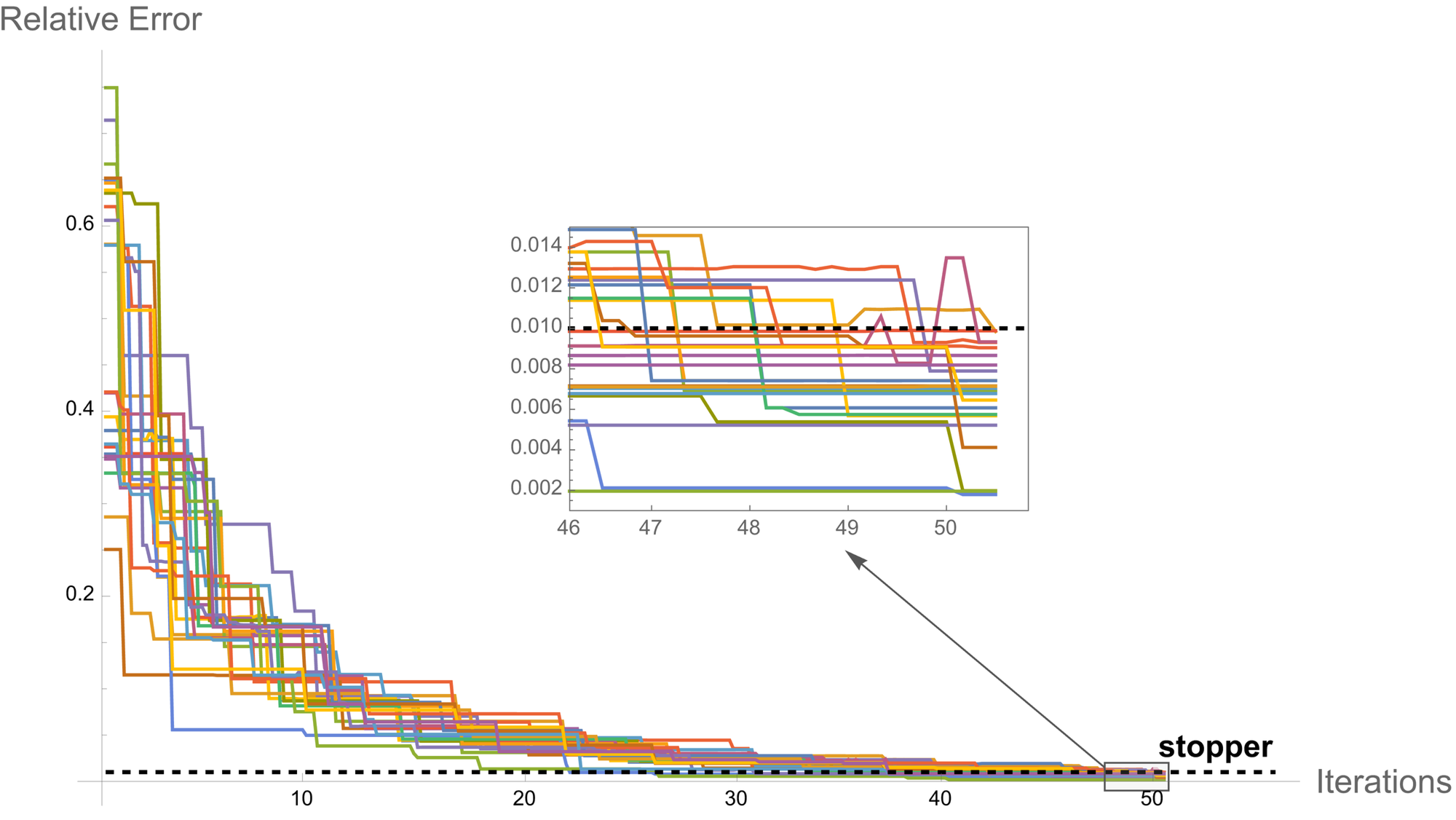}
\caption{The evolution of the relative error on the 24 \textit{shadows} of a 12D dataset for $g=1/2$, $\Lambda=60$ and $N_{\text{der}}=30$. The error computation starts as soon as the algorithm finds a \texttt{True} point and it stops as soon as all the lines reach the tolerance parameter \texttt{stopper} that in this case is 1/100.
}
\label{fig:stopper}
\end{figure}

At each iteration of the code, the algorithm presented in section \ref{sec:scanningalg} is used to select points of the $D$-dimensional space to be tested by the oracle. In the current case, six points are collected in parallel. This choice represents the most effective balance between computational performance and execution time. In practice, this choice optimises the search for the first allowed point by dividing the region into 6 portions and searching for the centre within each subregion. Testing revealed that parallelising the code to search for more than six points simultaneously increases the overall search time, as the subregions become too small, leading to several redundant inequalities.
Iterations are terminated when the precision of the \textit{shadows} (projections of the $D$-dimensional region on the axes) reaches the pre-defined threshold \texttt{stopper}. For the datasets presented in this section, we impose \texttt{stopper}$=1/100$ assuring an error for the bounds of less than 1$\%$ with respect to the difference between the upper and lower bounds, estimated by the \textit{shadows}. This value ensures that the computational process is both accurate and efficient.

In Figure \ref{fig:stopper}, we present the relative error of upper and lower bounds for a 12D region. The error is computed from the moment in which the first allowed point is identified and it ends when all the bounds pass the \texttt{stopper} threshold. In the example of Figure \ref{fig:reduce}, this process starts in D3 and it ends in G9. 

\begin{figure}[t]
\centering
    \begin{subfigure}[]{0.331\columnwidth}
    \centering
    \includegraphics[width=\columnwidth]{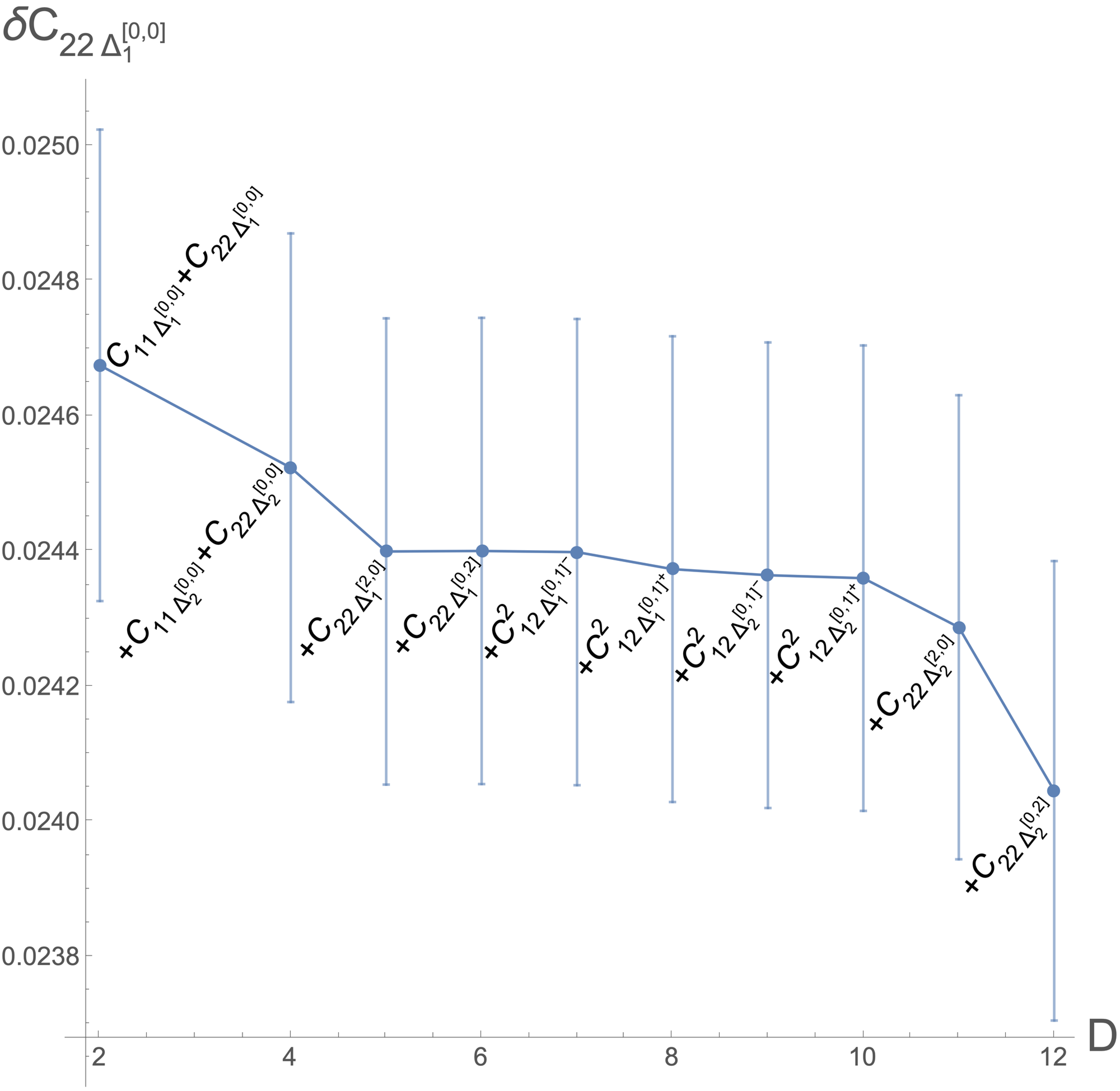}
    \end{subfigure}
    \begin{subfigure}[]{0.331\columnwidth}
    \centering
    \includegraphics[width=\columnwidth]{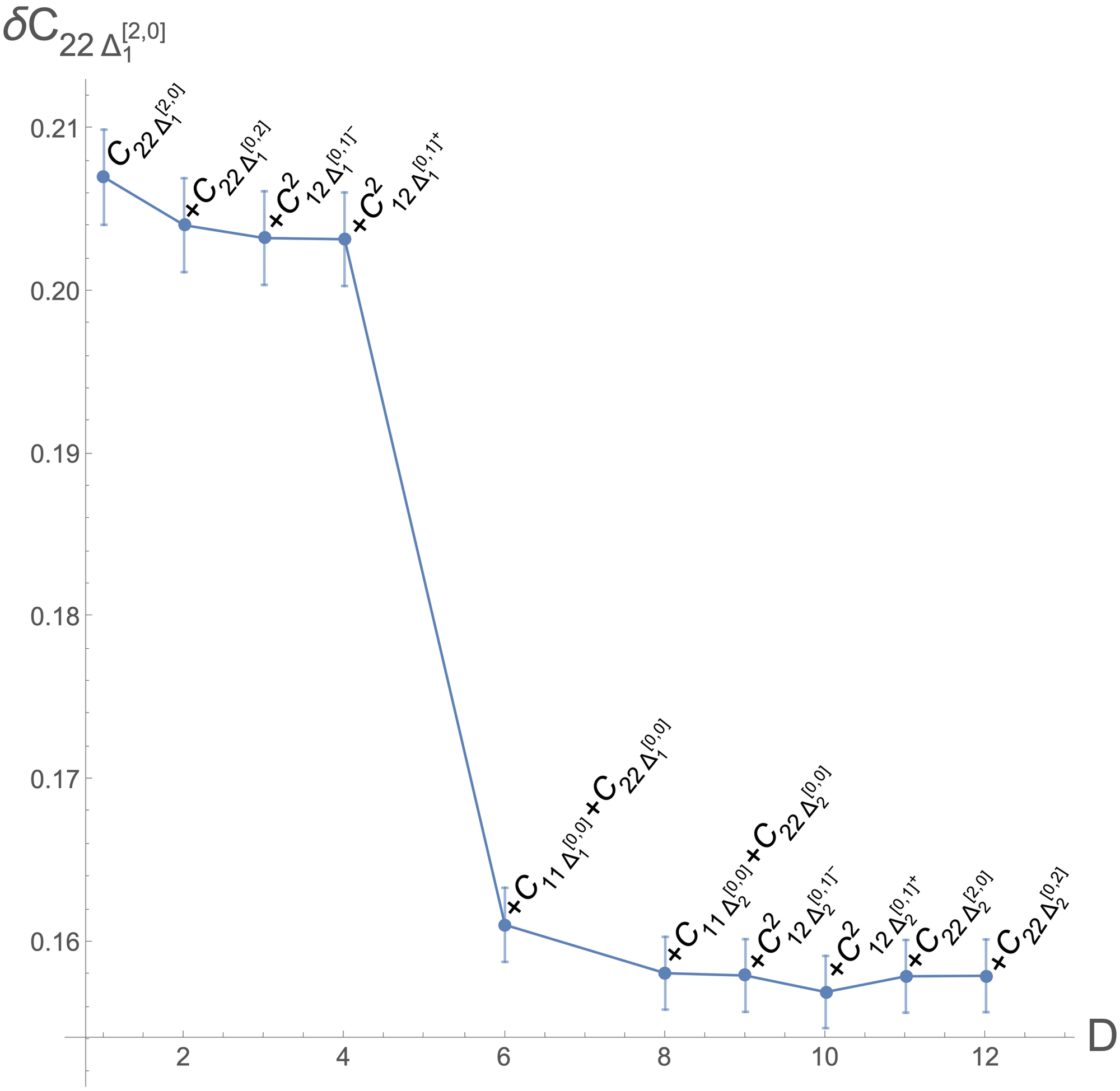}
    \end{subfigure}
    \begin{subfigure}[]{0.32\columnwidth}
    \centering
    \includegraphics[width=\columnwidth]{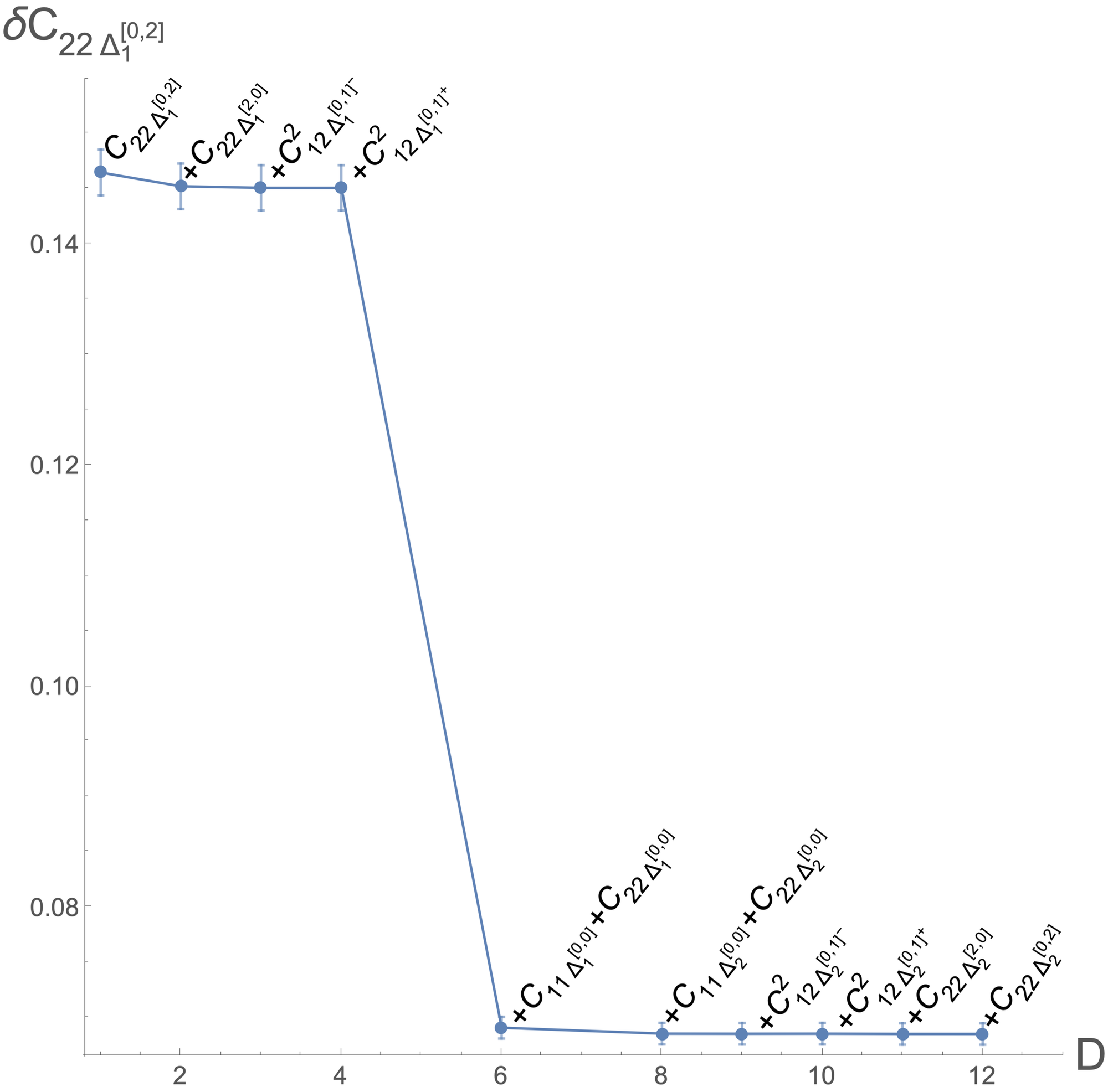}
    \end{subfigure}
\caption{Bound thickness of $C_{22\Delta_1}$ with $\Delta_1$ being the first state of $\texttt{S}^{[0,0]}$, $\texttt{S}^{[2,0]}$ and $\texttt{S}^{[0,2]}$ spectra defined by \eqref{deltaC} as a function of the dimension of the scanned space of OPE coefficients for $g=1/2$, $\Lambda=60$ and $N_{\text{poles}}=30$. At any point, the label indicates which OPE coefficient is added to the system respect to the previous point. Error bars represent the propagated uncertainty of data obtained with \texttt{stopper}$=1/100$.}
\label{fig:C(D)}
\end{figure}

Studying the bounds on various structure constants varying the dimension $D$ of the scanned space, we observe an interesting feature. Defining the thickness of the allowed region as
\begin{equation}\label{deltaC}
\delta C_{ij\Delta}= \frac{1}{2}(C^{\text{upper bound}}_{ij\Delta}-C^{\text{lower bound}}_{ij\Delta})
\end{equation}
we find that it gradually reduces as the size of the scanned space increases.

In Figure \ref{fig:C(D)}, we analyse how the thickness of the bounds  on $C_{22\Delta_1}$ evolves as additional unknowns are introduced into the system, up to $D=12$. The order in which variables are introduced can vary, leading to different paths to the target $D=12$ space. The three plots in Figure \ref{fig:C(D)} illustrate three distinct paths, distinguished by the order of the labels appearing on the data points.
All curves exhibit a decreasing trend, highlighting how a higher-dimensional space results in narrower bounds. We compare the discrepancy between consecutive points with the associated error bars. We notice that $\delta C$ jumps when we add OPE coefficients depending on $\Delta_1^{[0,0]}$ (which are very narrow due to the integrated correlators) to the scanned variables. So even though the integrated correlators do not impose constraints directly on $C_{22\Delta_1}$, they do still impact considerably their bounds. 
Notice also that the $\texttt{S}^{[0,0]}$ spectrum is also the one entering the bootstrap system with a non-trivial matrix structure for the blocks, thus scanning over its OPE coefficients is expected to have an impact on the bounds~\cite{ Kos:2016ysd,Chester:2019ifh,Chester:2020iyt,Rychkov:2023wsd}, consistently with what we see here.

Following the previous arguments, data collection follows a carefully designed order to optimise the performance of our computational runs. Although the sequence in which data are collected does not affect the final output, following a specific order reduces the computation time. In particular, we choose the following one 
\begin{equation}\begin{split}\label{orderC}
&D=2\qquad  C_{11\Delta_1^{[0,0]}}+C_{22\Delta_1^{[0,0]}}\\
&D=4\qquad\quad \hookrightarrow +C_{11\Delta_2^{[0,0]}}+C_{22\Delta_2^{[0,0]}}\\
&D=5\qquad \qquad\qquad\hookrightarrow +C_{22\Delta_1^{[2,0]}}\\
&D=6\qquad \qquad\qquad\qquad\quad\hookrightarrow +C_{22\Delta_1^{[0,2]}}\\
&D=7\qquad \qquad\qquad\qquad\qquad\quad\quad\hookrightarrow +C^2_{12\Delta_1^{[0,1]^{-}}}\\
&D=8\qquad \qquad\qquad\qquad\qquad\qquad\quad\quad\quad\hookrightarrow +C^2_{12\Delta_1^{[0,1]^{+}}}\\
&D=9\qquad\qquad\qquad\qquad\qquad \qquad\qquad\quad\quad\quad\quad\hookrightarrow +C^2_{12\Delta_2^{[0,1]^{-}}}\\
&D=10\qquad\qquad\qquad\qquad\qquad\qquad \qquad\qquad\quad\quad\quad\quad\quad\hookrightarrow +C^2_{12\Delta_2^{[0,1]^{+}}}\\
&D=11\qquad\qquad\qquad\qquad\qquad\qquad\qquad \qquad\qquad\quad\quad\quad\quad\quad\quad\hookrightarrow +C_{22\Delta_2^{[2,0]}}\\
&D=12\qquad\qquad\qquad\qquad\qquad\qquad\qquad\qquad \qquad\qquad\quad\quad\quad\quad\quad\quad\quad\hookrightarrow +C_{22\Delta_2^{[0,2]}}
\end{split}\end{equation}
that corresponds to the one of Figure \ref{fig:C(D)} on the left.
We begin by scanning structure constants involving operators from the $\texttt{S}^{[0,0]}$ spectrum, followed by those from the other spectra. This order is optimal considering that $D$ increases by 2 rather than 1 for any state in the $\texttt{S}^{[0,0]}$ sector. Therefore, it is more efficient to handle these states early in the process, as the algorithm becomes more computationally expensive when $D$ grows larger. An additional motivation to choose a specific order is that it enables the initialisation of systems with additional variables from the bounds identified in the previous iteration, accelerating convergence toward the final results.

\subsection{Data analysis}

\paragraph{Comparison with known data.}
A subset of the OPE coefficients analysed in this paper overlaps with those previously computed by the Bootstrability program. Specifically, these coefficients are the structure constants appearing in the OPE decomposition of the single correlator $\langle\langle O_{\mathcal{B}_1}O_{\mathcal{B}_1}O_{\mathcal{B}_1}O_{\mathcal{B}_1}\rangle\rangle$, namely $C_{11\Delta^{[0,0]}_1}$ and $C_{11\Delta^{[0,0]}_2}$.   
It is therefore natural to compare the tightness of their bounds, as computed in \cite{Cavaglia:2023mmu} with parity symmetry included, against our current dataset\footnote{In \cite{Cavaglia:2023mmu}, $C_{11\Delta^{[0,0]}_1}$ and $C_{11\Delta^{[0,0]}_2}$ are referred to as  $C_1$ and $C_2$ respectively}.
This comparison provides a direct measure of the effectiveness of the earlier single-correlator optimisation approach versus our current mixed-correlator setup, while also assessing the consistency of our data.

\begin{figure}[t]
\centering
    \begin{subfigure}[]{0.48\columnwidth}
    \centering
    \includegraphics[width=\columnwidth]{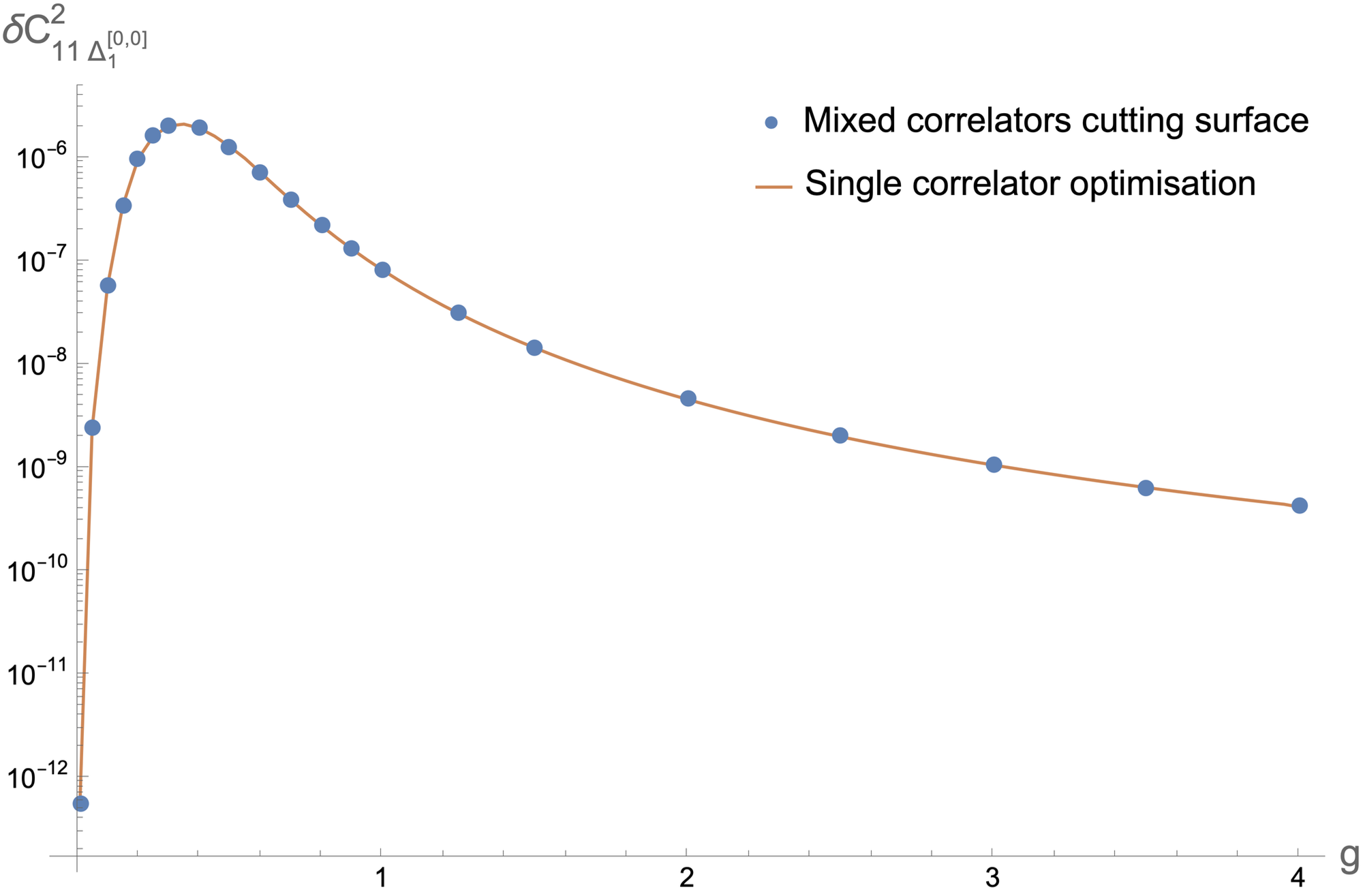}
    \end{subfigure}
\quad
    \begin{subfigure}[]{0.48\columnwidth}
    \centering
    \includegraphics[width=\columnwidth]{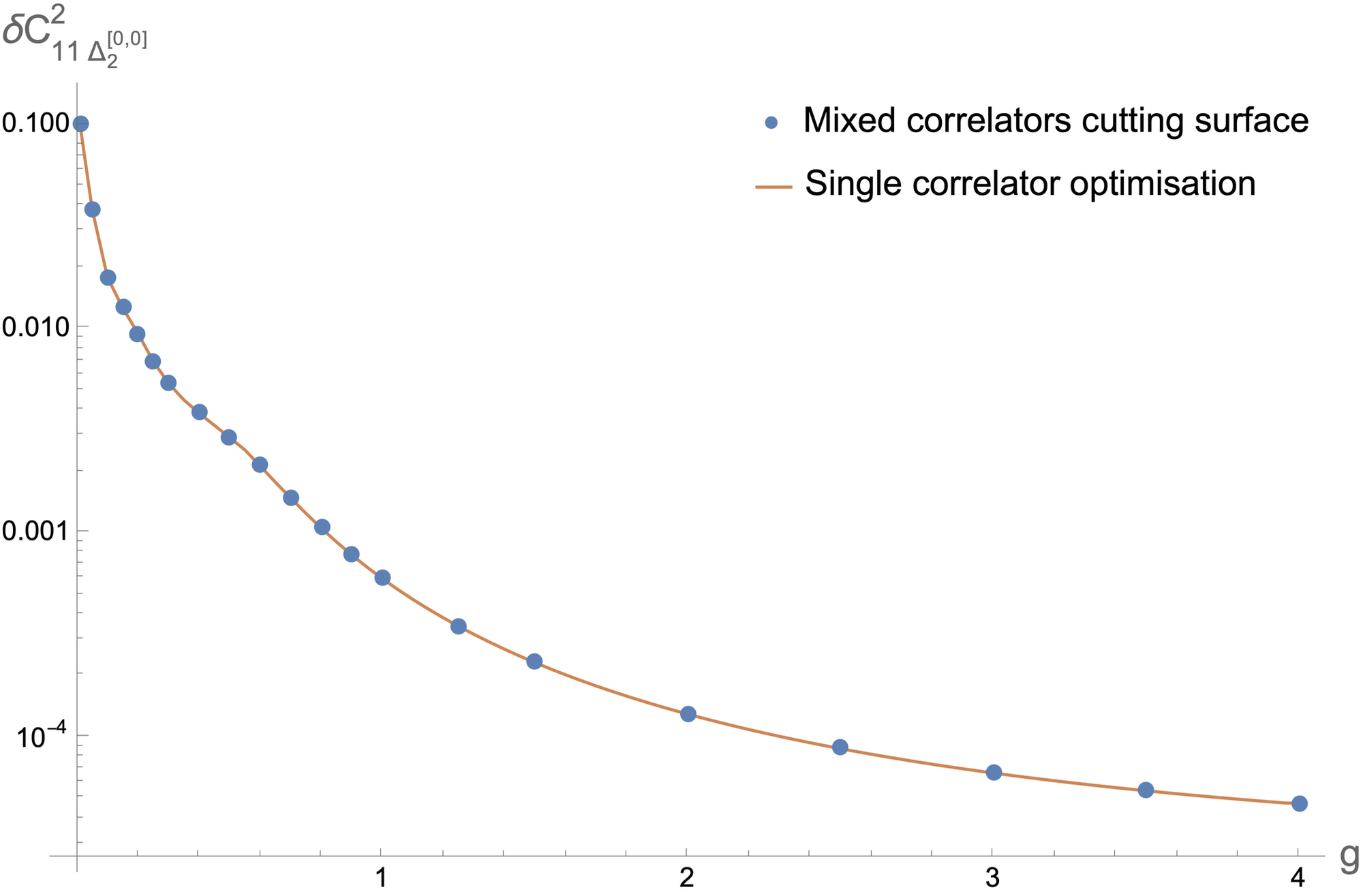}
    \end{subfigure}
\caption{Bound thickness for $C^2_{11\Delta_1^{[0,0]}}$ (left) and $C^2_{11\Delta_2^{[0,0]}}$ (right). Data points correspond to the current dataset while the line represents the same quantity computed with a single correlator and the old optimisation algorithm in \cite{Cavaglia:2023mmu}. We omit error-bars on data points since their size is negligible.}
\label{fig:errC1errC2}
\end{figure}

Data are presented in Figure \ref{fig:errC1errC2}. 
In order to compare the two datasets, we need to square our current results, as opposed to the previous computations, since now the structure constants of interest appear linearly in the crossing equations.
This analysis reveals that the bound thickness of the mixed-correlator data precisely matches that of the single correlator setup.
On one hand, that means that the current method does reproduce the correct result very precisely, which indicates we have good control over systematic errors. On the other hand, one would like to see further improvement of the bounds when the information about other correlators and the corresponding constraints from the spectrum and conformal symmetry are incorporated, like in the prototypical example in the three-dimensional Ising model~\cite{Kos:2014bka, Kos:2016ysd}. 

 At the same time, as discussed in the previous section, incorporating $C_{11\Delta_1^{[0,0]}}$ as a parameter within the full mixed-correlator setup offers a clear advantage in generating bounds for structure constants in other channels.

There could be various reasons why the bounds for $C_{11\Delta_1^{[0,0]}}$ do not 
 shrink further. First of all, its bounds are already extremely sharp, due to the existence of the integrated correlator constraints.
The introduction of  more correlators in the current setup also introduces additional spectra into the game, and thus both a larger number of unknown parameters and a correspondingly larger number of constraints on the functional searched by \texttt{SDPB}. 
In particular, this happens due to the presence of different R-charges of the external operators. 

An interesting future exploration, given the current algorithm which can tackle a multidimensional space of structure constants, would be considering a system of mixed-correlator involving unprotected states with the same quantum numbers as external operators. In this way, the structure constants would be more tightly interconnected since the matrix structure would increase its rank. We also expect that a large number of integrated correlators identities can be used to inject   additional information from outside the 1D CFT.

\paragraph{New numerical bounds.}

Expanding on the analysis of $C_{11\Delta^{[0,0]}}$ in the previous section, the bounds for the remaining OPE coefficients examined in the $D=12$ run are presented below.
In Figure \ref{fig:C22Delta1} we give bounds for the $C_{22\Delta_1}$ structure constants with $\Delta_1$ being the first operator (smaller dimension) in 
 the $\texttt{S}^{[0,0]}$, $\texttt{S}^{[0,2]}$ and $\texttt{S}^{[2,0]}$ spectra. Figure \ref{fig:C12Delta1} shows bounds for $C^2_{12\Delta_1}$ where $\Delta_1$ here stands for the first operator in $\texttt{S}^{[0,1]^-}$ and $\texttt{S}^{[0,1]^+}$. All the remaining bounds correspond to structure constants $C_{22\Delta_2}$ and $C^2_{12\Delta_2}$ with $\Delta_2$ referring to the second operators of the spectra. They are reported in Figure \ref{fig:CDelta2} left and right respectively.

\begin{figure}[!t]
    \centering
    \includegraphics[width=0.75\columnwidth]{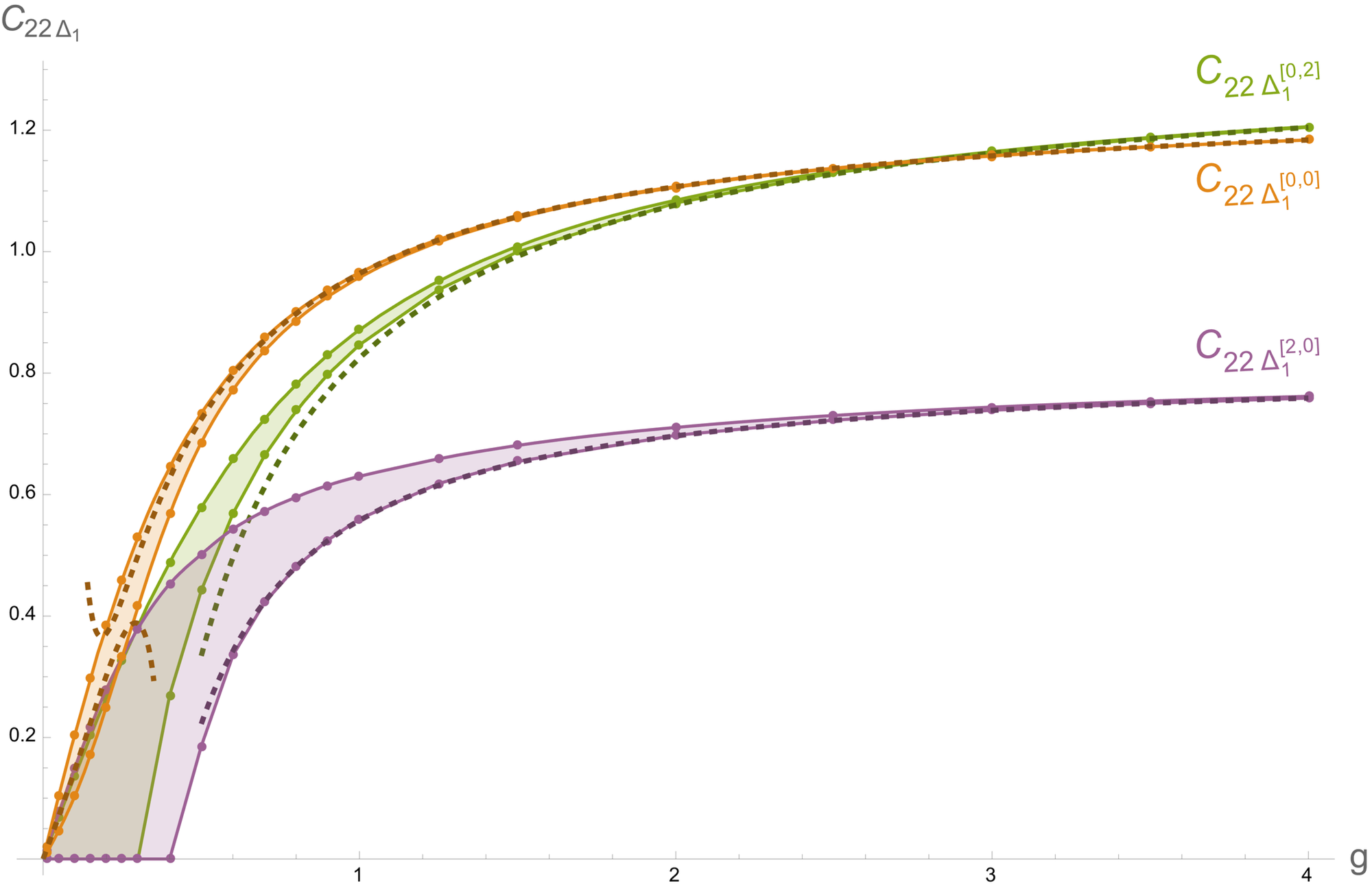}
    \caption{Bounds for $C_{22\Delta_1}$ with $\Delta_1$ being the first operator in $\texttt{S}^{[0,0]}$, $\texttt{S}^{[0,2]}$ and $\texttt{S}^{[2,0]}$ spectra. Data are obtained with $\Lambda=60$ derivatives, $N_{\text{poles}} = 30$ and \texttt{stopper}$=1/100$. Dots represents the sampling of data obtained by our algorithm. Dashed lines correspond to weak and strong coupling predictions \eqref{Cweak} and \eqref{Cstrong}.}
    \label{fig:C22Delta1}
\end{figure}

Together with the numerical bounds, we provide analytic weak and strong coupling expansions for some of the studied structure constants to compare with. At weak coupling, we exploit the four-point functions $\langle\langle O_{\mathcal{B}_1} O_{\mathcal{B}_2}O_{\mathcal{B}_1}O_{\mathcal{B}_2}\rangle\rangle$ and $\langle\langle O_{\mathcal{B}_1}O_{\mathcal{B}_1} O_{\mathcal{B}_2}O_{\mathcal{B}_2}\rangle\rangle$ computed analytically in \cite{Artico:2024wut}. 
Incorporating analytic expansions of the spectral data given in \eqref{eqn:perturbativespectrum1}, \eqref{eqn:perturbativespectrum2} and \cite{Cavaglia:2022qpg}, one can extract structure constants, leveraging the analytic form of the correlators through their OPE decomposition. Similarly to what was observed in \cite{Cavaglia:2022qpg}, the integrability input allows us to untangle the averages of OPE coefficients obtaining
\begin{equation}
\begin{aligned}
\label{Cweak}
   C_{22\Delta_1^{[0,0]}}^2 &= 2g^2 + 4(\pi^2 - 6) g^4 - \frac{4}{45} (19 \pi^4 + 420 \pi^2 - 3600) g^6 +  \mathcal{O}(g^8), \\
   C_{22\Delta_2^{[0,0]}}^2 &= \frac{1}{10} + \frac{10 \pi^2 - 3 \sqrt{5}-25}{50} g^2 + \mathcal{O}(g^4), \\
   C_{22\Delta_3^{[0,0]}}^2 &=\frac{1}{10} + \frac{10 \pi^2 + 3 \sqrt{5}-25}{50} g^2  + \mathcal{O}(g^4),\\
      C_{12\Delta_1^{[0,1]^{-}}}^2 &= g^2 + \frac{2}{3}(\pi^2-9)g^4 + b_1 g^6 + \mathcal{O}(g^8),\\
   C_{12\Delta_1^{[0,1]^{+}}}^2&= -g^2 - \frac{2}{3}(\pi^2 - 24)g^4 + b_2\, g^6 + \mathcal{O}(g^8),
\end{aligned}
\end{equation}
where $3 b_1 + b_2 = -\frac{4}{9}(279 + 6 \pi^2 + \pi^4)$. Notice that in \eqref{Cweak}, all structure constants are squared as they naturally appear in the OPE decomposition of a single correlator while $C_{22\Delta^{[0,0]}}$ show up linearly in our mixed system. Analytic data for $C^2_{22\Delta^{[0,2]}}$ and $C^2_{22\Delta^{[2,0]}}$ are missing due to the lack of available data for $\langle\langle O_{\mathcal{B}_2} O_{\mathcal{B}_2} O_{\mathcal{B}_2} O_{\mathcal{B}_2}\rangle\rangle$. Similarly, given the amount of  spectral data  we have computed in the $\texttt{S}^{[0,1]}$ sector, given  the increasing degeneracy of states at weak coupling we cannot completely untangle averages for $C^2_{12\Delta^{[0,1]}_n}$ with $n>1$.

\begin{figure}[!t]
    \centering
    \includegraphics[width=0.75\columnwidth]{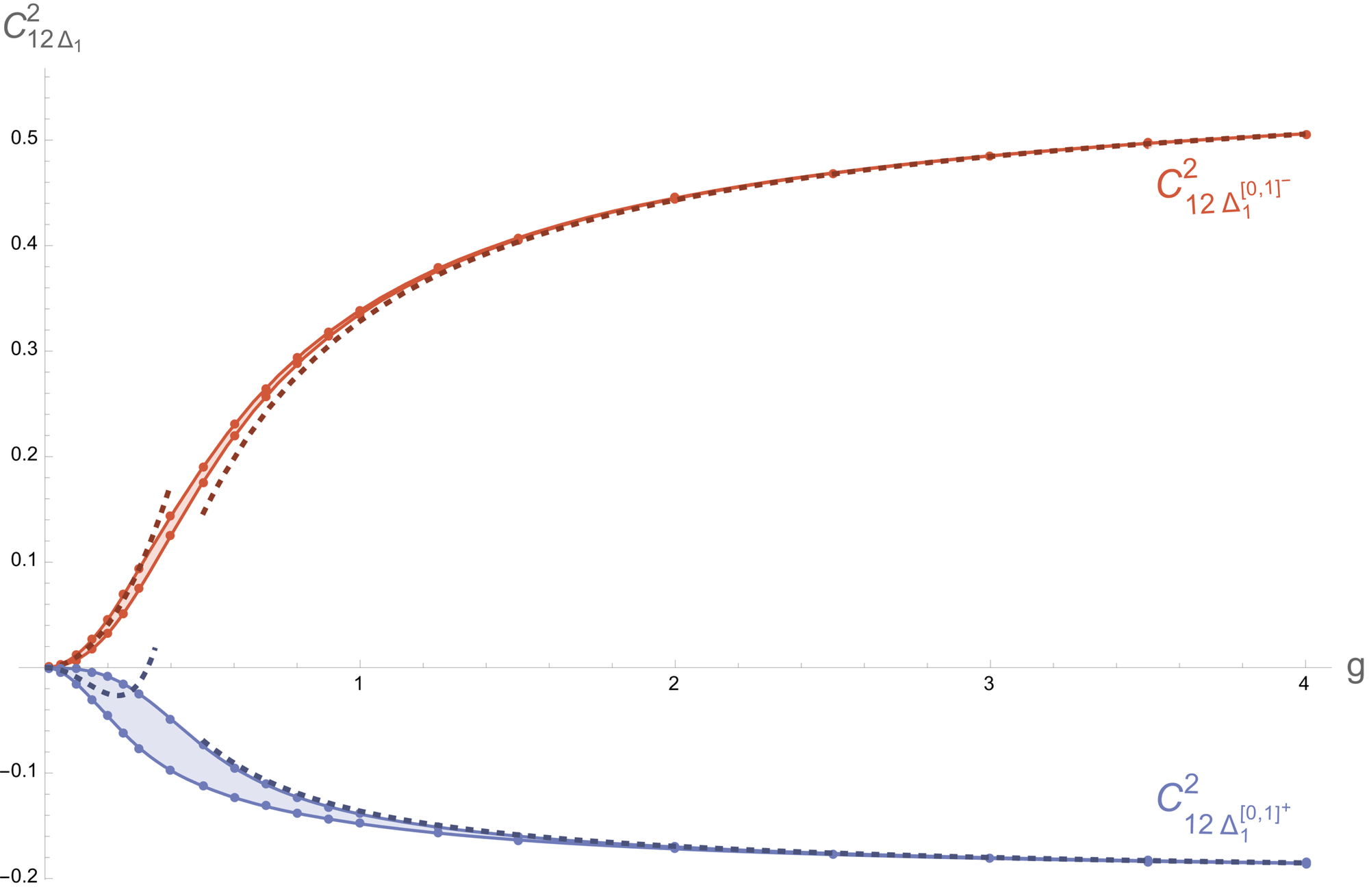}
    \caption{Bounds for $C^2_{12\Delta_1}$ with $\Delta_1$ being the first operator in $\texttt{S}^{[0,1]^-}$ and $\texttt{S}^{[0,1]^+}$ spectra. Data are obtained with $\Lambda=60$ derivatives, $N_{\text{poles}} = 30$ and \texttt{stopper}$=1/100$. Dots represents the sampling of data obtained by our algorithm. Dashed lines correspond to weak and strong coupling predictions \eqref{Cweak} and \eqref{Cstrong}.}
    \label{fig:C12Delta1}
\end{figure}

Another set of structure constants can be expanded at strong coupling following the arguments of \cite{Ferrero:2023gnu}. There, the authors reconstruct a family of four-point functions exploiting a set of stringent constraints at strong couplings. As a by-product, they obtain analytic expansions for the averages of structure constants. However, in some cases where the degeneracy of states at strong coupling is 1 (see Figure \ref{fig:spectrumplot}), the averages coincide with the OPE coefficients themselves.
In our setup this leads to the following expansions\footnote{We thank Pietro Ferrero for sharing with us the results of~\cite{Ferrero:2023gnu} priori to the publication.}
\begin{equation}
\small
\begin{aligned}\label{Cstrong}
  C_{qq\Delta^{[0,0]}_1}&=  \sqrt{\frac{2}{5}} q-\frac{q (36 q+7)}{12 \sqrt{10} \sqrt{\lambda }}+\frac{q \left(864 q^2+2448 q-3961\right)}{576 \sqrt{10} \lambda }+\mathcal{O}(\lambda^{-\frac{3}{2}})\\
  C_{qq\Delta^{[0,2]}_1}&=  \frac{2}{3} (q-1) q-\frac{(q-1) (8 q-3) q}{4 \sqrt{\lambda }}+\frac{(q-1) \left(160 q^2-88 q-419\right) q}{64 \lambda }+\mathcal{O}(\lambda^{-\frac{3}{2}})\\
  C_{qq\Delta^{[2,0]}_1}&=  \frac{(q\!-\!1) q}{\sqrt{6}}\!-\!\frac{(q\!-\!1) (144 q\!-\!125) q}{48 \sqrt{6} \sqrt{\lambda }}+\frac{(q\!-\!1) \left(1920 q^2\!-\!1024 q\!-\!9417\right) q}{512 \sqrt{6} \lambda }+\mathcal{O}(\lambda^{-\frac{3}{2}})\\
  C_{12\Delta^{[0,1]^{-}}_1}&=  \frac{2}{\sqrt{7}}-\frac{71}{12 \sqrt{7} \sqrt{\lambda }}-\frac{1153}{576 \sqrt{7} \lambda }+\mathcal{O}(\lambda^{-\frac{3}{2}})\\
  C_{12\Delta^{[0,1]^{+}}_1}&= \frac{1}{\sqrt{5}} -\frac{9}{5 \sqrt{5} \sqrt{\lambda }}-\frac{126}{25 \sqrt{5} \lambda }+\mathcal{O}(\lambda^{-\frac{3}{2}}),
\end{aligned}
\end{equation}
where $\lambda = 16 \pi^2 g^2$. Part of these expansions holds for generic external operators belonging to the short multiplet $\mathcal{B}_q$. In our case, we specialise them for $q = 2$. The structure constants in \eqref{Cstrong} appear linearly as they are naturally extracted from the analysis of \cite{Ferrero:2023gnu}.

In both cases, weak and strong coupling expansions \eqref{Cweak} and \eqref{Cstrong} nicely match with the numerical bounds. It is interesting to notice that strong coupling predictions in Figures \ref{fig:C22Delta1} and \ref{fig:C12Delta1} approach very closely either the upper or the lower bound of the allowed regions, even at middle range coupling. This is similar to the single-correlator setup, where the upper bound was significantly more precise, as it  was given by a functional which the algorithm did not force to change sign, thus making the optimisation problem less tightly  constrained.
The available weak coupling data align well with our results and lie inside the allowed regions, even though in this region the bound thickness is larger.

\begin{figure}[t]
\centering
    \begin{subfigure}[]{0.48\columnwidth}
    \centering
    \includegraphics[width=\columnwidth]{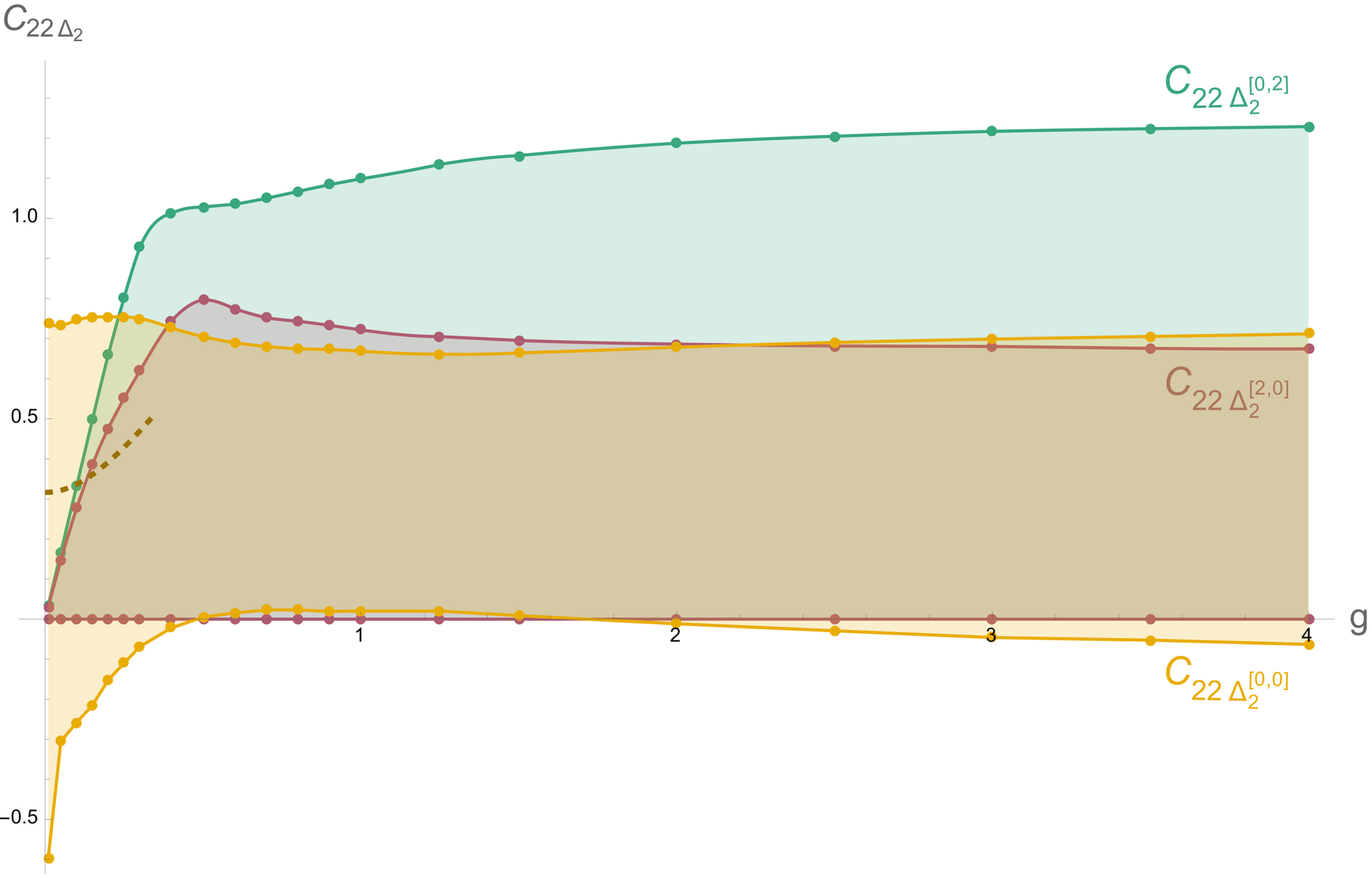}
    \end{subfigure}
\quad
    \begin{subfigure}[]{0.48\columnwidth}
    \centering
    \includegraphics[width=\columnwidth]{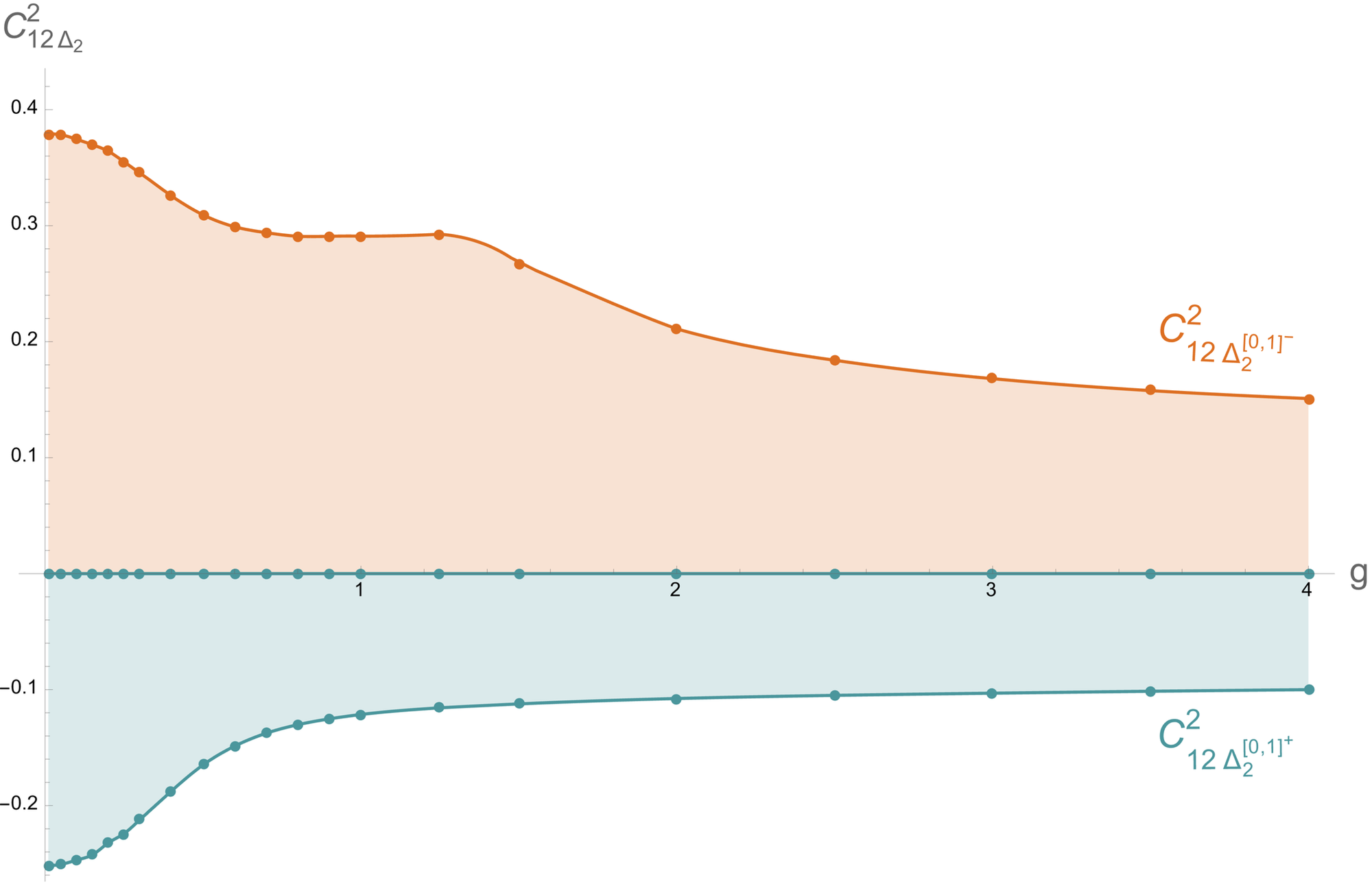}
    \end{subfigure}
\caption{Bounds for $C_{22\Delta_2}$ (left) and $C^2_{12\Delta_2}$ (right) with $\Delta_2$ being the second operator of the analysed spectra. Data are obtained with $\Lambda=60$ derivatives, $N_{\text{poles}} = 30$ and \texttt{stopper}$=1/100$. Dots represents the sampling of data obtained by our algorithm. Dashed line on the left correspond to the square root of the weak coupling prediction in the second line of \eqref{Cweak}. As those states are nearly degenerate with other states with the same quantum numbers, we do not expect narrow bounds.}
\label{fig:CDelta2}
\end{figure}

Comparing Figures \ref{fig:C22Delta1} and \ref{fig:C12Delta1} with Figure \ref{fig:CDelta2}, one striking feature becomes immediately apparent: the difference in the width of the bounds. The difference reflects the fact that the lowest lying operator dimensions in any spectra in Figure \ref{fig:spectrumplot} are very isolated from the rest, whereas the excited states are very nearly degenerate with some other states with the same quantum numbers. 
The degeneracy of higher states increases rapidly, with spectral lines remaining close to one another even at intermediate coupling, giving rise to a demanding mixing problem for the bootstrap. 
We notice that, the shape of the upper bounds on the right side of Figure \ref{fig:CDelta2} resembles the one computed in \cite{Cavaglia:2022qpg} (Figure 6) for similar structure constants precisely when integral relations were not included.  
In the single correlator case, the excited states were only possible to disentangle with the help of the integrated correlators, so it is not surprising that the bounds for those are not narrow. However, one should be able to set much better bounds on some linear combinations of these structure constants.

\paragraph{Bounds thickness.}

\begin{figure}[t]
\centering
    \begin{subfigure}[]{0.48\columnwidth}
    \centering
    \includegraphics[width=\columnwidth]{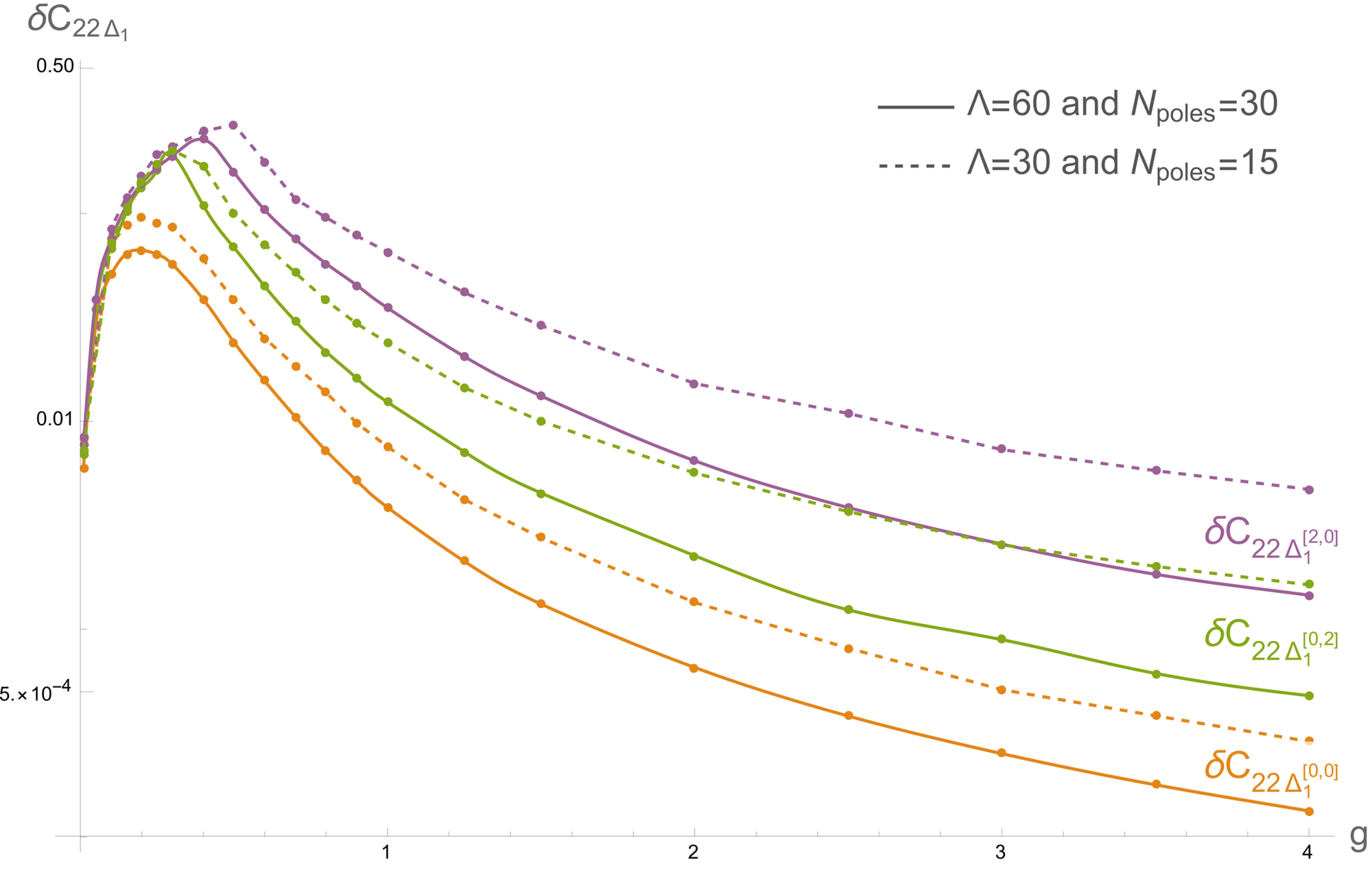}
    \end{subfigure}
\quad
    \begin{subfigure}[]{0.48\columnwidth}
    \centering
    \includegraphics[width=\columnwidth]{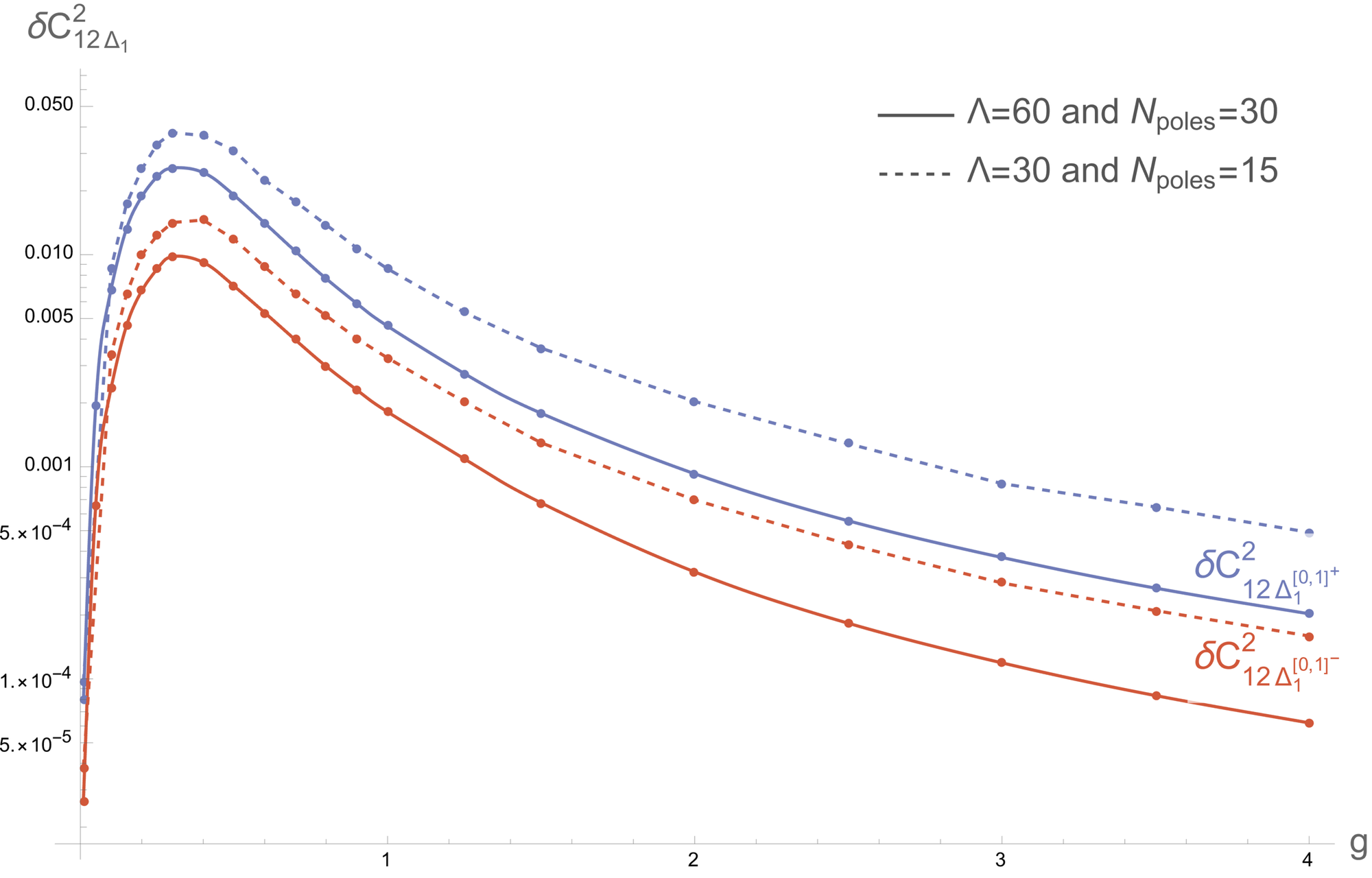}
    \end{subfigure}
\caption{Bound thickness $\delta C_{22\Delta_1}$ (left) and $\delta C^2_{12\Delta_1}$ (right) with $\Delta_1$ being the first operator of the analysed spectra. Solid lines corresponds to data obtained with $\Lambda=60$ derivatives and $N_{\text{poles}} = 30$ while dashed lines to a similar run with $\Lambda=30$ and $N_{\text{poles}} = 15$. Both datasets are computed with \texttt{stopper}$=1/100$. Dots represents the sampling of data obtained by our algorithm.}
\label{fig:errC}
\end{figure}

In Figure \ref{fig:errC}, we present the bound thickness computed using the definition in \eqref{deltaC} for two different datasets, obtained with different truncations in the approximations of the superconformal blocks.  We only display data for the bounds in Figures \ref{fig:C22Delta1} and \ref{fig:C12Delta1}, since for most of the bounds in Figure \ref{fig:CDelta2}, $\delta C$ corresponds to half the value of the upper bound. 
As observed for the OPE coefficient bounds, the precision improves rapidly with stronger coupling, yielding, for example, 4-5 digits of precision for 
$C^2_{12\Delta_1^{[0,1]^{-}}}$. We note that the curve peaks around $g=0.2-0.4$ as previously observed in single-correlator computations in \cite{Cavaglia:2021bnz, Cavaglia:2022qpg, Cavaglia:2023mmu} (see also Figure \ref{fig:errC1errC2}, left). 

\section{Conclusions and outlook}
\label{sec:Conclusions}

In this paper we explored conformal bootstrap methods in combination with  integrability data 
for the Wilson line 1D defect CFT in the $\mathcal{N}=4$ SYM theory at finite coupling. We considered a mixed system of correlators, originally introduced in \cite{Liendo:2018ukf}, which contains the correlator of four tilt operators previously studied in \cite{Cavaglia:2021bnz,Cavaglia:2022qpg,Cavaglia:2023mmu}. 
The current mixed system presents a number of challenges,
related to the multi-dimensionality of the space of OPE coefficients and the non-convexity of the problem.
We developed an efficient algorithm to scan over the $D$-dimensional space of structure constants and we exploit it to obtain upper and lower bounds for up to $12$ OPE coefficients at finite coupling.
This produces results for novel structure constants
which did not appear in the previous Bootstrability studies in agreement with analytical predictions at weak and strong coupling.

However, expanding the system of crossing equations has not led to any improvement in the precision of the bounds for the OPE coefficients associated with the decomposition of the simplest correlator of four tilt operators. For these coefficients, the bounds remain consistent with the results obtained in earlier works.
The benefit of expanding the system of constraints appears to be entirely compensated by the introduction of additional sets of intermediate states. However, for the lowest lying states in the other channels, the existence of the narrow bound for the previously studied structure constants does produce a serious gain in precision. For the excited states, the problem is, as usual, more complicated due to the mixing with other states nearby.

We hope that the methods developed in this paper will be useful for further studies of this system. In particular, it would be interesting to consider the case of non-protected external operators, which allows us to build a bigger system where all external states have the same quantum numbers. 
This is a situation that has not been studied in the literature so far, particularly not in conformal bootstrap studies of 
${\cal N}=4$ SYM or the related 1D defects. 
This setup would be interesting as it would allow one to increase the number of correlators without increasing the set of intermediate states, and could potentially help to resolve bounds for OPE operators which have very close dimensions.

Another future direction would be to add into the system additional integrated correlator constraints: we expect that there should be new integrated identities both for the correlators 
from the charged sectors we consider here, as well as for other situations such as correlators with external non-protected states. This could be a crucial step, as in various setups integrated correlators allow us to shrink quite dramatically the bounds on the OPE coefficients~\cite{Cavaglia:2022qpg,Chester:2023ehi,Caron-Huot:2024tzr}.

\acknowledgments
We thank Pietro Ferrero for enlightening discussions, pointing us to the mixed-correlator setup considered in this paper, and sharing with us the strong coupling CFT-data of the states studied here. 
We are grateful to Marco Meineri, Carlo Meneghelli, Marten Reehorst and Ning Su for valuable discussions, to Emilio Trevisani for very useful advice related to this project, and to Daniele Artico, Julien Barrat and Giulia Peveri for communications and sharing analytic weak coupling data before the publication of their paper. 
AC was supported by the INFN SFT specific initiative. 
The work of NG and NS was supported by the European Research Council (ERC) under the European
Union’s Horizon 2020 research and innovation program – 60 – (grant agreement No. 865075)
EXACTC. 
The work of MP was supported by Marie Skłodowska-Curie Global Fellowship
(HORIZON-MSCA-2022-PF-01) BOOTSTRABILITY-101109934 and by the INFN
SFT specific initiative.
The work of JJ was supported by the European Union (ERC, FUNBOOTS, project number
101043588).  NG’s research is supported in part by the Science Technology \& Facilities Council (STFC) under the grants ST/P000258/1 and ST/X000753/1.

\appendix

\section{Details of OPE decomposition of correlators}
\label{app:decomposition}
The OPE decompositions of the correlators entering our system can be written as follows, taking into account the parity  selection rules \eqref{eq:parityselection} and positivity properties \eqref{eq:positivesimple} discussed in the main text: 
\beq
\label{eq:OPE expansion}
\mathcal{A}_{1111}(x,\zeta_1, \zeta_2)= 1 + C_{112}^2 \; \mathcal{F}_{\mathcal{B}_2}^{0,0}(x, \zeta_1, \zeta_2) + \sum_{{ \scriptsize \begin{array}{cc}\Delta_n \in \texttt{S}^{[0,0]^+}
\end{array}}} C_{11\Delta_n^{[0,0]}}^2  \; \mathcal{F}^{0,0}_{\Delta^{[0,0]}_n}(x,\zeta_1, \zeta_2 ),
\eeq

 \beq
 \begin{aligned}
\mathcal{A}_{2222}(x,\zeta_1, \zeta_2)&= 1 + C_{2 2 2 }^2 \; \mathcal{F}_{\mathcal{B}_2}^{0,0}(x, \zeta_1, \zeta_2) +
C_{2 2 4 }^2 \; \mathcal{F}_{\mathcal{B}_4}^{0,0}(x, \zeta_1, \zeta_2) \\
&
+\sum_{{\scriptsize  \begin{array}{cc}\Delta_n \in \texttt{S}^{[0,0]^+}
\end{array}} }  C_{22\Delta_n^{[0,0]}}^2  \; \mathcal{F}^{0,0}_{\Delta_n^{[0,0]}}(x,\zeta_1, \zeta_2 )\\&+\sum_{{\scriptsize  \begin{array}{cc}\Delta_n \in \texttt{S}^{[0,2]^+}
\end{array}} }  C_{22\Delta_n^{[0,2]}}^2  \; \mathcal{F}^{0,0}_{\Delta_n^{[0,2]}}(x,\zeta_1, \zeta_2 )\\
&+\sum_{{\scriptsize  \begin{array}{cc}\Delta_n \in \texttt{S}^{[2,0]^-}
\end{array}} }  C_{22\Delta_n^{[2,0]}}^2  \; \mathcal{F}^{0,0}_{\Delta_n^{[2,0]}}(x,\zeta_1, \zeta_2 ),
\end{aligned}
\eeq
and for the mixed channels
\beq
\begin{aligned}
\mathcal{A}_{1212}(x,\zeta_1, \zeta_2)&=  C_{112 }^2  \; \mathcal{F}_{\mathcal{B}_1}^{1,1}(x, \zeta_1, \zeta_2) +
C_{123 }^2 \; \mathcal{F}_{\mathcal{B}_3}^{1,1}(x, \zeta_1, \zeta_2) \\
&+\sum_{{\scriptsize  \begin{array}{cc}\Delta_n \in \texttt{S}^{[0,1]^{\pm}}
\end{array}} }
(- \mathbb{P}_{\Delta_n} )\;  |C_{12\Delta_n^{[0,1]}}|^2  \; \mathcal{F}^{1,1}_{\Delta_n^{[0,1]}}(x,\zeta_1, \zeta_2 ) ,
\end{aligned}
\eeq
and 
\beq
\begin{aligned}
\mathcal{A}_{1221}(x,\zeta_1, \zeta_2)&=  C_{112}^2  \; \mathcal{F}_{\mathcal{B}_1}^{1,-1}(x, \zeta_1, \zeta_2) +
C_{123}^2 \; \mathcal{F}_{\mathcal{B}_3}^{1,-1}(x, \zeta_1, \zeta_2) \\
&+\sum_{{\scriptsize  \begin{array}{cc}\Delta_n \in \texttt{S}^{[0,1]^{\pm}}
\end{array}} }
 |C_{12\Delta_n^{[0,1]}} |^2  \; \mathcal{F}^{1,-1}_{\Delta_n^{[0,1]}}(x,\zeta_1, \zeta_2 ) ,
\end{aligned}
\eeq
\beq
\begin{split}
\mathcal{A}_{1122}(x,\zeta_1, \zeta_2)&= 1 + C_{112 }  C_{222 } \; \mathcal{F}_{\mathcal{B}_2}^{0,0}(x, \zeta_1, \zeta_2) \\&+ \sum_{{\scriptsize  \begin{array}{cc}\Delta_n \in \texttt{S}^{[0,0]^+}
\end{array}} } C_{11\Delta_n^{[0,0]}} \, 
 C_{22\Delta_n^{[0,0]}}   \; \mathcal{F}^{0,0}_{\Delta_n^{[0,0]}}(x,\zeta_1, \zeta_2 ).
 \end{split}
 \eeq
 
Above, the sums run over long multiplets in the spectrum denoted as $\texttt{S}^{[0,0]^{+}}$, $\texttt{S}^{[0,1]^{\pm}}$, $\texttt{S}^{[0,2]^{+}}$ and $\texttt{S}^{[2,0]^{-}}$. Furthermore,  $\mathcal{F}^{m,n}_{\Delta^{\textbf{r}}}(x,\zeta_1, \zeta_2 )$ and $\mathcal{F}_{\mathcal{B}_k}(x,\zeta_1, \zeta_2 )$  are superconformal blocks, which depend on dimensions of operators of the correlator $m,n \in \{\{0,0\},\{1,1\}, \{1,-1\}\}$ and on the R-symmetry representation of the superprimary $\mathbf{r}\in \{ [0,0], [0,2], [2,0], [0,1] \}$. The superblocks were determined in \cite{Liendo:2018ukf}, and are explicit functions of the cross ratios and of the scaling dimension of the exchanged multiplets. 
It was shown in \cite{Liendo:2018ukf} that
 the long blocks can each be written in terms of three functions $f_{i, \Delta^{\mathbf{r}}}^{m,n}(x)$, $i=1,2,3$, which depend only on the line cross ratio. Below, we give more details on the structure of superconformal blocks and on the reduced functions  \ref{app:Superblocks}. 
 
We recall that, as discussed in the previous sections, the OPE coefficients $C_{11\Delta^{\mathbf{r}}}$ and  $C_{22\Delta^{\mathbf{r}}}$ are real. The OPE coefficients $C_{12\Delta^{[0,1]}}$ are either real or imaginary depending on the parity of the exchanged state, and we have explicitly introduced this grading and written the equations in terms of the absolute values of these OPE coefficients. 

Finally, let us also comment that one of the correlators can be written in a simplified form \cite{Liendo:2016ymz}
\beq
\mathcal{A}_{1111}(x, \zeta_1, \zeta_2) = \frac{x^2}{\zeta_1 \zeta_2} \mathbb{F} + \mathbb{D}(x, \zeta_1, \zeta_2) f(x),  
\eeq
where $f(x)$ is a reduced correlator, the differential operator $\mathbb{D}$ is given as follows
\beq
\mathbb{D}(x, \zeta_1, \zeta_2) = \left( 2 x^{-1} - \zeta_1^{-1} - \zeta_2^{-1}\right) - x^2 (\zeta_1^{-1} - x^{-1})(\zeta_2^{-1}-x^{-1}) \partial_{x},  
\eeq
and the constant is given as $\mathbb{F} = 1 + (C^{\rm BPS}_{112})^2$. The reduced correlator has the following block expansion
\beq
\label{eq:reducedcor}
f(x) = x + (C_{112}^{\rm BPS})^2 \,F_{\mathcal{B}_2}(x) + \sum_{{ \scriptsize \begin{array}{cc}\Delta_n \in \texttt{S}^{[0,0]^+}
\end{array}}} C_{11 \Delta^{[0,0]}_n}^2\, F_{\Delta_n}(x),
\eeq
where the relation between the blocks of the reduced correlator and the superconformal blocks introduced above \eqref{eq:OPE expansion} is given as

\beq
\begin{aligned}
\mathcal{F}_{B_2}^{0,0}(x, \zeta_1, \zeta_2) &= \frac{x^2}{\zeta_1 \zeta_2} + \mathbb{D}(x, \zeta_1, \zeta_2)\,F_{\mathcal{B}_2}(x),\\
\mathcal{F}_{\Delta^{[0,0]}}^{0,0}(x, \zeta_1, \zeta_2) &= \frac{x^2}{\zeta_1 \zeta_2} + \mathbb{D}(x, \zeta_1, \zeta_2)\,F_{\Delta}(x).
\end{aligned}
\eeq

\subsection{Superblocks}
\label{app:Superblocks}

Here we give more details on the superblocks provided in \cite{Liendo:2018ukf}. The superblocks depend on the dimensions of external operators, and for the correlator $\mathcal{A}_{m_1 m_2 m_3 m_4}$ they are labelled by the difference between the dimensions of external operators as $\{m, n\} \equiv \{m_2 - m_1, m_4 - m_3\}$. For our system, these labels can have possible values  $\{m, n\} = \{ \{1,1\}, \{1,-1\}, \{0,0\} \}$.  Furthermore, the blocks are labelled by the scaling dimension of the superprimary operator $\Delta$ and by its R-charge $\mathbf{r}$. For instance, the superblocks for long multiplets read
\begin{equation}
\label{eq:blockdefinition}
    \mathcal{F}_{\Delta^{\mathbf{r}}}^{m,n}(x, \zeta_1, \zeta_2) = \sum_{h = \Delta}^{\Delta + 4} \sum_{R} c_{h, R}\ g_{h}^{m, n}(x) \mathbb{B}_{R}^{m, n }(\zeta_1, \zeta_2), 
\end{equation}
where the sum runs over primary operators in the multiplets. Above, $g_h^{ {m, n}}(x)$ is the 1D conformal block given as 
\begin{equation} 
g_{\Delta}^{{m, n}}(x) = \chi^{\Delta} {}_2F_1 \left( \Delta + \frac{m}{2}, \Delta - \frac{n}{2}; 2\Delta; x \right), 
\end{equation}
and  $\mathbb{B}_{R}^{ {m, n}}(\zeta_1, \zeta_2)$ are SP(4) R-symmetry structures of the conformal primaries of the multiplet. The coefficients $c_{h, R}$ are determined in \cite{Liendo:2018ukf} by imposing the superconformal Ward identities.

The superblocks for short multiplets, which contain less conformal primaries, are introduced as

\begin{equation}
\label{eq:blockdefinition}
    \mathcal{F}_{\mathcal{B}_k}^{m,n}(x, \zeta_1, \zeta_2) = \sum_{h = k}^{k + 2} \sum_{R} c_{h, R}\ g_{h}^{m, n}(x) \mathbb{B}_{R}^{m, n}(\zeta_1, \zeta_2), 
\end{equation}
where the superprimary operators of $\mathcal{B}_k$ operators have the scaling dimension $k$ and the $[0,k]$ R-charge.

Finally, we introduce reduced functions following \cite{Liendo:2018ukf}. Firstly, the action $\mathbb{G}$ on the superconformal blocks is defined as follows
\begin{equation}
    \mathbb{G}[\mathcal{F}]_{p,q} \equiv \left( \frac{\partial^p \partial^q}{\partial \zeta_1^p \partial \zeta_2^q}\frac{\mathcal{F}(x, \zeta_1, \zeta_2)}{\mathfrak{X}^2} \right)|_{\zeta_1=\zeta_2 = \chi}.
\end{equation}
Then, by acting on the superconformal blocks \eqref{eq:blockdefinition}, we get the reduced functions which enter the crossing equations \eqref{eq: define V and tilde V}

\begin{equation}
\begin{aligned}
f_{1, \Delta_{\mathbf{r}}}^{0,0} &= \frac{1}{2} \mathbb{G} [\mathcal{F}_{\Delta^{\mathbf{r}}}^{0,0}]_{0,0},\\
f_{2, \Delta_{\mathbf{r}}}^{0,0} &= 2 \mathbb{G} [\mathcal{F}_{\Delta^{\mathbf{r}}}^{0,0}]_{0,1} + \frac{1}{2}(1-x) \mathbb{G} [\mathcal{F}_{\Delta^{\mathbf{r}}}^{0,0}]_{0,2}, \\
f_{3, \Delta_{\mathbf{r}}}^{0,0}&= \frac{\mathbb{G} [\mathcal{F}_{\Delta^{\mathbf{r}}}^{0,0}]_{2,2}}{4} - \frac{\mathbb{G} [\mathcal{F}_{\Delta^{\mathbf{r}}}^{0,0}]_{1,2} + \partial_x \mathbb{G} [\mathcal{F}_{\Delta^{\mathbf{r}}}^{0,0}]_{0,2}}{x},\\ &\ \ \ \ \ \ \ \ \ \ \ \ \ \text{where}\ \mathbf{r}\in \{[0,0], [0,2], [2,0] \},\\
f_{\Delta_{[0,1]}}^{1,1} &= \frac{x}{2}\mathbb{G} [\frac{1}{\sqrt{\mathfrak{X}}}\mathcal{F}_{\Delta_{[0,1]}}^{1,1}]_{0,2},\\
f_{\Delta_{[0,1]}}^{1,-1} &= \frac{x}{2}\mathbb{G} [\frac{\tilde{\mathfrak{X}}}{\sqrt{\mathfrak{X}}}\mathcal{F}_{\Delta_{[0,1]}}^{1,-1}]_{0,2}.
\end{aligned}
\end{equation}
Analogously, for short multiplets, reduced superblock are defined as

\begin{equation}
\begin{aligned}
f_{1, \mathcal{B}_k}^{0,0} &= \frac{1}{2} \mathbb{G} [\mathcal{F}_{\mathcal{B}_k}^{0,0}]_{0,0},\\
f_{2, \mathcal{B}_k}^{0,0} &= 2 \mathbb{G} [\mathcal{F}_{\mathcal{B}_k}^{0,0}]_{0,1} + \frac{1}{2}(1-x) \mathbb{G} [\mathcal{F}_{\mathcal{B}_k}^{0,0}]_{0,2}, \\
f_{3, \mathcal{B}_k}^{0,0}&= \frac{\mathbb{G} [\mathcal{F}_{\mathcal{B}_k}^{0,0}]_{2,2}}{4} - \frac{\mathbb{G} [\mathcal{F}_{\mathcal{B}_k}^{0,0}]_{1,2} + \partial_x \mathbb{G} [\mathcal{F}_{\mathcal{B}_k}^{0,0}]_{0,2}}{x},\\ &\ \ \ \ \ \ \ \ \ \ \ \ \ \text{where}\ \mathbf{r}\in \{[0,0], [0,2], [2,0] \},\\
f_{\mathcal{B}_k}^{1,1} &= \frac{x}{2}\mathbb{G} [\frac{1}{\sqrt{\mathfrak{X}}}\mathcal{F}_{\mathcal{B}_k}^{1,1}]_{0,2},\\
f_{\mathcal{B}_k}^{1,-1} &= \frac{x}{2}\mathbb{G} [\frac{\tilde{\mathfrak{X}}}{\sqrt{\mathfrak{X}}}\mathcal{F}_{\mathcal{B}_k}^{1,-1}]_{0,2},
\end{aligned}
\end{equation}
and finally, for the identity operator, the reduced superblocks are defined only for the $\{0,0\}$ channel:
\begin{equation}\label{eqn:reducedblockI}
\begin{aligned}
f^{0,0}_{1, \mathcal{I}} = \frac{1}{x^2}\ \ \ \ \ \ 
f^{0,0}_{2, \mathcal{I}} = \frac{1}{x^2} + \frac{2}{x}\ \ \ \ \ \ 
f^{0,0}_{3, \mathcal{I}} = \frac{1}{x^4}.
\end{aligned}
\end{equation}

\subsection{Contributions of superblocks to the integrated constraints}
\label{app:blocksintconstraints}

The building blocks $b_{a}(\Delta)$, which are the part of the constraints \eqref{eq:intsumrule}, are given below. Here the conformal blocks of the reduced correlator \eqref{eq:reducedcor} are integrated with the  measures \eqref{eq:measures} producing the following exact expressions: 
\begin{equation}\begin{split}
\int_{0}^{1/2} \!\!\!\!\!\mu_1(x)  F_\Delta (x)  dx &=\!
\frac{-2^{-\Delta -1} 3 \Delta (\Delta(\Delta \!+\! 3) \!+\! 4)}{\Delta(\Delta + 2)(\Delta + 4)(\Delta^2 - 1)^2}
\bigg[
{}_2F_1\! \left(\!\Delta \!+\! 1, \Delta \!+\! 3; 2(\Delta \!+\! 2); \frac{1}{2} \right) \\
& \!\!\!\!\!+ \frac{8 + \Delta (\Delta + 3)(\Delta(\Delta + 3) + 4)}{\Delta (\Delta(\Delta + 3) + 4)}
{}_2F_1 \left(\Delta, \Delta + 3; 2(\Delta + 2); \frac{1}{2} \right)
\bigg], \\
\int_{0}^{1/2} \mu_2(x) F_\Delta(x) \, dx &=
\frac{2^{-\Delta}}{\Delta (\Delta^2 - 1)}
\bigg[
(\Delta + 1) {}_2F_1 \left(\Delta, \Delta + 2; 2(\Delta + 2); \frac{1}{2} \right) \\
&\quad - \Delta {}_2F_1 \left(\Delta - 1, \Delta + 1; 2(\Delta + 2); \frac{1}{2} \right)
\bigg].
\end{split}\end{equation}

\section{Details of the system of crossing equation}
\label{app:details crossing}

We clarify contributions from the BPS operators and the identity \eqref{eq:BPScompactcrossing} to the crossing equation \eqref{eq: final crossing equations 1a}. The similar vectors \eqref{eq: define V and tilde V} are introduced as for the long multiplets but with reduced superblocks corresponding to short multiplets (see  Appendix \ref{app:Superblocks}): 
\beq
V_{\mathcal{B}_k}=\left(\begin{array}{c}
\theta(\mathbf{r})\left(\begin{array}{cc}[x f_{1,\mathcal{B}_k}(x)]_s& 0\\ 0 & 0\end{array}\right)\\
\left(\begin{array}{cc} 0 & 0\\ 0 & [ f_{1,\mathcal{B}_k}(x)]_a\end{array}\right)\\
\left(\begin{array}{cc} 0 & 0\\ 0 & [f_{2,\mathcal{B}_k}(x)]_s\end{array}\right)\\
\left(\begin{array}{cc} 0 & 0\\ 0 & \left[f_{3,\mathcal{B}_k}(x)\right]_a\end{array}\right)\\
\left(\begin{array}{cc}0& 0\\ 0 & 0\end{array}\right)\\
\frac{\theta(\mathbf{r})}{2}\left(\begin{array}{cc}0& [ x f_{1,\mathcal{B}_k}(x)]_s\\ {}[ x f_{1,\mathcal{B}_k}(x)]_s & 0\end{array}\right)\\
\frac{\theta(\mathbf{r})}{2}\left(\begin{array}{cc}0& [ x f_{1,\mathcal{B}_k}(x)]_a\\ {}[ x f_{1,\mathcal{B}_k}(x)]_a & 0\end{array}\right)
\end{array}\right)\,, 
\quad
 \tilde{V}_{\mathcal{B}_l}=\left(\begin{array}{c}
  0 \\
 0 \\
0 \\
0 \\
\;\;[ f^{1,1}_{\mathcal{B}_l} ]_s \\
\;\;[ f^{1,-1}_{\mathcal{B}_l} ]_s \\
-[ f^{1,-1}_{\mathcal{B}_l} ]_a
\end{array}\right)\, ,
\eeq
where $\mathcal{B}_k$ with $k \in \{2,4\}$ and $\mathcal{B}_l$ with $l \in \{1,3 \}$ are short multiplets exchanged in the system. The parity of $\mathcal{B}_l$ multiplet is $\mathbb{P}=(-1)^l = -1$, such as the vector for these multiplets is introduced only for contributions with negative parity. Finally, let us clarify that in  \eqref{eq:BPScompactcrossing} the notation $(\tilde{V}_{\mathcal{B}_l})_{11}$ means that only a particular $\{1,1\}$ element of matrices in the vector is taken, and the same notation is used for $(\tilde{V}_{\mathcal{B}_k})_{22}$.

The contribution of the identity to \eqref{eq:BPScompactcrossing} looks as follows:

\beq
V_{\mathcal{I}}=\left(\begin{array}{c}
\theta(\mathbf{r})\left(\begin{array}{cc}[\frac{1}{x}]_s& 0\\ 0 & 0\end{array}\right)\\
\left(\begin{array}{cc} 0 & 0\\ 0 & [ \frac{1}{x^2}]_a\end{array}\right)\\
\left(\begin{array}{cc} 0 & 0\\ 0 & [\frac{1}{x^2}+ \frac{2}{x}]_s\end{array}\right)\\
\left(\begin{array}{cc} 0 & 0\\ 0 & \left[\frac{1}{x^4}\right]_a\end{array}\right)\\
\left(\begin{array}{cc}0& 0\\ 0 & 0\end{array}\right)\\
\frac{\theta(\mathbf{r})}{2}\left(\begin{array}{cc}0& [\frac{1}{x}]_s\\ {}[ \frac{1}{x}]_s & 0\end{array}\right)\\
\frac{\theta(\mathbf{r})}{2}\left(\begin{array}{cc}0& [\frac{1}{x}]_a\\ {}[\frac{1}{x}]_a & 0\end{array}\right)
\end{array}\right)\,, 
\eeq
where the reduced functions of the identity operator \eqref{eqn:reducedblockI} are used.

\section{Spectral data}
\label{app:spectraldata}
Here we collect spectral data for the charged sectors, obtained with the QSC.
This data was produced by generalising the methods developed in~\cite{Grabner:2020nis,Julius:2021uka,Cavaglia:2021bnz}.
Details of the QSC description used to obtain this data will be presented in the upcoming work~\cite{QSClineToAppear}. 
We will go by multiplet-type. For the $\mathcal{L}^{\Delta}_{[0,0]}$ multiplet, all the data is already published in~\cite{Cavaglia:2022qpg} and can be accessed in Appendix D of that paper. 

Therefore, we begin with the $\mathcal{L}^{\Delta}_{[0,1]}$ type multiplets. The scaling dimensions of the superconformal primaries of the first nine such multiplets for the first few orders in perturbation theory is given below. We have

\begin{align}
    \label{eqn:perturbativespectrum1}
    {}_2[\gr0\;0\;1\;\gr2]_1: \Delta_1^{[0,1]^-} &= 2 + 2\,g^2 - 4\,g^4 + \left(16 - \frac{4}{3}\pi^2\right) \,g^6 + \mathrm{O}(g^8)\;, \\
    \label{eqn:perturbativespectrum2}
    {}_2[\gr0\;0\;1\;\gr2]_2: \Delta_1^{[0,1]^+} &= 2 + 6\,g^2 -24\,g^4 + (168-4\pi^2)\,g^6+\mathrm{O}(g^8)\;, \\
    {}_3[\gr0\;0\;1\;\gr3]_{1}: \Delta_2^{[0,1]^-} &= 3 + z_1\,g^2 + a_1\,g^4 + \mathrm{O}(g^6)\;, \\
    {}_3[\gr0\;0\;1\;\gr3]_2: \Delta_2^{[0,1]^+} &= 3 + \frac{11 - \sqrt{17}}{2}\,g^2  - \frac{1207-225 \sqrt{17}}{68}\,g^4  + \mathrm{O}(g^6)\;, \\
    {}_3[\gr0\;0\;1\;\gr3]_3: \Delta_3^{[0,1]^-} &= 3 + 4\,g^2 - 12\,g^4 + \mathrm{O}(g^6)\;, \\
    {}_3[\gr0\;0\;1\;\gr3]_4: \Delta_3^{[0,1]^+} &= 3 + \frac{11 + \sqrt{17}}{2}\,g^2 - \frac{1207 + 225 \sqrt{17}}{68}\,g^4 + \mathrm{O}(g^6) \\
    {}_3[\gr0\;0\;1\;\gr3]_{5}: \Delta_4^{[0,1]^-} &= 3 + z_5\,g^2 + a_5\,g^4 + \mathrm{O}(g^6)\;, \\
    {}_3[\gr0\;0\;1\;\gr3]_{6}: \Delta_5^{[0,1]^-} &= 3 + z_6\,g^2 + a_6\,g^4 + \mathrm{O}(g^6)\;,\\
    {}_4[\gr0\;0\;1\;\gr4]_{1}: \Delta_6^{[0,1]^-} &= 4 + w_1\,g^2 + b_1\,g^4+\mathrm{O}(g^6)
    \;. 
\end{align}
Here $z_i$, $i = 1,5,6$ are solutions of the equation
\begin{align}
    z^3 - 21\,z^2 + 126\,z - 168 = 0
    \;,
\end{align}
arranged in increasing order of magnitude, 
$a_i$, $i = 1,5,6$ are solutions of
\begin{align}
    567\,a^3 + 44226\,a^2 + 824544\,a + 570016 = 0\;.
\end{align}
in increasing order of absolute value, $w_1$ is the solution to the equation
\begin{multline}
    w^8 - 62\,w^7 + 1618\,w^6 - 23116\,w^5 + 196644\,w^4 - 1011840\,w^3 \\+ 3039600\,w^2 - 4790400\,w + 2959488 = 0\;,
\end{multline}
with lowest magnitude, and $b_1$ is the solution of
\begin{multline}
    29045897561620224\,b^8 + 6408891967861027072\,b^7 + 579482092904176351152\,b^6 
    \\
    + 27742806477747634595596\,b^5 + 757061839838707131304283\,b^4 
    \\
    + 11731093466662770554932380\,b^3
    + 95371750707579774337996884\,b^2 
    \\
    + 318992809404389038136720288\,b 
    + 67201030403179647033207584
    =0\;.
\end{multline}
with lowest absolute value.

Next we consider the $\mathcal{L}^\Delta_{[0,2]}$ superprimaries. There are four multiplets in total that we consider here. For the first three of them, we have
\begin{align}
    {}_3[\gr0\;0\;2\;\gr3]_{1}: \Delta_1^{{[0,2]}^+} &= 3 + (4 - 2\,\sqrt{2})\,g^2 -(14 - 9\,\sqrt{2})\,g^4+ \mathrm{O}(g^6)\;, \\
    {}_3[\gr0\;0\;2\;\gr3]_{2}: \Delta_1^{{[0,2]}^-} &= 3 + 4\,g^2 - 12\,g^4 + \mathrm{O}(g^6)\;, \\
    {}_3[\gr0\;0\;2\;\gr3]_{3}: \Delta_2^{{[0,2]}^+} &= 3 + (4 + 2\,\sqrt{2})\,g^2 -(14 + 9\,\sqrt{2})\,g^4 + \mathrm{O}(g^6)\;, \\
    {}_4[\gr0\;0\;2\;\gr4]_{1}: \Delta_3^{{[0,2]}^+} &= 4 + w_1\,g^2 + b_1\,g^4 + \mathrm{O}(g^6)\;, 
\end{align}
where $w_1$ is the solution to the equation
\begin{align}
    w^7 - 46\,w^6 + 852\,w^5 - 8144\,w^4 + 42784\,w^3 - 121536\,w^2 + 170496\,w - 89856 = 0\;,
\end{align}
with least magnitude, and $b_1$ is the solution of
\begin{align}
\begin{split}
    &768717064\,b^7 + 121457296112\,b^6 \\&+ 7280058966424\,b^5 + 204908723870632\,b^4
    + 2674078437270158\,b^3 \\&+ 12476257788725865\,b^2 + 24571867339368970\,b + 8490367768144676
    =0
    \;.
\end{split}
\end{align}
with the least absolute value.

Finally, we consider the $\mathcal{L}^{\Delta}_{[2,0]}$ multiplets. There are three such multiplets which we consider. The one loop scaling dimensions of the first two are given by
\begin{align}
    {}_3[\gr0\;2\;0\;\gr3]_{1}: \Delta_1^{[2,0]^-} &= 3 + (6 - 2\,\sqrt{3})\,g^2 - (20 - 10\,\sqrt{3})g^4 + \mathrm{O}(g^6)\;, \\
    {}_3[\gr0\;2\;0\;\gr3]_{2}: \Delta_2^{[2,0]^-} &= 3 + (6 + 2\,\sqrt{3})\,g^2 - (20 + 10\,\sqrt{3})g^4 + \mathrm{O}(g^6)\;,\\
    {}_4[\gr0\;2\;0\;\gr4]_{1}: \Delta_3^{[2,0]^-} &= 4 + w_1\,g^2 + b_1\,g^4 +  \mathrm{O}(g^6)\;, \\
\end{align}
where $w_1$ is the solution to the equation
\begin{align}
    w^4 - 28\,w^3 + 272\,w^2 - 1040\,w + 1184 = 0\;,
\end{align}
with least magnitude and $b_1$ is the solution of
\begin{align}
    20228\,b^4 + 1885872\,b^3 + 54962272\,b^2 + 494844174\,b + 325918891
    =0\;,
\end{align}
with lowest absolute value.
In Tables~\ref{tab:states00} --~\ref{tab:states20} below, we summarise the field content of the one-loop wavefunctions of the highest-twist states in the four types of supermultiplets.

\newpage
\begin{table}[ht!]
\scriptsize
\begin{adjustwidth}{}{}
\begin{tabular}{|C|C|C|C|C|C|C|}
\hline
    \multicolumn{7}{|C|}{\rule{0pt}{3.5ex}\mathcal{L}^\Delta_{[0,0]}}  \\[2ex]
    \hline\hline 
    \rule{0pt}{3.5ex}\texttt{State ID} & \texttt{Tag} & \Delta_0 &  \#\,g^2 & \Delta_{\infty} & \text{Wavefunction for highest twist states} & \mathbb{P}\\[2ex] 
    \hline\hline 
    \rule{0pt}{3.5ex}{}_ 1 {[\gr0\; 0\; 0\;\gr 1]}_ 1 & \Delta_1^{[0,0]^+} & 1 & 4 & 2 & \Phi_{||} & +1 \\[2ex]\hline
    \rule{0pt}{3.5ex}{}_ 2 {[\gr0\; 0\; 0\;\gr 2]}_ 1 & \Delta_2^{[0,0]^+} & 2 & 2.764 & 4 & \Phi_{||}^2 - \frac{1}{\sqrt{5}}\Phi^i_\perp\,\Phi^i_\perp & +1 \\[2ex]\hline
    \rule{0pt}{3.5ex}{}_ 2 {[\gr0\; 0\; 0\;\gr 2]}_ 2 & \Delta_3^{[0,0]^+} & 2 & 7.236 & 4 & \Phi_{||}^2 + \frac{1}{\sqrt{5}}\Phi^i_\perp\,\Phi^i_\perp & +1 \\[2ex]\hline
    \rule{0pt}{3.5ex}{}_ 3 {[\gr0\; 0\; 0\;\gr 3]}_ 1 & \Delta_4^{[0,0]^+} & 3 & 2.132 & 6 & \Phi_{||}^3 + a_{3,1;1}\,(\Phi_{||}\,\Phi_\perp^i\,\Phi_\perp^i + \Phi_\perp^i\,\Phi_\perp^i\,\Phi_{||}) + a_{3,1;2}\,\Phi_\perp^i\,\Phi_{||}\,\Phi_\perp^i & +1 \\[2ex]\hline
    \rule{0pt}{3.5ex}{}_ 3 {[\gr0\; 0\; 0\;\gr 3]}_ 2 & \Delta_5^{[0,0]^+} & 3 & 5.103 & 6 & \Phi_{||}^3 + a_{3,2;1}\,(\Phi_{||}\,\Phi_\perp^i\,\Phi_\perp^i + \Phi_\perp^i\,\Phi_\perp^i\,\Phi_{||}) + a_{3,2;2}\,\Phi_\perp^i\,\Phi_{||}\,\Phi_\perp^i & +1 \\[2ex]\hline
    \rule{0pt}{3.5ex}{}_ 3 {[\gr0\; 0\; 0\;\gr 2]}_ 1 & \Delta_6^{[0,0]^+} & 3 & 5.639 & 6 & \text{N/A} & \\[2ex]\hline
    \rule{0pt}{3.5ex}{}_ 3 {[\gr0\; 0\; 0\;\gr 3]}_ 3 & \Delta_1^{[0,0]^-} & 3 & 9 & 7 & \Phi_{||}\,\Phi_\perp^i\,\Phi_\perp^i - \Phi_\perp^i\,\Phi_\perp^i\,\Phi_{||} & -1 \\[2ex]\hline
    \rule{0pt}{3.5ex}{}_ 3 {[\gr0\; 0\; 0\;\gr 2]}_ 2 & \Delta_7^{[0,0]^+} & 3 & 9.694 & 6 & \text{N/A} & \\[2ex]\hline
    \rule{0pt}{3.5ex}{}_ 3 {[\gr0\; 0\; 0\;\gr 3]}_ 4 & \Delta_8^{[0,0]^+} & 3 & 11.765 & 8 & \Phi_{||}^3 + a_{3,4;1}\,(\Phi_{||}\,\Phi_\perp^i\,\Phi_\perp^i + \Phi_\perp^i\,\Phi_\perp^i\,\Phi_{||}) + a_{3,4;2}\,\Phi_\perp^i\,\Phi_{||}\,\Phi_\perp^i & +1 \\[2ex]\hline
    \rule{0pt}{3.5ex}{}_ 4 {[\gr0\; 0\; 0\;\gr 4]}_{1} & \Delta_{9}^{[0,0]^+} & 4 & 1.742 & 8 &
    \shortstack{ \rule{0pt}{3.5ex}
    $\Phi_{||}^{4} + a_{4,1;1}(\Phi_{||}^{2}\,\Phi_{\perp}^{i}\,\Phi_{\perp}^{i} + \Phi_{\perp}^{i}\,\Phi_{\perp}^{i}\,\Phi_{||}^{2})$
    \\
    $+ a_{4,1;2}\,\Phi_{||}\,\Phi^{i}_{\perp}\,\Phi^{i}_{\perp}\,\Phi_{||} + a_{4,1;3}\,\Phi_\perp^i\,\Phi_{||}^2\,\Phi_\perp^i$
    \\
    $+ a_{4,1;4}(\Phi_{||}\,\Phi_\perp^i\,\Phi_{||}\,\Phi_\perp^i + \Phi_\perp^i\Phi_{||}\,\Phi_\perp^i\,\Phi_{||})$
    \\
    $+ a_{4,1;5}\,\Phi_\perp^i\,\Phi_\perp^i\,\Phi_\perp^j\,\Phi_\perp^j + a_{4,1;6}\,\Phi_\perp^i\,\Phi_\perp^j\,\Phi_\perp^i\,\Phi_\perp^j$
    \\
    $+ a_{4,1;7}\,\Phi_\perp^i\,\Phi_\perp^j\,\Phi_\perp^j\,\Phi_\perp^i$
    } & +1 \\[2ex]
    \hline
\end{tabular}
\captionof{table}{
\label{tab:states00}
The states belonging to the $\mathcal{L}^\Delta_{[0,0]}$ multiplets. We display the \texttt{State ID}, the \texttt{Tag}, which can be used to identify the state in the bootstrap implementation, the bare and one-loop scaling dimension at weak coupling, and scaling dimension at infinite coupling. Then, we display the one-loop wavefunction and Parity charge. In some cases the explicit numerical coefficients of various combinations of fields are cumbersome to write. In such cases we denote them using the letters $a_{\Delta_0,\mathtt{sol},i}$, ensuring still that the charge under Parity can be inferred.} 
\end{adjustwidth}
\end{table}

\newpage
\begin{table}[ht!]
\centering
\scriptsize
\begin{tabular}{|C|C|C|C|C|C|C|}
\hline
\multicolumn{7}{|C|}{\rule{0pt}{3.5ex} \mathcal{L}^\Delta_{[0,1]}} \\[2ex]
\hline\hline
\rule{0pt}{3.5ex}\texttt{State ID} & \texttt{Tag} & \Delta_0 & \#\,g^2 & \Delta_{\infty} & \text{Wavefunction for highest twist states} & \mathbb{P}\\[2ex] 
    \hline\hline
    \rule{0pt}{3.5ex}{}_ 2 {[\gr0\; 0\; 1\;\gr 2]}_ 1 & \Delta_1^{[0,1]^-} & 2 & 2 & 3 & \Phi_{||} \Phi_{\perp}^i + \Phi_{\perp}^i \Phi_{||} & -1 \\[2ex]\hline
    \rule{0pt}{3.5ex}{}_ 2 {[\gr0\; 0\; 1\;\gr 2]}_ 2 & \Delta_1^{[0,1]^+} & 2 & 6 & 4 & \Phi_{||} \Phi_{\perp}^i - \Phi_{\perp}^i \Phi_{||} & +1 \\[2ex]\hline
    \rule{0pt}{3.5ex}{}_ 3 {[\gr0\; 0\; 1\;\gr 3]}_ 1 & \Delta_2^{[0,1]^-} & 3 & 1.858 & 5 & 
    \shortstack{ \rule{0pt}{3.5ex} $\Phi_{||}^2\Phi_i + \Phi_{i}\Phi_{||}^2 - b_{3,1;1} \Phi_{||}\Phi_{\perp}^i\Phi_{||}$
    \\
    $+b_{3,1;2}(\Phi_{\perp}^{j}\Phi_{\perp}^{j}\Phi_{\perp}^{i} +\Phi_{\perp}^{i}\Phi_{\perp}^{j}\Phi_{\perp}^{j})  + b_{3,1;3} \Phi_{\perp}^{j}\Phi_{\perp}^{i}\Phi_{\perp}^{j}$} & -1 \\[2ex]\hline
    \rule{0pt}{3.5ex}{}_ 3 {[\gr0\; 0\; 1\;\gr 3]}_ 2 & \Delta_2^{[0,1]^+} & 3 & 3.438 & 6 & (\Phi_{||}^2 \Phi_{\perp}^i - \Phi_{\perp}^i \Phi_{||}^2) + \frac{1 - \sqrt{17}}{8} (\Phi_{\perp}^j \Phi_{\perp}^j \Phi_{\perp}^i - \Phi_{\perp}^i \Phi_{\perp}^j \Phi_{\perp}^j)  & +1 \\[2ex]\hline
    \rule{0pt}{3.5ex}{}_ 3 {[\gr0\; 0\; 1\;\gr 3]}_ 3 & \Delta_3^{[0,1]^-} & 3 & 4 & 5 & \Phi_{\perp}^j \Phi_{\perp}^i \Phi_{\perp}^j + \Phi_{||}^2 \Phi_{\perp}^i + \Phi_{\perp}^i \Phi_{||}^2    & -1 \\[2ex]\hline
    \rule{0pt}{3.5ex}{}_ 3 {[\gr0\; 0\; 1\;\gr 3]}_ 4 & \Delta_3^{[0,1]^+} & 3 & 7.562 & 6 &  (\Phi_{||}^2 \Phi_{\perp}^i - \Phi_{\perp}^i \Phi_{||}^2) + \frac{1 + \sqrt{17}}{8} (\Phi_{\perp}^j \Phi_{\perp}^j \Phi_{\perp}^i - \Phi_{\perp}^i \Phi_{\perp}^j \Phi_{\perp}^j)  & +1 \\[2ex]\hline
    \rule{0pt}{3.5ex}{}_ 3 {[\gr0\; 0\; 1\;\gr 3]}_ 5 & \Delta_4^{[0,1]^-} & 3 & 8.491 & 9  & 
    \shortstack{\rule{0pt}{3.5ex}
    $\Phi_{||}^2\Phi_\perp^i + \Phi_\perp^{i}\Phi_{||}^2 - b_{3,5;1} \Phi_{||}\Phi_{\perp}^i\Phi_{||} $
    \\
    $+b_{3,5;2}(\Phi_{\perp}^{j}\Phi_{\perp}^{j}\Phi_{\perp}^{i} +\Phi_{\perp}^{i}\Phi_{\perp}^{j}\Phi_{\perp}^{j})  + b_{3,5;3} \Phi_{\perp}^{j}\Phi_{\perp}^{i}\Phi_{\perp}^{j}$
    }
    & -1 \\[2ex]\hline
    \rule{0pt}{3.5ex}{}_ 3 {[\gr0\; 0\; 1\; \gr3]}_ 6 & \Delta_5^{[0,1]^-} & 3 & 10.651 & 9 &  
    \shortstack{ \rule{0pt}{3.5ex}
    $\Phi_{||}^2\Phi_\perp^i + \Phi_\perp^{i}\Phi_{||}^2 - b_{3,6;1} \Phi_{||}\Phi_{\perp}^i\Phi_{||}$ 
    \\
    $+b_{3,6;2}(\Phi_{\perp}^{j}\Phi_{\perp}^{j}\Phi_{\perp}^{i} +\Phi_{\perp}^{i}\Phi_{\perp}^{j}\Phi_{\perp}^{j})  + b_{3,6;3} \Phi_{\perp}^{j}\Phi_{\perp}^{i}\Phi_{\perp}^{j}$
    }
    & -1 \\[2ex]\hline
    \rule{0pt}{3.5ex}{}_ 4 {[\gr0\; 0\; 1\;\gr 4]}_ 1 & \Delta_6^{[0,1]^-} & 4 & 1.613 & 9 & 
    \shortstack{
    \rule{0pt}{3.5ex}
    $ \Phi_{||}^3\Phi_\perp^i +
    \Phi_\perp^i\Phi_{||}^3 + 
    b_{4,1;1}(\Phi_{||}^2\Phi_\perp^i\Phi_{||} + 
    \Phi_{||}\Phi_\perp^i\Phi_{||}^2) $ \\ $ +
    b_{4,1;2}(\Phi_{||}\Phi_\perp^i\Phi_\perp^j\Phi_\perp^j +
    \Phi_\perp^j\Phi_\perp^j\Phi_\perp^i\Phi_{||}) $\\$ + 
    b_{4,1;3}(\Phi_{||}\Phi_\perp^j\Phi_\perp^i\Phi_\perp^j +
    \Phi_\perp^j\Phi_\perp^i\Phi_\perp^j\Phi_{||}) $ \\ $ + 
    b_{4,1;4}(\Phi_{||}\Phi_\perp^j\Phi_\perp^j\Phi_\perp^i +
    \Phi_\perp^i\Phi_\perp^j\Phi_\perp^j\Phi_{||}) $ \\ 
    $ 
    b_{4,1;5}(\Phi_\perp^i\Phi_{||}\Phi_\perp^j\Phi_\perp^j + 
    \Phi_\perp^j\Phi_\perp^j \Phi_{||}\Phi_\perp^i) $\\$ +
    b_{4,1;6}(\Phi_\perp^i\Phi_\perp^j\Phi_{||}\Phi_\perp^j + \Phi_\perp^j\Phi_{||}\Phi_\perp^j\Phi_\perp^i) $\\$
    b_{4,1;7}(\Phi_\perp^j\Phi_\perp^i\Phi_{||}\Phi_\perp^j  
    \Phi_\perp^j\Phi_{||}\Phi_\perp^i\Phi_\perp^j )
    $
    }  & -1 \\[2ex]
    \hline
\end{tabular}
\captionof{table}{
\label{tab:states01}
The states belonging to the $\mathcal{L}^\Delta_{[0,1]}$ multiplets. We display the \texttt{State ID}, the \texttt{Tag}, which can be used to identify the state in the bootstrap implementation, the bare and one-loop scaling dimension at weak coupling, and scaling dimension at infinite coupling. Then, we display the one-loop wavefunction and Parity charge. In some cases the explicit numerical coefficients of various combinations of fields are cumbersome to write. In such cases we denote them using the letters $b_{\Delta_0,\mathtt{sol},i}$, ensuring still that the charge under Parity can be inferred.} 
\end{table}

\newpage
\begin{table}[ht!]
\centering
\scriptsize
\begin{tabular}{|C|C|C|C|C|C|C|}
\hline
\multicolumn{7}{|C|}{\rule{0pt}{3.5ex}\mathcal{L}^\Delta_{[0,2]}} \\[2ex]
    \hline\hline
\rule{0pt}{3.5ex}\texttt{State ID} & \texttt{Tag} & \Delta_0 & \#\,g^2 & \Delta_{\infty} & \text{Wavefunction for highest twist states} & \mathbb{P}\\[2ex] 
    \hline\hline
    \rule{0pt}{3.5ex} {}_3{[\gr 0\; 0\; 2\;\gr 3]}_1 & \Delta_1^{[0,2]^+} & 3 & 1.172 & 4 & 
    \shortstack{ \rule{0pt}{3.5ex}
    $\sqrt{2} (\Phi_\perp^i \Phi_{||}\Phi_\perp^j + \Phi_\perp^j \Phi_{||}\Phi_\perp^i  )$ \\ 
    $+ \Phi_{||} ( \Phi_\perp^i \Phi_\perp^j + \Phi_\perp^j \Phi_\perp^i) + ( \Phi_\perp^i \Phi_\perp^j + \Phi_\perp^j \Phi_\perp^i) \Phi_{||} $ \\
    $-\delta^{ij}\frac{2}{5} \bigg[\sqrt{2} \Phi_\perp^k \Phi_{||}\Phi_\perp^k + \Phi_{||} \Phi_\perp^k \Phi_\perp^k +  \Phi_\perp^k \Phi_\perp^k  \Phi_{||} \bigg]$ 
    } 
    & 1 \\[2ex]
    \hline
    \rule{0pt}{3.5ex} {}_3{[\gr 0\; 0\; 2\;\gr 3]}_2 & \Delta_1^{[0,2]^-} & 3 & 4 & 5 & 
    \shortstack{  \rule{0pt}{3.5ex}
    $\Phi_{||} (\Phi_\perp^i \Phi_\perp^j + \Phi_\perp^j \Phi_\perp^i) - (\Phi_\perp^j \Phi_\perp^i + \Phi_\perp^i \Phi_\perp^j) \Phi_{||}$ \\ 
    $- \delta^{ij}\frac{2}{5} (\Phi_{||}\Phi_\perp^k\Phi_\perp^k - \Phi_\perp^k\Phi_\perp^k\Phi_{||})$ 
    } & -1 \\[2ex]
    \hline
    \rule{0pt}{3.5ex} {}_3{[\gr 0\; 0\; 2\;\gr 3]}_3 & \Delta_2^{[0,2]^+} & 3 & 6.828 & 6 & 
    \shortstack{\rule{0pt}{3.5ex}
    $-\sqrt{2} (\Phi_\perp^i \Phi_{||}\Phi_\perp^j + \Phi_\perp^j \Phi_{||}\Phi_\perp^i  )$ \\ 
    $+ \Phi_{||} ( \Phi_\perp^i \Phi_\perp^j + \Phi_\perp^j \Phi_\perp^i) + ( \Phi_\perp^i \Phi_\perp^j + \Phi_\perp^j \Phi_\perp^i) \Phi_{||} $ \\ 
    $-\delta^{ij}\frac{2}{5}\bigg[-\sqrt{2} \Phi_\perp^k \Phi_{||}\Phi_\perp^k + \Phi_{||}  \Phi_\perp^k \Phi_\perp^k + \Phi_\perp^k \Phi_\perp^k  \Phi_{||}\bigg]$
    } & 1 \\[2ex]
    \hline
    \rule{0pt}{3.5ex} {}_4{[\gr 0\; 0\; 2\;\gr 4]}_{1} & \Delta_3^{[0,2]^+} & 4 & 1.293 & 6 & 
    \shortstack{\rule{0pt}{3.5ex}  
    $\Phi_{||}^2(\Phi_\perp^i\Phi_\perp^j + \Phi_\perp^j\Phi_\perp^i ) + (\Phi_\perp^i\Phi_\perp^j + \Phi_\perp^j\Phi_\perp^i )\Phi_{||}^2$ \\ 
    $+ c_{4,1;1} (\Phi_{||}\Phi_\perp^i\Phi_{||}\Phi_\perp^j + \Phi_{||}\Phi_\perp^j\Phi_{||}\Phi_\perp^i$ \\ 
    $\Phi_\perp^i\Phi_{||}\Phi_\perp^j\Phi_{||} + \Phi_\perp^j\Phi_{||}\Phi_\perp^i\Phi_{||})$ \\ 
    $+ c_{4,1;2} \Phi_{||}(\Phi_\perp^i\Phi_\perp^j + \Phi_\perp^j\Phi_\perp^i)\Phi_{||}$ \\ 
    $+ c_{4,1;3} (\Phi_\perp^i\Phi_{||}^2\Phi_\perp^j + \Phi_\perp^j\Phi_{||}^2\Phi_\perp^i)$ \\ 
    $c_{4,1;4} (\Phi_\perp^l\Phi_\perp^l(\Phi_\perp^i\Phi_\perp^j + \Phi_\perp^j\Phi_\perp^i)$ \\ 
    $+ (\Phi_\perp^i\Phi_\perp^j + \Phi_\perp^j\Phi_\perp^i)\Phi_\perp^l\Phi_\perp^l)$ \\ 
    $+ c_{4,1;5} (\Phi_\perp^i\Phi_\perp^l\Phi_\perp^j\Phi_\perp^l + \Phi_\perp^j\Phi_\perp^l\Phi_\perp^i\Phi_\perp^l$ \\ 
    $+ \Phi_\perp^l\Phi_\perp^i\Phi_\perp^l\Phi_\perp^j + \Phi_\perp^l\Phi_\perp^j\Phi_\perp^l\Phi_\perp^i)$ \\ 
    $+ c_{4,1;6} (\Phi_\perp^i\Phi_\perp^l\Phi_\perp^l\Phi_\perp^j + \Phi_\perp^j\Phi_\perp^l\Phi_\perp^l\Phi_\perp^i)$ \\ 
    $+ c_{4,1;7} (\Phi_\perp^l\Phi_\perp^i\Phi_\perp^j\Phi_\perp^l + \Phi_\perp^l\Phi_\perp^j\Phi_\perp^i\Phi_\perp^l)$ \\ 
    $-\delta^{ij}\frac{2}{5}\bigg[\Phi_{||}^2\Phi_\perp^k\Phi_\perp^k + \Phi_\perp^k\Phi_\perp^k\Phi_{||}^2$ \\ 
    $+ c_{4,1;1} (\Phi_{||}\Phi_\perp^k\Phi_{||}\Phi_\perp^k + \Phi_\perp^k\Phi_{||}\Phi_\perp^k\Phi_{||})$ \\ 
    $+ c_{4,1;2} \Phi_{||}(\Phi_\perp^k\Phi_\perp^k + \Phi_\perp^k\Phi_\perp^k)\Phi_{||}$ \\ 
    $+ c_{4,1;3} \Phi_\perp^k\Phi_{||}^2\Phi_\perp^k$ \\ 
    $+ c_{4,1;4} (\Phi_\perp^l\Phi_\perp^l\Phi_\perp^k\Phi_\perp^k + \Phi_\perp^k\Phi_\perp^k\Phi_\perp^l\Phi_\perp^l)$ \\ 
    $+ c_{4,1;5} (\Phi_\perp^k\Phi_\perp^l\Phi_\perp^k\Phi_\perp^l + \Phi_\perp^l\Phi_\perp^k\Phi_\perp^l\Phi_\perp^k)$ \\ 
    $+ c_{4,1;6} \Phi_\perp^k\Phi_\perp^l\Phi_\perp^l\Phi_\perp^k + c_{4,1;7}\Phi_\perp^l\Phi_\perp^k\Phi_\perp^k\Phi_\perp^l \bigg]$ 
    } & 1 \\[2ex]
    \hline 
\end{tabular}
\captionof{table}{
\label{tab:states02}
The states belonging to the $\mathcal{L}^\Delta_{[0,2]}$ multiplets. We display the \texttt{State ID}, the \texttt{Tag}, which can be used to identify the state in the bootstrap implementation, the bare and one-loop scaling dimension at weak coupling, and scaling dimension at infinite coupling. Then, we display the one-loop wavefunction and Parity charge. In some cases the explicit numerical coefficients of various combinations of fields are cumbersome to write. In such cases we denote them using the letters $c_{\Delta_0,\mathtt{sol},i}$, ensuring still that the charge under Parity can be inferred.} 
\end{table}

\newpage
\begin{table}[ht!]
\centering
\scriptsize
\begin{tabular}{|C|C|C|C|C|C|C|}
\hline
    \multicolumn{7}{|C|}{\rule{0pt}{3.5ex} \mathcal{L}^\Delta_{[2,0]}} \\[2ex]
    \hline\hline
    \rule{0pt}{3.5ex}\texttt{State ID} & \texttt{Tag} & \Delta_0 & \#\,g^2 & \Delta_{\infty} & \text{Wavefunction for highest twist states} & \mathbb{P}\\[2ex] 
    \hline\hline
    \rule{0pt}{3.5ex}{}_ 3 {[\gr0\; 2\; 0\;\gr 3]}_ 1 & \Delta_1^{[2,0]^-} & 3 & 2.536 & 5 & 
    \shortstack{\rule{0pt}{3.5ex} $(\sqrt{3}+1)(\Phi^i_{\perp}\Phi_{||}\Phi^j_{\perp} - \Phi^j_{\perp}\Phi_{||}\Phi^i_{\perp})$\\  $+(\Phi^i_{\perp}\Phi^j_{\perp} \Phi_{||} - \Phi^j_{\perp}\Phi^i_{\perp} \Phi_{||})  $ \\ $ + (\Phi_{||} \Phi^i_{\perp}\Phi^j_{\perp} -  \Phi_{||}\Phi^j_{\perp}\Phi^i_{\perp})$}  & -1 \\[2ex]\hline 
    \rule{0pt}{3.5ex}{}_ 3 {[\gr0\; 2\; 0\;\gr 3]}_ 2 & \Delta_2^{[2,0]^-} & 3 & 9.464 & 7 & 
    \shortstack{\rule{0pt}{3.5ex}$(\sqrt{3}-1)(\Phi^i_{\perp}\Phi_{||}\Phi^j_{\perp} - \Phi^j_{\perp}\Phi_{||}\Phi^i_{\perp})$ \\  $ - (\Phi^i_{\perp}\Phi^j_{\perp} \Phi_{||} - \Phi^j_{\perp}\Phi^i_{\perp} \ \Phi_{||}) $ \\ $-    (\Phi_{||} \Phi^i_{\perp}\Phi^j_{\perp}  -  \Phi_{||}\Phi^j_{\perp}\Phi^i_{\perp})$}  & -1 \\[2ex]\hline
    {}_ 4 {[\gr0\; 2\; 0\;\gr 4]}_ 1 & \Delta_3^{[2,0]^-} & 4 & 1.939 & 7 & 
    \shortstack{\rule{0pt}{3.5ex} $\Phi_{||}^2(\Phi^i_{\perp}\Phi^j_{\perp} - \Phi^j_{\perp}\Phi^i_{\perp}) $\\$ + d_{4,1;1} (\Phi_{||}\Phi^i_{\perp}\Phi_{||}\Phi^j_{\perp} - \Phi_{||}\Phi^j_{\perp}\Phi_{||}\Phi^i_{\perp}) $ \\  $+ d_{4,1;2} (\Phi_{||}\Phi^i_{\perp}\Phi^j_{\perp} \Phi_{||} - \Phi_{||}\Phi^j_{\perp}\Phi^i_{\perp} \Phi_{||})$ \\  $+ d_{4,1;3} (\Phi^i_{\perp}\Phi_{||}^2\Phi^j_{\perp}  - \Phi^j_{\perp} \Phi_{||}^2 \Phi^i_{\perp})$ \\ \small  $+ d_{4,1;4} (\Phi^i_{\perp}\Phi_{||}\Phi^j_{\perp} \Phi_{||} - \Phi^j_{\perp} \Phi_{||} \Phi^i_{\perp} \Phi_{||})$\\ $ + d_{4,1;5} (\Phi^i_{\perp}\Phi^j_{\perp} \Phi_{||}^2 - \Phi^j_{\perp}\Phi^i_{\perp} \ \Phi_{||}^2) $  }   & -1 \\[2ex]\hline 
\end{tabular}
\captionof{table}{
\label{tab:states20}
The states belonging to the $\mathcal{L}^\Delta_{[2,0]}$ multiplets. We display the \texttt{State ID}, the \texttt{Tag}, which can be used to identify the state in the bootstrap implementation, the bare and one-loop scaling dimension at weak coupling, and scaling dimension at infinite coupling. Then, we display the one-loop wavefunction and Parity charge. In some cases the explicit numerical coefficients of various combinations of fields are cumbersome to write. In such cases we denote them using the letters $d_{\Delta_0,\mathtt{sol},i}$, ensuring still that the charge under Parity can be inferred.} 
\end{table}

\section{Some details on the \texttt{SDPB} algorithm}

To complete the description of the method let us give the explicit form of $\mathcal{F}$ introduced schematically: this is a vector of 7 components representing the contribution of all the states below the gaps to the crossing equations, and it is parametrized in terms of the (known) spectral levels and (unknown) OPE coefficients of those states. Given the candidate data, this is a  vector of functions of the cross ratio defined as
\beq\label{eq:Fdef}
\mathcal{F} \equiv \mathcal{B}_{\text{BPS}}  +\!\!\!\!\!\!\!\! \sum_{ \Delta \in  \texttt{S}^{[0,0]^+}\cup \texttt{S}^{[2,0]^-}\cup \texttt{S}^{[0,2]^+}}^{ \Delta < \Delta_{\text{gap}}^{\mathbf{r}}  }
\!\!\!\!\!\!\!\!\!\!\left(\begin{array}{cc} C_{11 \Delta}^{\mathbf{r}} & C_{22 \Delta}^{\mathbf{r}} \end{array}\right)V_{\Delta^{\mathbf{r}}}\left(\begin{array}{c}C_{11 \Delta}^{\mathbf{r}} \\C_{22 \Delta}^{\mathbf{r}} \end{array}\right) + \!\!\!\! \sum_{\Delta \in \texttt{S}^{[0,1]^{\pm}}}^{\Delta < \Delta_{\text{gap}\pm}^{[0,1]}} |C_{12\Delta}^{[0,1]} |^2 \; \tilde{V}_{\Delta^{[0,1]^\pm}} .
\eeq

\subsection{Approximation used in \texttt{SDPB}}\label{app:Deltaapprox}

The \texttt{SDPB} algorithm works by  using a certain polynomial-based approximation for the dependence of the blocks on $\Delta$ above the gaps. 

Specifically, since we consider derivatives of the crossing equations we are interested to build approximations of the objects
\beqa
v_{n}^{\mathbf{r}}(\Delta)&\equiv &\left. \left(\frac{d}{dx} \right)^{n}V_{\Delta^{\mathbf{r}}}(x) \right|_{x = \frac{1}{2}} , \;\;\;\mathbf{r} = [0,0], [2,0], [0,2],\label{eq:derblock1}\\
\tilde{v}_{n,\pm}(\Delta)&\equiv &\left. \left(\frac{d}{dx} \right)^{n}\tilde{V}_{\Delta^{[0,1]^\pm}}(x) \right|_{x = \frac{1}{2}}  ,\label{eq:derblock2}
\eeqa
for $ n \leq \Lambda$, $n \in \mathbb{N}$. The approximation depends on a parameter called $N_{\text{poles}}$  and takes the form
\beqa
v_{n}^{\mathbf{r}}(\Delta )&\simeq &\texttt{pos}(\Delta) \times P_{n}^{\mathbf{r}}(\Delta),\\
\tilde{v}_{n,\pm}(\Delta)&\simeq &\texttt{pos}(\Delta) \times \tilde{P}_{n,\pm}(\Delta) .
\eeqa
where the positive prefactor $\texttt{pos}(\Delta)$ is defined as
\beq\label{eq:pos}
\texttt{pos}(\Delta) \equiv  \frac{(4 \rho_0 )^{\Delta}}{  \prod_{p \in \texttt{poles}}(\Delta - p )} ,
\eeq
with $\rho_0 \equiv 3 - 2 \sqrt{2}$,
\beq
\texttt{poles} \equiv \left\{ -5,-4,-2,-1,0,1,1,2,2,3,3,5,5\right\}\cup \left\{ -\frac{n+4}{2} \, \biggl|\,  n= 1,2, \dots N_{\text{poles}} + 4 \right\} ,
\eeq
and 
with $P_n(\Delta)$, $\tilde{P}_n(\Delta)$ polynomials in $\Delta$ of degree growing linearly with $N_{\text{poles}}/2$. A canonical way to derive such approximation is to first expand the full superblocks as a fastly convergent series in the reparametrisation of the cross ratio given by the variable $\rho(x) = \frac{x}{( 1 + \sqrt{1-x^2} )^2}$ (see \cite{Hogervorst:2013sma}), truncating the series to $N_{\text{poles}}$ terms, and then taking the derivatives in (\ref{eq:derblock1})-(\ref{eq:derblock2}). The precise form of the prefactor above comes from this computation. 
Such approximation is valid in a semi-infinite region $\Delta \in [\Delta_{\text{gap}}, +\infty] $. Notice that in order to find this common approximation it is important that all the blocks in the crossing equations have a common exponential growth at large $\Delta$,  i.e.  $\mathcal{F}_{\Delta} \sim \frac{(4 \rho_0 )^{\Delta}}{\Delta^k}$ for $\Delta \rightarrow +\infty$, where $k$ is a finite integer depending on the block. 

Finally, as explained in  \cite{Cavaglia:2022qpg}, with the same trick one can build an analogous approximation  also for the quantities $b_a(\Delta)$ entering the sum rules (\ref{eq:intsumrule}), which are defined as integrals of a superblock. The approximation for the integral constraints can be written with the same positive prefactor  as (\ref{eq:pos}), which allows one to treat these constraints together with the crossing equations  within the same \texttt{SDPB} setup. 

Taking a larger value of  $N_{\text{poles}}$ makes the approximation more precise, on the other hand this parameter strongly affects the running times of \texttt{SDPB}. We experimented with values of $N_{\text{poles}}=15$ and $N_{\text{poles}} = 30$, and estimated that raising  this parameter further would not affect the order of magnitude of our bounds.

\section{Bounds data}\label{app:bounds}

In this appendix we report the data obtained by the 12D run described in section \ref{sec:Results}. Each table corresponds to data obtained for a single OPE coefficient with a sampling in the coupling $g$. We provide the significant digits of the upper and lower bounds together with their error computed as explained in section \ref{sec:Algorithm}. Lastly, we include the bound thickness computed using \eqref{deltaC}.

\begin{table}[h!]
\centering
\scriptsize
\begin{tabular}{|c|c|c|c||c|c|c|c|}
\hline
\multicolumn{8}{|c|}{$\boldsymbol{C_{11\Delta_1^{[0,0]}}}$}\\
\hline
\multirow{2}{*}{$\boldsymbol{g}$} & \textbf{Lower bound} & \multirow{2}{*}{\textbf{Error}} & \multirow{2}{*}{\textbf{Thickness}} & \multirow{2}{*}{$\boldsymbol{g}$} & \textbf{Lower bound} & \multirow{2}{*}{\textbf{Error}} & \multirow{2}{*}{\textbf{Thickness}} \\
 & \textbf{Upper bound} & & & & \textbf{Upper bound} & & \\
\hline
\multirow{2}{*}{0.01} & 0.01413830433461 & \multirow{2}{*}{$3.9 \times 10^{-13}$} & \multirow{2}{*}{$2.0 \times 10^{-11}$} & \multirow{2}{*}{0.80} & 0.5198908444 & \multirow{2}{*}{$4.2 \times 10^{-9}$} & \multirow{2}{*}{$2.1 \times 10^{-7}$} \\
                      & 0.01413830437405 &                                        &                                        &                      & 0.5198912628 &                                        &                       \\
\hline
\multirow{2}{*}{0.05} & 0.07023603993    & \multirow{2}{*}{$3.4 \times 10^{-10}$} & \multirow{2}{*}{$1.7 \times 10^{-8}$}  & \multirow{2}{*}{0.90} & 0.5322954903 & \multirow{2}{*}{$2.4 \times 10^{-9}$} & \multirow{2}{*}{$1.2 \times 10^{-7}$} \\
                      & 0.07023607429    &                                        &                                        &                      & 0.5322957334 &                                        &                       \\
\hline
\multirow{2}{*}{0.10} & 0.1377295758     & \multirow{2}{*}{$4.1 \times 10^{-9}$}  & \multirow{2}{*}{$2.1 \times 10^{-7}$}  & \multirow{2}{*}{1.00} & 0.5422313021 & \multirow{2}{*}{$1.5 \times 10^{-9}$} & \multirow{2}{*}{$7.5 \times 10^{-8}$} \\
\hline
\end{tabular}
\end{table}

\begin{table}[h!]
\centering
\scriptsize
\begin{tabular}{|c|c|c|c||c|c|c|c|} \hline \multicolumn{8}{|c|}{$\boldsymbol{C_{22\Delta_1^{[0,0]}}}$} \\ \hline \multirow{2}{*}{$\boldsymbol{g}$} & \textbf{Lower bound} & \multirow{2}{*}{\textbf{Error}} & \multirow{2}{*}{\textbf{Thickness}} & \multirow{2}{*}{$\boldsymbol{g}$} & \textbf{Lower bound} & \multirow{2}{*}{\textbf{Error}} & \multirow{2}{*}{\textbf{Thickness}} \\  & \textbf{Upper bound} & & & & \textbf{Upper bound} & & \\
\hline
\multirow{2}{*}{0.01} & 0.00920          & \multirow{2}{*}{0.00012} & \multirow{2}{*}{0.0060} & \multirow{2}{*}{0.80} & 0.88662          & \multirow{2}{*}{0.00014} & \multirow{2}{*}{0.0072 }         \\
                      & 0.02116          &                          &                         &                       & 0.90106          &                          &                 \\
\hline
\multirow{2}{*}{0.05} & 0.04718          & \multirow{2}{*}{0.00058} & \multirow{2}{*}{0.0289} & \multirow{2}{*}{0.90} & 0.92599          & \multirow{2}{*}{0.00010} & \multirow{2}{*}{0.0052  }        \\
                      & 0.10501          &                          &                         &                       & 0.93633          &                          &                 \\
\hline
\multirow{2}{*}{0.10} & 0.1034           & \multirow{2}{*}{0.0010}  & \multirow{2}{*}{0.051}  & \multirow{2}{*}{1.00} & 0.957829         & \multirow{2}{*}{0.000077} & \multirow{2}{*}{0.00385 }        \\
\hline
\end{tabular}
\end{table}

\begin{table}[h!]
\scriptsize
\centering
\begin{tabular}{|c|c|c|c||c|c|c|c|} \hline \multicolumn{8}{|c|}{$\boldsymbol{C_{11\Delta_2^{[0,0]}}}$} \\ \hline \multirow{2}{*}{$\boldsymbol{g}$} & \textbf{Lower bound} & \multirow{2}{*}{\textbf{Error}} & \multirow{2}{*}{\textbf{Thickness}} & \multirow{2}{*}{$\boldsymbol{g}$} & \textbf{Lower bound} & \multirow{2}{*}{\textbf{Error}} & \multirow{2}{*}{\textbf{Thickness}} \\  & \textbf{Upper bound} & & & & \textbf{Upper bound} & & \\
\hline
\multirow{2}{*}{0.01} & 0                & \multirow{2}{*}{0.0045}  & \multirow{2}{*}{0.224}  & \multirow{2}{*}{0.80} & 0.208195         & \multirow{2}{*}{0.000050} & \multirow{2}{*}{0.00250  }       \\
                      & 0.4470           &                          &                          &                       & 0.213185         &                          &                 \\
\hline
\multirow{2}{*}{0.05} & 0                & \multirow{2}{*}{0.0012}  & \multirow{2}{*}{0.059}  & \multirow{2}{*}{0.90} & 0.202565         & \multirow{2}{*}{0.000038} & \multirow{2}{*}{0.00190  }       \\
                      & 0.3866           &                          &                          &                       & 0.206364         &                          &                 \\
\hline
\multirow{2}{*}{0.10} & 0                & \multirow{2}{*}{0.00056} & \multirow{2}{*}{0.0282} & \multirow{2}{*}{1.00} & 0.198065         & \multirow{2}{*}{0.000030} & \multirow{2}{*}{0.00149 }        \\
\hline
\end{tabular}
\end{table}

\begin{table}[h!]
\vspace{2cm}
\centering
\scriptsize
\begin{tabular}{|c|c|c|c||c|c|c|c|} \hline \multicolumn{8}{|c|}{$\boldsymbol{C_{22\Delta_2^{[0,0]}}}$} \\ \hline \multirow{2}{*}{$\boldsymbol{g}$} & \textbf{Lower bound} & \multirow{2}{*}{\textbf{Error}} & \multirow{2}{*}{\textbf{Thickness}} & \multirow{2}{*}{$\boldsymbol{g}$} & \textbf{Lower bound} & \multirow{2}{*}{\textbf{Error}} & \multirow{2}{*}{\textbf{Thickness}} \\  & \textbf{Upper bound} & & & & \textbf{Upper bound} & & \\
\hline
\multirow{2}{*}{0.01} & -0.595            & \multirow{2}{*}{0.013}  & \multirow{2}{*}{0.67}   & \multirow{2}{*}{0.80} & 0.0230           & \multirow{2}{*}{0.0065}  & \multirow{2}{*}{0.326 }          \\
                      & 0.738             &                          &                          &                       & 0.6749           &                          &                 \\
\hline
\multirow{2}{*}{0.05} & -0.305            & \multirow{2}{*}{0.010}  & \multirow{2}{*}{0.52}   & \multirow{2}{*}{0.90} & 0.0193           & \multirow{2}{*}{0.0065}  & \multirow{2}{*}{0.327  }         \\
                      & 0.733             &                          &                          &                       & 0.6727           &                          &                 \\
\hline
\multirow{2}{*}{0.10} & -0.258            & \multirow{2}{*}{0.010}  & \multirow{2}{*}{0.50}   & \multirow{2}{*}{1.00} & 0.0200           & \multirow{2}{*}{0.0065}  & \multirow{2}{*}{0.324 }          \\
\hline
\end{tabular}
\end{table}

\begin{table}[h!]
\centering
\scriptsize
\begin{tabular}{|c|c|c|c||c|c|c|c|} \hline \multicolumn{8}{|c|}{$\boldsymbol{C_{22\Delta_1^{[2,0]}}}$} \\ \hline \multirow{2}{*}{$\boldsymbol{g}$} & \textbf{Lower bound} & \multirow{2}{*}{\textbf{Error}} & \multirow{2}{*}{\textbf{Thickness}} & \multirow{2}{*}{$\boldsymbol{g}$} & \textbf{Lower bound} & \multirow{2}{*}{\textbf{Error}} & \multirow{2}{*}{\textbf{Thickness}} \\  & \textbf{Upper bound} & & & & \textbf{Upper bound} & & \\
\hline
\multirow{2}{*}{0.01} & 0                & \multirow{2}{*}{0.00015} & \multirow{2}{*}{0.0077} & \multirow{2}{*}{0.80} & 0.4811           & \multirow{2}{*}{0.0011}  & \multirow{2}{*}{0.057}           \\
                      & 0.01535          &                          &                          &                       & 0.5960           &                          &                 \\
\hline
\multirow{2}{*}{0.05} & 0                & \multirow{2}{*}{0.00077} & \multirow{2}{*}{0.0384} & \multirow{2}{*}{0.90} & 0.52503          & \multirow{2}{*}{0.00090} & \multirow{2}{*}{0.0451 }         \\
                      & 0.07678          &                          &                          &                       & 0.61521          &                          &                 \\
\hline
\multirow{2}{*}{0.10} & 0                & \multirow{2}{*}{0.0015}  & \multirow{2}{*}{0.075}  & \multirow{2}{*}{1.00} & 0.55969          & \multirow{2}{*}{0.00070} & \multirow{2}{*}{0.0351  }        \\
\hline
\end{tabular}
\end{table}

\begin{table}[h!]
\vspace{2cm}
\centering
\scriptsize
\begin{tabular}{|c|c|c|c||c|c|c|c|} \hline \multicolumn{8}{|c|}{$\boldsymbol{C_{22\Delta_2^{[2,0]}}}$} \\ \hline \multirow{2}{*}{$\boldsymbol{g}$} & \textbf{Lower bound} & \multirow{2}{*}{\textbf{Error}} & \multirow{2}{*}{\textbf{Thickness}} & \multirow{2}{*}{$\boldsymbol{g}$} & \textbf{Lower bound} & \multirow{2}{*}{\textbf{Error}} & \multirow{2}{*}{\textbf{Thickness}} \\  & \textbf{Upper bound} & & & & \textbf{Upper bound} & & \\
\hline
\multirow{2}{*}{0.01} & 0                & \multirow{2}{*}{0.00030} & \multirow{2}{*}{0.0149} & \multirow{2}{*}{0.80} & 0                 & \multirow{2}{*}{0.0074} & \multirow{2}{*}{0.372 }          \\
                      & 0.02970          &                          &                          &                       & 0.7436            &                          &                 \\
\hline
\multirow{2}{*}{0.05} & 0                & \multirow{2}{*}{0.0015}  & \multirow{2}{*}{0.073}  & \multirow{2}{*}{0.90} & 0                 & \multirow{2}{*}{0.0073} & \multirow{2}{*}{0.367  }         \\
                      & 0.1466           &                          &                          &                       & 0.7332            &                          &                 \\
\hline
\multirow{2}{*}{0.10} & 0                & \multirow{2}{*}{0.0028}  & \multirow{2}{*}{0.138}  & \multirow{2}{*}{1.00} & 0                 & \multirow{2}{*}{0.0072} & \multirow{2}{*}{0.361  }         \\
\hline
\end{tabular}
\end{table}

\begin{table}[h!]
\centering
\scriptsize
\begin{tabular}{|c|c|c|c||c|c|c|c|} \hline \multicolumn{8}{|c|}{$\boldsymbol{C_{22\Delta_1^{[0,2]}}}$} \\ \hline \multirow{2}{*}{$\boldsymbol{g}$} & \textbf{Lower bound} & \multirow{2}{*}{\textbf{Error}} & \multirow{2}{*}{\textbf{Thickness}} & \multirow{2}{*}{$\boldsymbol{g}$} & \textbf{Lower bound} & \multirow{2}{*}{\textbf{Error}} & \multirow{2}{*}{\textbf{Thickness}} \\  & \textbf{Upper bound} & & & & \textbf{Upper bound} & & \\
\hline
\multirow{2}{*}{0.01} & 0                & \multirow{2}{*}{0.00014} & \multirow{2}{*}{0.0069} & \multirow{2}{*}{0.80} & 0.73837          & \multirow{2}{*}{0.00043} & \multirow{2}{*}{0.0216}          \\
                      & 0.01381          &                          &                          &                       & 0.78157          &                          &                 \\
\hline
\multirow{2}{*}{0.05} & 0                & \multirow{2}{*}{0.00069} & \multirow{2}{*}{0.0345} & \multirow{2}{*}{0.90} & 0.79783          & \multirow{2}{*}{0.00032} & \multirow{2}{*}{0.0161}          \\
                      & 0.06892          &                          &                          &                       & 0.83002          &                          &                 \\
\hline
\multirow{2}{*}{0.10} & 0                & \multirow{2}{*}{0.0014}  & \multirow{2}{*}{0.068}  & \multirow{2}{*}{1.00} & 0.84655          & \multirow{2}{*}{0.00025} & \multirow{2}{*}{0.0124  }        \\
\hline
\end{tabular}
\end{table}

\begin{table}[h!]
\vspace{2cm}
\centering
\scriptsize
\begin{tabular}{|c|c|c|c||c|c|c|c|} \hline \multicolumn{8}{|c|}{$\boldsymbol{C_{22\Delta_2^{[0,2]}}}$} \\ \hline \multirow{2}{*}{$\boldsymbol{g}$} & \textbf{Lower bound} & \multirow{2}{*}{\textbf{Error}} & \multirow{2}{*}{\textbf{Thickness}} & \multirow{2}{*}{$\boldsymbol{g}$} & \textbf{Lower bound} & \multirow{2}{*}{\textbf{Error}} & \multirow{2}{*}{\textbf{Thickness}} \\  & \textbf{Upper bound} & & & & \textbf{Upper bound} & & \\
\hline
\multirow{2}{*}{0.01} & 0                & \multirow{2}{*}{0.00033} & \multirow{2}{*}{0.0167} & \multirow{2}{*}{0.80} & 0                & \multirow{2}{*}{0.011}  & \multirow{2}{*}{0.534 }           \\
                      & 0.03335          &                          &                          &                       & 1.067             &                          &                 \\
\hline
\multirow{2}{*}{0.05} & 0                & \multirow{2}{*}{0.0017}  & \multirow{2}{*}{0.083}  & \multirow{2}{*}{0.90} & 0                & \multirow{2}{*}{0.011}  & \multirow{2}{*}{0.542}            \\
                      & 0.1669           &                          &                          &                       & 1.084             &                          &                 \\
\hline
\multirow{2}{*}{0.10} & 0                & \multirow{2}{*}{0.0033}  & \multirow{2}{*}{0.166}  & \multirow{2}{*}{1.00} & 0                & \multirow{2}{*}{0.011}  & \multirow{2}{*}{0.549 }           \\
\hline
\end{tabular}
\end{table}

\begin{table}[h!]
\centering
\scriptsize
\begin{tabular}{|c|c|c|c||c|c|c|c|} \hline \multicolumn{8}{|c|}{$\boldsymbol{C^2_{12\Delta_1^{[0,1]^{-}}}}$} \\ \hline \multirow{2}{*}{$\boldsymbol{g}$} & \textbf{Lower bound} & \multirow{2}{*}{\textbf{Error}} & \multirow{2}{*}{\textbf{Thickness}} & \multirow{2}{*}{$\boldsymbol{g}$} & \textbf{Lower bound} & \multirow{2}{*}{\textbf{Error}} & \multirow{2}{*}{\textbf{Thickness}} \\  & \textbf{Upper bound} & & & & \textbf{Upper bound} & & \\
\hline
\multirow{2}{*}{0.01} & 0.00006931       & \multirow{2}{*}{5.3$\times 10^{-7}$} & \multirow{2}{*}{0.0000267}  & \multirow{2}{*}{0.80} & 0.287574         & \multirow{2}{*}{0.000060}  & \multirow{2}{*}{0.00298}         \\
                      & 0.00012268       &                                    &                             &                       & 0.293530         &                          &                 \\
\hline
\multirow{2}{*}{0.05} & 0.001760         & \multirow{2}{*}{0.000013}           & \multirow{2}{*}{0.00065}    & \multirow{2}{*}{0.90} & 0.313386         & \multirow{2}{*}{0.000046}  & \multirow{2}{*}{0.00230 }        \\
                      & 0.003068         &                                    &                             &                       & 0.317990         &                          &                 \\
\hline
\multirow{2}{*}{0.10} & 0.007404         & \multirow{2}{*}{0.000047}           & \multirow{2}{*}{0.00235}    & \multirow{2}{*}{1.00} & 0.335111         & \multirow{2}{*}{0.000036}  & \multirow{2}{*}{0.00182 }        \\
\hline
\end{tabular}
\end{table}

\begin{table}[h!]
\vspace{2cm}
\centering
\scriptsize
\begin{tabular}{|c|c|c|c||c|c|c|c|} \hline \multicolumn{8}{|c|}{$\boldsymbol{C^2_{12\Delta_2^{[0,1]^{-}}}}$} \\ \hline \multirow{2}{*}{$\boldsymbol{g}$} & \textbf{Lower bound} & \multirow{2}{*}{\textbf{Error}} & \multirow{2}{*}{\textbf{Thickness}} & \multirow{2}{*}{$\boldsymbol{g}$} & \textbf{Lower bound} & \multirow{2}{*}{\textbf{Error}} & \multirow{2}{*}{\textbf{Thickness}} \\  & \textbf{Upper bound} & & & & \textbf{Upper bound} & & \\
\hline
\multirow{2}{*}{0.01} & 0                & \multirow{2}{*}{0.0038}  & \multirow{2}{*}{0.189}  & \multirow{2}{*}{0.80}  & 0                & \multirow{2}{*}{0.0029} & \multirow{2}{*}{0.145}           \\
                      & 0.3786           &                          &                          &                       & 0.2908           &                          &                 \\
\hline
\multirow{2}{*}{0.05} & 0                & \multirow{2}{*}{0.0038}  & \multirow{2}{*}{0.189}  & \multirow{2}{*}{0.90}  & 0                & \multirow{2}{*}{0.0029} & \multirow{2}{*}{0.145 }          \\
                      & 0.3784           &                          &                          &                       & 0.2908           &                          &                 \\
\hline
\multirow{2}{*}{0.10} & 0                & \multirow{2}{*}{0.0037}  & \multirow{2}{*}{0.187}  & \multirow{2}{*}{1.00}  & 0                & \multirow{2}{*}{0.0029} & \multirow{2}{*}{0.145  }         \\
\hline
\end{tabular}
\end{table}

\begin{table}[h!]
\scriptsize
\centering
\begin{tabular}{|c|c|c|c||c|c|c|c|} \hline \multicolumn{8}{|c|}{$\boldsymbol{C^2_{12\Delta_1^{[0,1]^{+}}}}$} \\ \hline \multirow{2}{*}{$\boldsymbol{g}$} & \textbf{Lower bound} & \multirow{2}{*}{\textbf{Error}} & \multirow{2}{*}{\textbf{Thickness}} & \multirow{2}{*}{$\boldsymbol{g}$} & \textbf{Lower bound} & \multirow{2}{*}{\textbf{Error}} & \multirow{2}{*}{\textbf{Thickness}} \\  & \textbf{Upper bound} & & & & \textbf{Upper bound} & & \\
\hline
\multirow{2}{*}{0.01} &   -0.0001695       & \multirow{2}{*}{1.6$\times$10$^{-6}$} & \multirow{2}{*}{0.000079} & \multirow{2}{*}{0.80}  &  -0.13811          & \multirow{2}{*}{0.00015} &  \multirow{2}{*}{0.0077}        \\
                      &-0.0000113            &                                  &                          &                       &-0.12269          &                          &                 \\
\hline
\multirow{2}{*}{0.05} &    -0.004163         & \multirow{2}{*}{0.000039}         & \multirow{2}{*}{0.00193} & \multirow{2}{*}{0.90}  &   -0.14355         & \multirow{2}{*}{0.00012} & \multirow{2}{*}{0.0059}          \\
                      & -0.000296          &                                  &                          &                       &-0.13179          &                          &                 \\
\hline
\multirow{2}{*}{0.10} &    -0.01520        & \multirow{2}{*}{0.00014}          & \multirow{2}{*}{0.0069}  & \multirow{2}{*}{1.00}  &   -0.148126      & \multirow{2}{*}{0.000092} & \multirow{2}{*}{0.00460  }       \\
\hline
\end{tabular}
\end{table}

\begin{table}[h!]
\vspace{2cm}
\centering
\scriptsize
\begin{tabular}{|c|c|c|c||c|c|c|c|} \hline \multicolumn{8}{|c|}{$\boldsymbol{C^2_{12\Delta_2^{[0,1]^{+}}}}$} \\ \hline \multirow{2}{*}{$\boldsymbol{g}$} & \textbf{Lower bound} & \multirow{2}{*}{\textbf{Error}} & \multirow{2}{*}{\textbf{Thickness}} & \multirow{2}{*}{$\boldsymbol{g}$} & \textbf{Lower bound} & \multirow{2}{*}{\textbf{Error}} & \multirow{2}{*}{\textbf{Thickness}} \\  & \textbf{Upper bound} & & & & \textbf{Upper bound} & & \\
\hline
\multirow{2}{*}{0.01} & -0.2513                & \multirow{2}{*}{0.0025}  & \multirow{2}{*}{0.126}  & \multirow{2}{*}{0.80} & -0.1303                 & \multirow{2}{*}{0.0013}  & \multirow{2}{*}{0.065}           \\
                      & 0           &                          &                          &                       & 0          &                          &                 \\
\hline
\multirow{2}{*}{0.05} & -0.2507                & \multirow{2}{*}{0.0025}  & \multirow{2}{*}{0.125}  & \multirow{2}{*}{0.90} & -0.1252               & \multirow{2}{*}{0.0013}  & \multirow{2}{*}{0.063 }          \\
                      & 0          &                          &                          &                       & 0          &                          &                 \\
\hline
\multirow{2}{*}{0.10} & -0.2468                 & \multirow{2}{*}{0.0025}  & \multirow{2}{*}{0.123}  & \multirow{2}{*}{1.00} & -0.1214               & \multirow{2}{*}{0.0012}  & \multirow{2}{*}{0.061 }          \\
\hline
\end{tabular}
\end{table}

\bibliographystyle{JHEP.bst}
\bibliography{references}

\end{document}